\documentclass[pdftex,twocolumn,3]{jour3}          
\usepackage{comment}
\usepackage[utf8]{inputenc}
\usepackage[T1]{fontenc}
\usepackage[english]{babel}
\usepackage{epsf}
\usepackage{dirtytalk}
\usepackage{latexsym,amssymb,euscript}
\usepackage{widetext}
\usepackage[dvips]{graphicx}
\usepackage[numbers,sort&compress]{natbib}
\usepackage{amsmath}
\usepackage{nicefrac}
\usepackage{slashed}
\usepackage{booktabs}
\usepackage[linktocpage]{hyperref}
\usepackage{braket}
\usepackage{chngcntr}
\usepackage{bm}
\usepackage{bbold}
\usepackage{multicol}
\usepackage{graphics}
\usepackage{graphicx}
\usepackage{mciteplus}
\usepackage{caption}
\usepackage{subcaption}
\usepackage{pdfpages}
\usepackage[titletoc]{appendix}
\usepackage[normalem]{ulem}
\usepackage{microtype} 

\graphicspath{{./figures/}}
\hypersetup{
 linktocpage = false,
 urlcolor = linkred,
 colorlinks = true,
 linkcolor = urlblue,
 anchorcolor = citegreen,
 citecolor = citegreen,
 pdfstartview = {XYZ null null 1.25} 
           }
\usepackage[left=2cm, right=2cm]{geometry}
\usepackage{pstricks}
\usepackage{color}
\usepackage{xcolor}
\definecolor{urlblue}{rgb}{0.2,0.4,0.7}
\definecolor{citegreen}{rgb}{0,0.4,0.2}
\definecolor{linkred}{rgb}{0.9,0.2,0.1}
\usepackage{float}
\usepackage{academicons}
\definecolor{orcidlogocol}{HTML}{A6CE39}
\usepackage{fancyhdr}
\newcommand{\drv}{{\rm d}}

\newcommand{\LL}{{\rm LL/LO}}
\newcommand{\NLL}{{\rm NLL/NLO}}
\newcommand{\NLLp}{{\rm NLL/NLO^+}}
\newcommand{\NLLpp}{{\rm NLL/NLO^{(+)}}}
\newcommand{\HENLOp}{{\rm HE}\mbox{-}{\rm NLO^+}}

\newcommand{\CnNLLp}{{\cal C}_n^\NLLp}

\newcommand{\DY}{\Delta Y}

\newcommand{\E}{{\cal E}}

\newcommand{{\HFNRevo}}{\textsc{HF-NRevo}}

\newcommand{{\Jethad}}{\tt JETHAD}
\newcommand{{\symJethad}}{\tt symJETHAD}
\newcommand{{\bsymJethad}}{\tt (sym)JETHAD}
\newcommand{{\Hell}}{\tt HELL}
\newcommand{{\RadISH}}{\tt RadISH}
\newcommand{{\Pegasus}}{\tt QCD-PEGASUS}
\newcommand{{\HOPPET}}{\tt HOPPET}
\newcommand{{\QCDNUM}}{\tt QCDNUM}
\newcommand{{\APFEL}}{\tt APFEL}
\newcommand{{\APFELpp}}{\tt APFEL++}
\newcommand{{\APFELppp}}{\tt APFEL(++)}
\newcommand{{\EKO}}{\tt EKO}

\newcommand{\LHAPDF}{{\tt LHAPDF}}
\newcommand{{\HCFF}}{{\tt HCFF1.0}}
\newcommand{{\NRFF}}{{\tt NRFF1.0}}

\newcommand{\orcidFGC}{\href{https://orcid.org/0000-0002-2123-5883}{\includegraphics[scale=0.1]{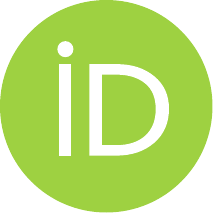}}}

\newcommand{\orcidFL}{\href{https://orcid.org/0009-0003-2052-4549}{\includegraphics[scale=0.1]{logo-orcid.pdf}}}

\smartqed  

\journalname{}

\begin{document}

\normalem
\title{The NLL/NLO$^+$ QCD spectrum from Higgs-plus-charm \\[0.15cm] correlations at the LHC 
}

\subtitle{}

\author{
Francesco Giovanni Celiberto
\and
\thanksref{e1,addr1} \orcidFGC
\; 
Francesca Lonigro
\thanksref{e2,addr1} \orcidFL
}

\thankstext{e1}{{\it e-mail}:
\href{mailto:francesco.celiberto@uah.es}{francesco.celiberto@uah.es} (corresponding author)}
\thankstext{e2}{{\it e-mail}:
\href{mailto:francesca.lonigro@uah.es}{lfrancesca.lonigro@uah.es}}

\institute{Universidad de Alcal\'a (UAH), E-28805 Alcal\'a de Henares, Madrid, Spain\label{addr1}
}

\date{\today}

\maketitle


\section*{Abstract}
We investigate the high-energy behavior of observables sensitive to the production of a Higgs boson in association with a singly charmed hadron at the $\sqrt{s}=14$ TeV LHC, with light-flavored hadrons considered for comparison. 
Our reference formalism is the hybrid factorization at NLL/NLO$^+$, where the standard collinear picture is enhanced by the resummation of energy logarithms at the next-to-leading level and beyond. 
We present a prime phenomenological application of the next-to-leading correction to the singly off-shell Higgs emission function, as implemented in the multimodular {\tt JETHAD} interface, a core element to test the stability of our high-energy distributions under energy-scale variations and to assess the weight of missing higher-order uncertainties.
The resulting natural stability traces our path towards precision studies of Higgs production within high-energy next-to-leading-order QCD.
\vspace{0.30cm} \hrule
\vspace{0.30cm}
{
 \setlength{\parindent}{0pt}
 \textsc{Keywords}: \vspace{0.00cm} \\ 
 Higgs, Heavy~flavor, LHC,  Precision~QCD, \\ High-energy resummation
}
\vspace{1.75cm}
\vspace{-0.15cm}

\setcounter{tocdepth}{3}
\renewcommand{\baselinestretch}{1.0}\normalsize
\tableofcontents
\renewcommand{\baselinestretch}{1.0}\normalsize
\clearpage


\section{Hors d'{\oe}uvre}
\label{sec:intro}

Since its discovery at the LHC~\cite{ATLAS:2012yve,CMS:2012qbp}, the Higgs boson has become a central laboratory for precision tests of the Standard Model.
On the theory side, this program has been accompanied by a continuous refinement of perturbative predictions for Higgs production and related precision observables~\cite{Bargiela:2022dla}, with gluon-fusion cross sections now known to very high fixed-order accuracy in the large-top-mass effective theory~\cite{Anastasiou:2015vya,Anastasiou:2016cez,Mistlberger:2018etf}, together with substantial progress in the treatment of finite heavy-quark mass effects~\cite{Spira:1995rr,Harlander:2009my,Harlander:2009mq,Davies:2019wmk,Czakon:2021yub}.

Fixed-order collinear factorization does not provide a uniform description of all corners of phase space.
Whenever kinematics generates logarithmic contributions capable of compensating the smallness of the strong coupling, an all-order treatment becomes necessary.
At low transverse momentum, Sudakov logarithms are organized through transverse-momentum resummation~\cite{Catani:2000vq,Bozzi:2005wk,Catani:2010pd,Catani:2013tia,Duhr:2022yyp}, whereas close to partonic threshold soft and collinear radiation gives rise to large-$x$ logarithms resummed within threshold approaches~\cite{Sterman:1986aj,Catani:1996yz,Bonciani:2003nt,deFlorian:2005fzc,Forte:2021wxe}.
A complementary limit is reached when the center-of-mass energy $\sqrt{s}$ becomes much larger than the characteristic transverse scales of the reaction.
In this \emph{semi-hard} or Regge--Gribov regime~\cite{Gribov:1983ivg,Celiberto:2017ius},
$s \gg \{Q\}^2 \gg \Lambda_{\rm QCD}^2$,
where $\{Q\}$ collectively denotes one or more characteristic hard scales set by the process kinematics.
Powers of logarithms of the ratio between the center-of-mass energy and the characteristic hard scales become increasingly enhanced along the perturbative series and must be resummed to all orders.

The complementarity of different logarithmic regimes has already been extensively explored in precision Higgs phenomenology.
Threshold-enhanced contributions have been studied both at the level of the gluon-fusion hard cross section and of the corresponding parton distribution functions (PDFs)~\cite{Bonvini:2012an,Bonvini:2014joa,Bonvini:2015ira,Bonvini:2016frm}, while small-$x$ resummation has been developed within approaches rooted in the Altarelli--Ball--Forte framework~\cite{Ball:1997vf,Altarelli:2005ni}.
This program has led to PDF determinations supplemented by small-$x$ resummation~\cite{Ball:2017otu}, and has recently been reinforced by a new formulation of resummed DGLAP splitting functions~\cite{Bonvini:2026cxp}.
Threshold and high-energy logarithms were further combined and matched to state-of-the-art fixed-order Higgs predictions~\cite{Ball:2013bra,Bonvini:2018ixe}.
While the net resummation effect remains at the few-percent level at current LHC energies, the high-energy component becomes increasingly important with increasing collision energy and was found to dominate the resummed correction at prospective FCC energies~\cite{Bonvini:2018ixe,Mangano:2016jyj,FCC:2025lpp,FCC:2025uan,FCC:2025jtd}.
The same small-$x$ resummation program has also been extended to differential observables, including heavy-quark pair production~\cite{Silvetti:2022hyc}.

The natural framework for this resummation is the Balitsky--Fadin--Kuraev--Lipatov (BFKL) approach~\cite{Fadin:1975cb,Kuraev:1976ge,Kuraev:1977fs}.
High-energy cross sections take a factorized form in transverse-momentum space, built from a universal Green's function, which governs the evolution across the large rapidity interval, and process-dependent impact factors associated with the tagged final states.
The latter can be regarded as singly off-shell hard coefficient functions, describing the interaction of an incoming collinear parton with an off-shell Reggeized gluon exchanged in the $t$ channel.
For hadron-initiated reactions, these hard coefficients are further convoluted in the longitudinal momentum fractions with collinear inputs associated with the parent hadrons and with the observed final states: PDFs for the incoming protons and, whenever an identified hadron is detected, fragmentation functions (FFs) describing its formation.
We refer to the resulting process-dependent objects as emission functions.
This interplay between transverse-momentum high-energy factorization and longitudinal collinear factorization defines the hybrid high-energy and collinear framework used throughout this work.

A genuine next-to-leading-logarithmic (NLL) description requires the NLL Green's function to be consistently combined with next-to-leading-order (NLO) impact factors.
Historically, this requirement has represented one of the main bottlenecks of high-energy phenomenology, since only a restricted class of process-dependent impact factors is known at NLO.
This limitation motivated several pure leading-logarithmic (LL) and partial-NLL studies, in which the (N)LL Green's function was combined with one or both emission vertices evaluated at leading order (LO), possibly supplemented by renormalization-group dictated terms.

Higgs production has played a distinctive role in this program.
The first high-energy resummed study of Higgs-plus-jet correlations at the LHC~\cite{Celiberto:2020tmb,Mohammed:2022gbk} showed that the large Higgs transverse mass substantially mitigates the perturbative instabilities encountered in light-jet and light-hadron BFKL observables.
It also identified Higgs transverse-momentum spectra as particularly sensitive probes of the separation between all-order high-energy dynamics and its fixed-order expansion.
A complementary stabilization mechanism emerged from reactions involving identified heavy-flavored hadrons, where the scale dependence of heavy-flavor FFs was found to reduce the sensitivity of resummed predictions to higher-order corrections and scale variations~\cite{Celiberto:2021dzy,Celiberto:2021fdp,Celiberto:2022rfj}.

These two mechanisms were brought together in the first study of Higgs production in association with a charmed hadron~\cite{Celiberto:2022zdg}.
There, the hard scale supplied by Higgs production and the collinear dynamics associated with heavy-flavor fragmentation generated a remarkably stable perturbative pattern, referred to as \emph{natural stability}~\cite{Celiberto:2022grc}.
This suggested Higgs-plus-heavy-flavor reactions as a favorable environment for precision-oriented applications of high-energy resummation.

The Higgs-charm connection also carries a broader phenomenological interest.
Direct information on the charm Yukawa coupling remains experimentally challenging, while increasingly stringent constraints are being obtained from searches for $H\to c\bar c$~\cite{ATLAS:2025pxp}.
At the same time, sizeable departures from the Standard-Model charm Yukawa pattern can arise in extended Higgs sectors while remaining compatible with stringent flavor constraints~\cite{Giannakopoulou:2024unn,Kopp:2014rva,DiLuzio:2023ndz}.
In particular, suitable flavor-protection mechanisms in two-Higgs-doublet realizations can accommodate enhanced charm Yukawa couplings while suppressing otherwise dangerous flavor-changing neutral currents~\cite{Giannakopoulou:2024unn}.
Associated Higgs-plus-charm production therefore provides a complementary handle on the Higgs coupling to charm~\cite{Brivio:2015fxa}, although the observables considered here are primarily designed to probe high-energy QCD dynamics rather than to extract the charm Yukawa coupling.

A decisive theoretical development has been the progressive construction of the forward-Higgs impact factor beyond leading order.
Its one-loop virtual correction was first determined in the large-top-mass effective theory~\cite{Nefedov:2019mrg}.
The complete NLO result in the same approximation was subsequently obtained within Lipatov's high-energy effective action~\cite{Hentschinski:2020tbi}, and an independent calculation in the standard BFKL framework provided the full NLO impact factor together with its Mellin projection onto the LO eigenfunctions of the BFKL kernel~\cite{Celiberto:2022fgx}.
The latter representation is directly suited for resummed phenomenology.

The first finite-$m_t$ extension followed with the calculation of the real component of the NLO Higgs impact factor at physical top mass~\cite{Celiberto:2024bbv}.
This step is phenomenologically important because finite-top-mass effects become increasingly relevant when the Higgs transverse momentum approaches the top-quark mass, while also providing a nontrivial test of high-energy factorization beyond the dimension-five Higgs-gluon effective interaction.
More recently, the two-loop Higgs impact factor was extracted in the large-top-mass effective theory from the Regge limit of Higgs-plus-jet amplitudes~\cite{DelDuca:2025vux}.
This result represents a first virtual ingredient toward a beyond-NLO description of forward-Higgs production, whereas a complete inclusive next-to-NLO emission function would additionally require the corresponding real-virtual and double-real sectors.

The present study brings this chain of developments to phenomenology.
We investigate the inclusive production of a Higgs boson in association with a singly charmed hadron at the $\sqrt{s}=14$~TeV LHC within hybrid factorization at $\NLLp$ accuracy.
For the first time in this class of Higgs-plus-heavy-flavor analyses, the complete NLO correction to the singly off-shell Higgs emission function is combined with the NLO identified-hadron emission vertex and the NLL BFKL Green's function.
Predictions are generated through the multimodular {\Jethad} interface~\cite{Celiberto:2020wpk,Celiberto:2022rfj,Celiberto:2023fzz,Celiberto:2024mrq,Celiberto:2024swu,Celiberto:2025csa,Celiberto:2026ooh}, with light-flavored hadrons included as a comparison channel.

Our analysis addresses three main questions.
We first test whether the natural-stability pattern identified in partial-NLL studies persists once the complete next-to-leading Higgs emission dynamics is included.
We then quantify the sensitivity of rapidity, azimuthal, and transverse-momentum observables to high-energy resummation and hard-factor missing-higher-order uncertainties.
Finally, by extending the Higgs transverse-momentum spectrum up to scales of the order of the top-quark mass, we assess finite heavy-quark effects and Born-improvement prescriptions.
The transverse-momentum spectrum is especially informative in this respect, since it probes both the interplay between resummed and high-energy fixed-order dynamics and the mass structure of the Higgs-production vertex.
We complement this analysis with a systematic assessment of the uncertainty sources entering our predictions, distinguishing the contributions propagated to the final error bands from complementary theory variations used as dedicated diagnostics.

Systematic matching of high-energy resummation with fixed-order calculations~\cite{Celiberto:2023uuk,Celiberto:2023eba,Celiberto:2023nym} represents the natural next step toward precision applications to Higgs- and electroweak-boson-associated final states.

The article is organized as follows.
In Section~\ref{sec:hybrid_factorization} we introduce the hybrid-factorization framework, discuss its high-energy and collinear ingredients, and specify our treatment of theoretical uncertainties.
Section~\ref{sec:pheno} presents predictions for rapidity-interval rates, azimuthal-correlation moments, normalized azimuthal distributions, and Higgs transverse-momentum spectra.
Our conclusions and perspectives are collected in Section~\ref{sec:conclusions}.

\section{Hybrid factorization at NLL/NLO$^+$}
\label{sec:hybrid_factorization}

Hybrid high-energy and collinear factorization has progressively developed into a versatile framework for the description of semi-inclusive reactions characterized by two tagged objects separated by a sizeable rapidity interval.
Its phenomenological applications at $\NLLpp$ accuracy span Mueller--Navelet jets~\cite{Mueller:1986ey,Ducloue:2013hia,Ducloue:2013bva,Caporale:2014gpa,Colferai:2015zfa,Celiberto:2015yba,Celiberto:2016ygs,Celiberto:2017ius,Caporale:2018qnm,Celiberto:2022gji}, dihadron~\cite{Celiberto:2016hae,Celiberto:2017ptm,Celiberto:2017ius,Celiberto:2020rxb,Celiberto:2022rfj} and hadron-jet production~\cite{Bolognino:2018oth,Celiberto:2020wpk,Celiberto:2020rxb,Mohammed:2022gbk,Celiberto:2022kxx}, as well as heavy-flavor channels~\cite{Celiberto:2021dzy,Celiberto:2021fdp,Celiberto:2022dyf,Celiberto:2022keu,Celiberto:2024omj,Celiberto:2025euy}.
Notable phenomenological studies have also been carried out at partial-NLL accuracy for multijet final states~\cite{Caporale:2016zkc} and Drell--Yan reactions~\cite{Celiberto:2018muu,Golec-Biernat:2018kem}.
This broad range of reactions has made the hybrid formalism a useful laboratory for testing high-energy resummation across markedly different hard scales and final-state configurations.

\begin{figure}[!t]
\centering
\includegraphics[width=0.475\textwidth]{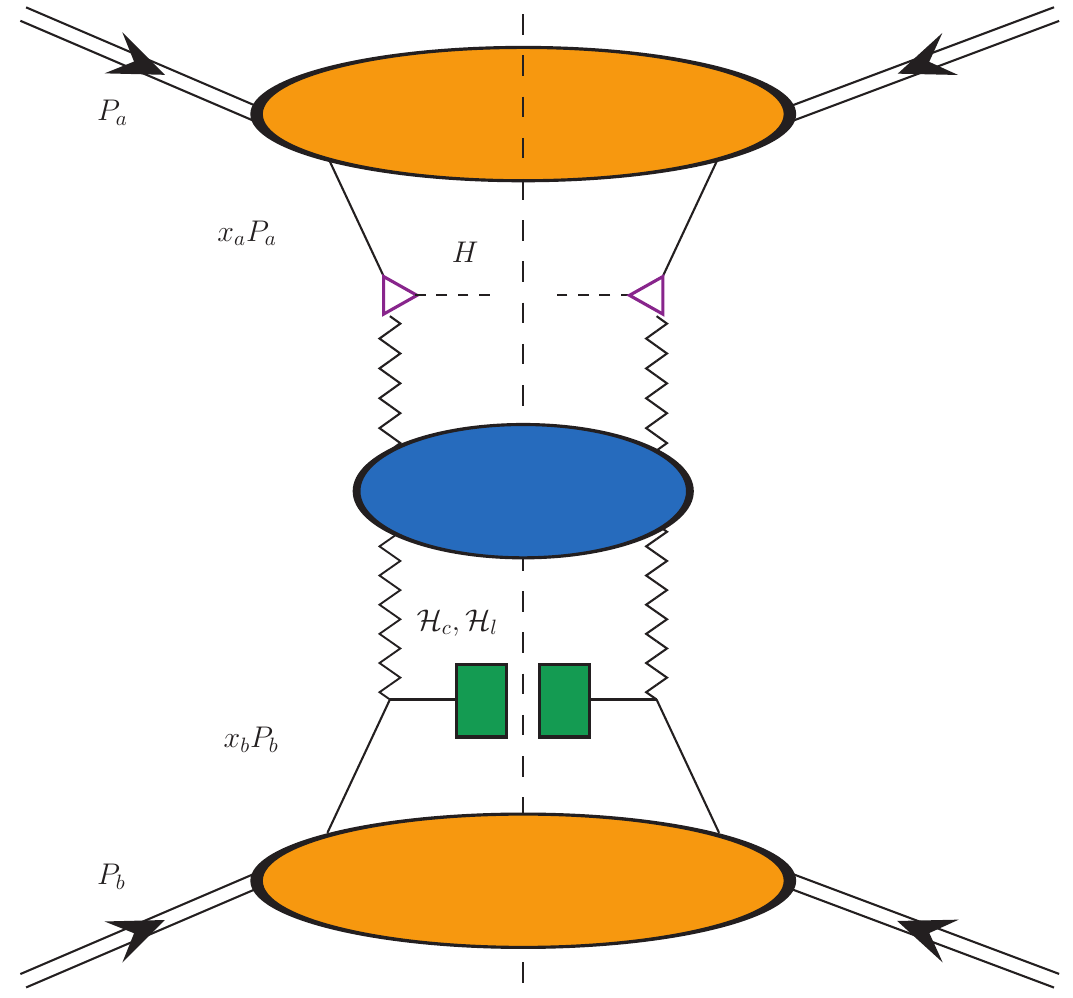}

\caption{Diagrammatic representation of the hybrid high-energy and collinear factorization used for semi-inclusive Higgs-plus-hadron production at hadron colliders.
Orange blobs account for the collinear parton densities of the incoming protons, while the purple triangles denote the heavy-quark fermion loops mediating Higgs production.
The green blocks encode the collinear FFs associated with the detected charmed or light-flavored hadron, ${\cal H}_c$ or ${\cal H}_l$.
The two tagged-particle emission sectors are linked by the universal BFKL Green's function, depicted by the central blue blob, through Reggeized-gluon exchanges.
The diagram was generated with {\tt JaxoDraw~2.0}~\cite{Binosi:2008ig}.}
\label{fig:reaction}
\end{figure}

Higgs and heavy-flavor emissions are particularly relevant in this context because they provide two distinct sources of perturbative stabilization.
In Higgs-associated reactions, the comparatively large transverse mass of the scalar boson sets a naturally hard scale and was found to substantially reduce the sensitivity of high-energy observables to higher-order corrections and scale variations~\cite{Celiberto:2020tmb,Celiberto:2022zdg}.
Heavy-flavored hadrons provide a complementary mechanism: within a zero-mass variable-flavor number scheme (ZM-VFNS)~\cite{Mele:1990cw,Cacciari:1993mq,Buza:1996wv}, the scale evolution of their FF can compensate part of the residual scale dependence of the hard subprocess, leading to the pattern known as \emph{natural stability}~\cite{Celiberto:2021dzy,Celiberto:2021fdp,Celiberto:2022grc}.
The simultaneous presence of these two mechanisms makes Higgs-plus-charmed-hadron production especially well suited for precision-oriented studies of high-energy resummation.
A more detailed discussion of their interplay and of the origin of the resulting stabilization is deferred to Sec.~\ref{ssec:natural_stability}.

The remainder of this section is organized as follows.
Section~\ref{ssec:NLL_cross_section} introduces the $\NLLp$ resummed cross section, while Secs.~\ref{ssec:HE_ingredients} and~\ref{ssec:collinear_ingredients} describe, respectively, the high-energy and collinear ingredients entering the calculation.
The mechanism behind the natural stabilization of the perturbative series is discussed in Sec.~\ref{ssec:natural_stability}.
Finally, Sec.~\ref{ssec:uncertainty} details the strategy adopted to quantify theoretical and numerical uncertainties.

\subsection{Resummed cross section}
\label{ssec:NLL_cross_section}

The semi-inclusive process considered throughout this work reads
\begin{equation*}
\label{reaction}
{\rm p}(P_a) + {\rm p}(P_b) \rightarrow H(p_H, y_H, \varphi_H) + {\cal X} + {\cal H}_{c,l}(p_h, y_h, \varphi_h) \;,
\end{equation*}
where ${\rm p}(P_{a,b})$ identifies an incoming proton carrying momentum $P_{a,b}$, while $H(p_H, y_H, \varphi_H)$ denotes the detected Higgs boson with four-momentum $p_H$, rapidity $y_H$, and azimuthal angle $\varphi_H$.
The second tagged state, ${\cal H}_{c,l}(p_h, y_h, \varphi_h)$, stands either for a singly charmed hadron, ${\cal H}_c$, or, for comparison, for a light-flavored hadron, ${\cal H}_l$, characterized by momentum $p_h$, rapidity $y_h$, and azimuthal angle $\varphi_h$, whereas ${\cal X}$ collectively denotes the unobserved hadronic radiation.
By ${\cal H}_c$, we mean an inclusive singly charmed hadron state obtained by summing over fragmentation channels into $D^\pm$, $D^0$, and $D^{*\pm}$ mesons, together with $\Lambda_c^\pm$ baryons.
Likewise, ${\cal H}_l$ refers to the inclusive sum over charged-conjugate pion, kaon, and proton species.
The relative azimuthal variable used in our analysis is defined as $\varphi = \varphi_H - \varphi_h - \pi$.

The semi-hard regime relevant for high-energy factorization requires both tagged objects to carry perturbatively large transverse momenta together with a sizeable rapidity separation, $\DY = y_H - y_h$.
The transverse-momentum cuts adopted in our phenomenology are also sufficiently hard to support the ZM-VFNS treatment of singly charmed hadrons through collinear fragmentation.

The incoming proton momenta are chosen as lightlike Sudakov vectors, with $P_a^2 = P_b^2 = 0$ and $2 (P_a \cdot P_b) = s$.
The four-momenta of the tagged final-state objects can accordingly be written as
\begin{equation}
\begin{split}
\label{Sudakov}
 p_H \,&=\, x_H \, P_a + \frac{m_{T,H}^2}{x_H s} \, P_b + p_{H \perp} \;,
 \\
 p_h \,&=\, x_h \, P_b - \frac{p_{h \perp}^2}{x_h s} \, P_a + p_{h \perp} \;,
\end{split}
\end{equation}
where $p_{H,h \perp}^2 \equiv - \vec p_{T_{H,h}}^{\,2}$, $m_H$ denotes the Higgs-boson mass, taken as $m_H = 125.08$~GeV~\cite{ParticleDataGroup:2024cfk}, and $m_{T,H}^2 \equiv m_H^2 + |\vec p_{T_H}|^2$.
The longitudinal fractions $x_H$ and $x_h$ are related to the corresponding rapidities through
$y_H = \ln(x_H \sqrt{s}/m_{T,H})$ and
$y_h = - \ln(x_h \sqrt{s}/|\vec p_{T_h}|)$.
Consequently, $\drv y_H = \drv x_H/x_H$ and $\drv y_h = - \drv x_h/x_h$, while the rapidity interval becomes
\begin{equation}
\label{Delta_Y}
 \DY = y_H - y_h \equiv \ln \frac{x_H x_h s}{m_{T,H} |\vec p_{T_h}|} \;.
\end{equation}

Within hybrid factorization, the high-energy evolution and the collinear description of the external states are consistently combined at the level of the differential cross section.
Its dependence on the relative azimuthal angle is conveniently organized through a Fourier decomposition in terms of the coefficients ${\cal C}_{n \, \ge \, 0}$,
\begin{equation}
\label{dsigma_Fourier}
\hspace{-0.22cm}
\frac{(2\pi)^2 \, \drv \sigma}{\drv y_H \drv y_h \drv |\vec p_{T_H}| \drv |\vec p_{T_h}| \drv \varphi_H \drv \varphi_h} =
\left[ {\cal C}_0 + 2 \sum_{n=1}^\infty \cos (n \varphi) \,
{\cal C}_n \right] \;,
\end{equation}
where $\varphi \equiv \varphi_H - \varphi_h - \pi$.
Working in the $\overline{\text{MS}}$ renormalization scheme~\cite{PhysRevD.18.3998}, the azimuthal coefficients at $\NLLp$ accuracy take the form
\begin{equation}
\label{Cn_NLLp_MSb}
\CnNLLp =
\frac{e^{\DY}}{s}
\int_{-\infty}^{+\infty} \drv \nu \,
{\cal G}_n(\DY,\nu;\mu_R)
\end{equation}
\[
\times \, \alpha_s^2(\mu_R) \biggl\{
\E_H^{\rm NLO}(n,\nu,|\vec p_{T_H}|,x_H)
[\E_h^{\rm NLO}(n,\nu,|\vec p_{T_h}|,x_h)]^*
\]
\[
+ \!
\left.
\alpha_s^2(\mu_R) \DY \frac{\beta_0}{4 \pi}
\chi(n,\nu) \!
\left[
\ln\left(|\vec p_{T_H}| |\vec p_{T_h}|\right)
\!+\!
\frac{i}{2} \frac{\drv}{\drv \nu}
\ln\frac{\E_H}{\E_h^*}
\right]
\right\}
.
\]

Here, ${\cal G}_n(\DY,\nu;\mu_R)$ denotes the BFKL Green's function in the $(n,\nu)$ representation, where $n$ labels the azimuthal harmonic and $\nu$ is the Mellin variable conjugate to transverse-momentum logarithms.
In this representation, ${\cal G}_n$ resums energy logarithms up to $\NLL$ accuracy.
The quantities $\E_H^{\rm NLO}$ and $\E_h^{\rm NLO}$ are, respectively, the singly off-shell emission functions associated with the Higgs boson and with the identified hadron.
Their explicit structure, together with the ingredients entering ${\cal G}_n$, will be discussed in Sec.~\ref{ssec:HE_ingredients}.
The specific choice of the renormalization scale $\mu_R$ entering ${\cal G}_n$ will also be specified there.

Here, $\beta_0 = 11N_c/3 - 2n_f/3$ denotes the first coefficient of the QCD $\beta$ function, and $N_c=3$ the color number.
The strong coupling is evolved at two loops, with $\alpha_s(m_Z)=0.118$ and a dynamically adjusted number of active quark flavors $n_f$.

The ``$+$'' qualifier in $\NLLp$ specifically indicates that the product of the two NLO emission functions is kept unexpanded.
Consequently, the contribution proportional to $\hat\E_H\hat\E_h$, which lies formally beyond a strict NLO truncation of the external emission blocks, is retained.
Formally beyond-NLL terms generated by the exponentiated representation of the NLL BFKL kernel are instead common to both the ``$+$'' and pure-NLO representations and are not part of this qualifier.

For comparison, we also consider the $\LL$ representation.
It is obtained by retaining only the leading high-energy evolution together with the lowest-order expressions for the process-dependent emission functions.
Accordingly, the first label refers to the logarithmic accuracy of the BFKL Green's function, whereas the second one identifies the fixed-order accuracy of the external emission blocks.
This distinction is particularly useful in the present process, since the perturbative structure of the Higgs and hadron emission functions is intrinsically process dependent, while the BFKL kernel remains universal.

A complementary fixed-order benchmark is provided by the $\HENLOp$ representation.
Following the strategy originally developed for Mueller--Navelet~\cite{Celiberto:2015yba,Celiberto:2015mpa} and hadron-jet~\cite{Celiberto:2020wpk} observables, it is constructed by expanding the high-energy resummed expression through NLO in the energy-logarithmic series rather than keeping the all-order BFKL evolution.
It should therefore be understood as the high-energy projection of a fixed-order NLO calculation, rather than as a complete collinear NLO prediction.
The ``$+$'' label has the same meaning as in the resummed case: the NLO emission functions are kept in their unexpanded product, so that formally higher-order contributions beyond a strict NLO truncation are retained.
Comparing $\HENLOp$ with $\NLLp$ then offers a direct way to isolate the numerical impact of the all-order high-energy resummation within the same underlying factorization setup.

The precise meaning of the $\LL$ limit and of the $\HENLOp$ high-energy fixed-order truncation will become transparent in Sec.~\ref{ssec:HE_ingredients}, where the relevant high-energy ingredients are introduced explicitly and their perturbative structure is analyzed in detail.

\subsection{High-energy ingredients}
\label{ssec:HE_ingredients}

The first building block entering Eq.~\eqref{Cn_NLLp_MSb} is the BFKL Green's function ${\cal G}_n$, responsible for the universal high-energy evolution.
In the exponentiated representation adopted throughout this work, its $(n,\nu)$ projection is written as
\begin{equation}
\label{Gn_NLL}
{\cal G}_n(\DY,\nu;\mu_{R_c}) =
e^{
\DY \frac{N_c}{\pi} \alpha_s(\mu_{R_c})
\chi^{\rm NLO}(n,\nu)
}
\;,
\end{equation}
where the kernel eigenvalue collects the $\LL$ contribution together with its next-to-leading correction,
\begin{eqnarray}
\label{chi}
\chi^{\rm NLO}(n,\nu) = \chi(n,\nu) + \frac{N_c}{\pi}\alpha_s(\mu_{R_c}) \hat \chi(n,\nu) \;.
\end{eqnarray}
The LO characteristic function is
\begin{eqnarray}
\label{chi_LO}
\chi\left(n,\nu\right) = -2 \gamma_{\rm E} - 2 \, {\rm Re} \left\{ \psi\left(\frac{1+n}{2} + i \nu \right) \right\} \;,
\end{eqnarray}
where $\gamma_{\rm E}$ is the Euler--Mascheroni constant and $\psi(z) \equiv \Gamma^\prime(z)/\Gamma(z)$ denotes the digamma function.
The NLO contribution entering Eq.~\eqref{chi} is organized as
\begin{eqnarray}
\label{chi_NLO}
&\hat \chi&\left(n,\nu\right) = \bar\chi(n,\nu)+\frac{\beta_0}{8 N_c}\chi(n,\nu)
\\ \nonumber &\times& \left\{\chi(n,\nu)+\frac{10}{3}+2\ln\left[\mu_{R_c}^2/(|\vec p_{T_H}| |\vec p_{T_h}|)\right]\right\} \;,
\end{eqnarray}
where $\bar\chi(n,\nu)$ represents the scale-independent part of the NLO BFKL characteristic function~\cite{Kotikov:2000pm}.

Its explicit expression reads
\begin{align}
\label{kernel_NLO}
&\bar\chi(n,\nu)
=
-\frac{1}{4}
\Bigg\{
\frac{\pi^2-4}{3}\chi(n,\nu)
-6\zeta(3)
-\frac{\drv^2\chi}{\drv\nu^2}
\nonumber
\\
&
+2\phi_\chi(n,\nu)
+2\phi_\chi(n,-\nu)
\nonumber
\\
&
+\frac{\pi^2\sinh(\pi\nu)}
{2\nu\cosh^2(\pi\nu)}
\Bigg[
\left(
3+
\left(1+\frac{n_f}{N_c^3}\right)
\frac{11+12\nu^2}{16(1+\nu^2)}
\right)
\delta_{n0}
\nonumber
\\
&
-
\left(1+\frac{n_f}{N_c^3}\right)
\frac{1+4\nu^2}{32(1+\nu^2)}
\delta_{n2}
\Bigg]
\Bigg\}
\; .
\end{align}
where $\delta_{nm}$ is the Kronecker delta, while the auxiliary function $\phi_\chi(n,\nu)$ can be represented as
\begin{equation}
\label{kernel_NLO_phi}
 \phi_\chi(n,\nu)\,=\,-\int\limits_0^1 \drv x
 \frac{x^{-1/2+i\nu+n/2}}{1+x}
\end{equation}
\[
 \times\,
 \left\{\frac{1}{2}\left(\psi^\prime\left(\frac{n+1}{2}\right)-\zeta(2)\right)+\mbox{Li}_2(x)+\mbox{Li}_2(-x)\right.
\]
\[
\,+\,\ln x\left[\psi(n+1)-\psi(1)+\ln(1+x)
+\sum_{k=1}^\infty\frac{(-x)^k}{k+n}\right]
\]
\[
\left.
+\,
\sum_{k=1}^\infty\frac{x^k}{(k+n)^2}\left[1-(-1)^k\right]\right\}
\]
\[
 =\,\sum_{k=0}^\infty\frac{(-1)^{k+1}}{k+(n+1)/2+i\nu}
\]
\[
\times\,
\left\{\psi^\prime(k+n+1)-\psi^\prime(k+1)\right.
\]
\[
 +\,(-1)^{k+1}\left[\lambda_\psi(k+n+1)+\lambda_\psi(k+1)\right]
\]
\[
 \left.
 -\,\frac{\psi(k+n+1)-\psi(k+1)}{k+(n+1)/2+i\nu}\right\} \; ,
\]
with
\begin{equation}
\label{kernel_NLO_phi_beta_psi}
 \lambda_\psi(z)=\frac{1}{4}\left[\psi^\prime\left(\frac{z+1}{2}\right)
 -\psi^\prime\left(\frac{z}{2}\right)\right] \; .
\end{equation}
Finally,
\begin{equation}
\label{dilog}
\mbox{Li}_2(x) = - \int\limits_0^x \drv \zeta \,\frac{\ln(1-\zeta)}{\zeta}
\end{equation}
denotes the dilogarithm function.

The $\HENLOp$ limit can now be obtained straightforwardly by truncating the high-energy expansion of Eq.~\eqref{Cn_NLLp_MSb} through first order in $\alpha_s$.
In practice, this amounts to expanding the BFKL Green's function of Eq.~\eqref{Gn_NLL} as
${\cal G}_n \to 1 + \frac{N_c}{\pi}\alpha_s\DY\chi(n,\nu)$,
while retaining the NLO emission functions in their unexpanded product.

Then, the two process-dependent building blocks are the NLO emission functions
\begin{eqnarray}
\label{EFs}
\E_{H,h}^{\rm NLO}(n,\nu,|\vec p_{T_{H,h}}|,x_{H,h}) =
\E_{H,h} +
\alpha_s(\mu_R) \, \hat \E_{H,h}
\end{eqnarray}
which describe, respectively, the forward and backward production of the Higgs boson and of the identified hadron.
They are formulated in Mellin space and projected onto the eigenfunctions of the LO BFKL kernel.

By neglecting the NLO corrections $\hat{\E}_{H,h}$ in Eq.~\eqref{Cn_NLLp_MSb}, and simultaneously removing the NLO contribution $\hat{\chi}(n,\nu)$ from the kernel in Eq.~\eqref{chi}, so that only the LO characteristic function of Eq.~\eqref{chi_LO} is retained, one recovers the pure $\LL$ limit.

Then, the corresponding representation without the ``$+$'' prescription is obtained by truncating only the fixed-order content carried by the product of the two NLO emission functions.
More precisely, this product is expanded through NLO, so that the term proportional to $\hat{\E}_H\hat{\E}_h$ is discarded, while the NLL BFKL kernel is kept in its exponentiated form.
Consequently, formally beyond-NLL contributions generated by the exponentiated high-energy evolution are still present in both representations; the ``$+$'' qualifier specifically identifies the additional beyond-NLO terms originating from the unexpanded product of the NLO emission functions.

We first focus on the LO Higgs emission function.
Its projection onto the $(n,\nu)$ representation can be written as
\begin{equation}
\begin{split}
\label{LOHiggsEF}
\E_H&(n,\nu,|\vec p_{T_H}|,x_H)
\,=\,
\frac{\alpha_s(\mu_{R_H})}{128\pi^{3} v_{\rm ev}^2}
\frac{
\left|\mathcal{F}_H(|\vec p_{T_H}|^2)\right|^2
}{\sqrt{2(N_c^2-1)}
}
\\
&\times\,
\left(|\vec p_{T_H}|^2\right)^{i\nu+1/2}
f_g(x_H,\mu_{F_H})
\;.
\end{split}
\end{equation}
Here, $v_{\rm ev}=(\sqrt{2}G_F)^{-1/2}\simeq246.22$~GeV~\cite{ParticleDataGroup:2024cfk} denotes the Higgs vacuum expectation value, while $f_g(x_H,\mu_{F_H})$ is the collinear gluon PDF of the incoming proton, evaluated at longitudinal momentum fraction $x_H$ and factorization scale $\mu_{F_H}$.
The function $\mathcal{F}_H$ encodes the quark-loop dynamics of the gluon--Reggeon--Higgs coupling.
For uniformity with the perturbative bookkeeping adopted throughout this work, we include one explicit power of $\alpha_s(\mu_{R_H})$ in the definition of $\E_H$.\footnote{This convention differs from the normalization adopted in Refs.~\cite{Celiberto:2020tmb,Celiberto:2022zdg}, where the corresponding explicit factor of $\alpha_s$ was not included in the definition of the Higgs emission function.}

The finite-mass loop factor generalizes the expression originally employed in Ref.~\cite{Celiberto:2020tmb}, based on the result of Ref.~\cite{DelDuca:2003ba}, by retaining both top- and bottom-quark contributions,
\begin{equation}
\label{Higgs_loop_sum}
\mathcal{F}_H(|\vec p_{T_H}|^2)
=
\sum_{q=t,b}
\mathcal{F}_H^{(q)}(|\vec p_{T_H}|^2)
\;.
\end{equation}
The heavy-quark masses entering the loop amplitudes are taken as $m_t = 172.52$~GeV and $m_b = 4.18$~GeV.
For each heavy-quark flavor, we have
\begin{equation}
\begin{split}
\label{Higgs_loop_q}
\mathcal{F}_H^{(q)}(|\vec p_{T_H}|^2)
&=
\frac{4m_q^2}{m_{T,H}^2}
\Bigg\{
\left(
\frac{1}{2}
-
\frac{2m_q^2}{m_{T,H}^2}
\right)
\\
&\times
\left[
\Delta_h\!\left(\upsilon_2^{(q)}\right)^2
-
\Delta_h\!\left(\upsilon_1^{(q)}\right)^2
\right]
\\
&+
\frac{2|\vec p_{T_H}|^2}{m_{T,H}^2}
\Bigg[
\sqrt{\upsilon_1^{(q)}}\,
\Delta_h\!\left(\upsilon_1^{(q)}\right)
\\
&
-
\sqrt{\upsilon_2^{(q)}}\,
\Delta_h\!\left(\upsilon_2^{(q)}\right)
\Bigg]
+2
\Bigg\}
\;,
\end{split}
\end{equation}
where
$m_{T,H}^2=m_H^2+|\vec p_{T_H}|^2$ and
\begin{equation}
\label{Higgs_loop_variables}
\upsilon_1^{(q)}
=
1-\frac{4m_q^2}{m_H^2}
\;,
\qquad
\upsilon_2^{(q)}
=
1+\frac{4m_q^2}{|\vec p_{T_H}|^2}
\;.
\end{equation}
The analytic function entering Eq.~\eqref{Higgs_loop_q} is
\begin{equation}
\label{IF_W}
\Delta_h(\upsilon) =
\left\{
\begin{aligned}
&-2i\arcsin\frac{1}{\sqrt{1-\upsilon}} \;,
&&\upsilon<0 \;,
\\
&\ln\frac{1+\sqrt{\upsilon}}{1-\sqrt{\upsilon}}-i\pi \;,
&&0<\upsilon<1 \;,
\\
&\ln\frac{1+\sqrt{\upsilon}}{\sqrt{\upsilon}-1} \;,
&&\upsilon>1 \;.
\end{aligned}
\right.
\end{equation}
For negative values of $\upsilon_1^{(q)}$, the corresponding square root is understood through the analytic continuation
$\sqrt{\upsilon_1^{(q)}}=i\sqrt{|\upsilon_1^{(q)}|}$.

Equation~\eqref{Higgs_loop_sum} retains the top- and bottom-quark amplitudes coherently before taking the modulus squared in Eq.~\eqref{LOHiggsEF}, and therefore also includes their interference.
The finite-bottom-mass contribution constitutes an extension with respect to the setup employed in our earlier high-energy Higgs studies and will be activated only in dedicated phenomenological analyses discussed in Sec.~\ref{sec:pheno}.

In contrast to the LO emission function, which retains the finite quark-mass dependence, the complete NLO correction is currently known only in the infinite-top-mass limit~\cite{Hentschinski:2020tbi,Celiberto:2022fgx}.
In Ref.~\cite{Celiberto:2022fgx}, this correction was calculated within the standard BFKL approach and projected onto the eigenfunctions of the LO BFKL kernel.
The resulting $(n,\nu)$ representation provides precisely the ingredient required by our high-energy factorization formula.
In the infinite-top-mass approximation, the general
decomposition in Eq.~\eqref{EFs} becomes
\begin{equation}
\label{NLOHiggsEF_EFT}
\E_H^{{\rm NLO},{\rm EFT}}
=
\E_H^{\rm EFT}
+
\alpha_s(\mu_{R_H})\,\hat\E_H^{\rm EFT}
\;.
\end{equation}
Given the length of the full expression, we summarize
the convolution structure of the NLO correction
schematically as
\begin{equation}
\begin{split}
\label{NLOHiggsEF_scheme}
\hat\E_H^{\rm EFT}
\,&\sim\,
\frac{\mathcal N_H^{\rm EFT}}{2\pi}
\\
&\times\,
\Bigg[
a_H f_g
+c_g^H\otimes f_g
+
\sum_{a=q,\bar q}
c_a^H\otimes f_a
\Bigg]
\;,
\end{split}
\end{equation}
where $\E_H^{\rm EFT}$ denotes the LO Higgs emission
function in Eq.~\eqref{LOHiggsEF}, evaluated in the
infinite-top-mass limit and with the bottom-quark loop
omitted.
Its PDF-independent prefactor, $\mathcal N_H^{\rm EFT}$,
is defined via
$\E_H^{\rm EFT}=\mathcal N_H^{\rm EFT}
f_g(x_H,\mu_{F_H})$.
The factor $1/(2\pi)$ is displayed explicitly following
the normalization of the NLO result in
Ref.~\cite{Celiberto:2022fgx}; the coefficients
$a_H$ and $c_a^H$ are defined with this factor extracted.
The longitudinal convolutions are understood as
\begin{equation}
\label{Higgs_convolution}
c_a^H\otimes f_a
=
\int_{x_H}^{1}\frac{\drv \xi}{\xi}\,
c_a^H(\xi)\,
f_a\left(\frac{x_H}{\xi},\mu_{F_H}\right)
\;,
\end{equation}
where $a=g,q,\bar q$.
Besides the gluon distribution $f_g$, we have introduced the quark and antiquark PDFs, $f_q$ and $f_{\bar q}$, entering the NLO correction.
The dependence of the coefficient functions on the remaining kinematic variables and scales is left implicit.

The local coefficient $a_H$ collects the finite virtual contribution, the coupling-renormalization term, and the finite remainder associated with the BFKL subtraction.
The coefficients $c_g^H$ and $c_a^H$ encode the gluon- and quark-initiated real contributions, respectively, together with the corresponding initial-state collinear subtractions and associated endpoint terms.
The common prefactor $\mathcal N_H^{\rm EFT}$ reflects the
shared effective Higgs-gluon operator and the normalization
adopted for the projected coefficient functions.
In the quark-initiated channels, the incoming quark couples
to the effective Higgs-production vertex through a gluon,
while its PDF enters through an independent longitudinal
convolution.
Consequently, extracting the PDF-independent Born
normalization does not imply that these contributions
are proportional to the gluon distribution.

Factoring out the complete LO emission function instead
would leave ratios of the quark-PDF convolutions to
$f_g(x_H,\mu_{F_H})$ in the relative NLO correction.
Thus, although the LO Higgs emission function contains only the gluon PDF, its NLO correction also involves convolutions with the PDFs of all active quarks and antiquarks.
The combined result is finite after coupling renormalization, PDF factorization, and subtraction of the contribution assigned to the BFKL evolution.

Finite-mass information can additionally be incorporated into the radiative correction through a Born-improvement prescription, referred to as ``iBorn''.
Defining the relative NLO EFT correction as
\begin{equation}
\label{Higgs_relative_NLO}
\mathcal R_H^{\rm EFT}
=
\alpha_s(\mu_{R_H})
\,
\frac{
\hat\E_H^{\rm EFT}
}{
\E_H^{\rm EFT}
}
\;,
\end{equation}
the iBorn emission function reads
\begin{equation}
\begin{split}
\label{NLOHiggsEF_BI}
\E_H^{{\rm NLO},{\rm iBorn}}
\,=\,{}&
\E_H
+
\frac{\E_H}{\E_H^{\rm EFT}}\,
\alpha_s(\mu_{R_H})\,\hat\E_H^{\rm EFT}
\\
\,\equiv\,{}&
\E_H\left[1+\mathcal R_H^{\rm EFT}\right]
\;.
\end{split}
\end{equation}
The Born ratio $\E_H/\E_H^{\rm EFT}$ depends only on the external Higgs kinematics and quark masses.
The iBorn reweighting therefore preserves the Regge-factorized structure while leaving the BFKL Green's function unchanged.
In particular, the universal energy-scale dependence is rescaled consistently with the finite-mass Born contribution, as can be checked from the subtraction structure derived in Sec.~5.3, Eq.~(5.23), of
Ref.~\cite{Celiberto:2022fgx}.

This prescription replaces $\mathcal N_H^{\rm EFT}$ by the corresponding finite-mass prefactor in Eq.~\eqref{NLOHiggsEF_scheme}, thereby reweighting the entire NLO correction, including the quark-initiated channels.
When the bottom-quark loop is included in $\E_H$, the reweighting also retains the top-bottom interference present in the Born contribution.
The radiative coefficient functions retain their EFT form;
finite-quark-mass effects in the NLO correction are therefore
approximated through the Born reweighting.

Unless otherwise stated, our NLO predictions employ the iBorn
prescription with finite top- and bottom-quark masses retained
in the Born contribution.
Accordingly, the Higgs emission function entering
Eq.~\eqref{EFs} is understood as
\begin{equation}
\label{NLOHiggsEF_default}
\E_H^{\rm NLO}(n,\nu,|\vec p_{T_H}|,x_H)
\equiv
\left.
\E_H^{{\rm NLO},{\rm iBorn}}
\right|_{t+b}
\;,
\end{equation}
where the arguments on the right-hand side are left implicit.
The notation $t+b$ specifies that the top- and bottom-quark
loop amplitudes are combined coherently in the Born input,
including their interference.
Alternative EFT and finite-mass configurations are considered
in dedicated studies in Sec.~\ref{sec:pheno}.

We now turn our attention to the the identified-hadron emission function.
Its NLO expression is taken from Ref.~\cite{Ivanov:2012iv}.
Although the original derivation was formulated for light-flavored hadrons, the same perturbative structure can be consistently employed for singly charmed hadrons within our ZM-VFNS setup, provided that the characteristic transverse momenta remain sufficiently above the heavy-quark thresholds governing DGLAP evolution.

At LO, the hadron emission function takes the form
\begin{equation}
\begin{split}
\label{LOHEF}
\E_h(n,\nu,|\vec p_{T_h}|,x_h) =
\delta_c \, |\vec p_{T_h}|^{2i\nu-1}
\int_{x_h}^1 \frac{\drv \xi}{\xi} \;
\hat{x}^{1-2i\nu}
\\
\times \,
\left[
\rho_c f_g(\xi)D_g^h\left(\hat{x}\right)
+
\sum_{a=q,\bar q}
f_a(\xi)D_a^h\left(\hat{x}\right)
\right]
\;,
\end{split}
\end{equation}
where $\hat{x}=x_h/\xi$, $\delta_c=2\sqrt{C_F/C_A}$, and $\rho_c=C_A/C_F$.
Here, $C_F=(N_c^2-1)/(2N_c)$ and $C_A=N_c$ denote the fundamental and adjoint QCD Casimir operators, respectively.
The relative enhancement of the gluon term by $C_A/C_F$ originates entirely from the color structure of the LO forward-parton emission vertex and is therefore independent of the nonperturbative fragmentation dynamics.

The NLO correction $\hat\E_h(n,\nu,|\vec p_{T_h}|,x_h)$ is obtained from the identified-hadron impact factor of Ref.~\cite{Ivanov:2012iv}, specifically from Eq.~(4.58) together with the coefficient functions given in Eqs.~(4.59)--(4.62).
Its convolution structure can be summarized schematically as
\begin{equation}
\E_h^{\rm NLO}
\sim
\sum_\alpha
f_\alpha\otimes D_\alpha^h
+
\frac{\alpha_s}{2\pi}
\sum_{\alpha,\beta}
f_\alpha\otimes c^h_{\alpha\beta}\otimes D_\beta^h
\;,
\end{equation}
where $\alpha$ and $\beta$ independently span the active gluon, quark, and antiquark species, while $c^h_{\alpha\beta}$ are the corresponding NLO coefficient functions.

The LO term is flavor diagonal: the partonic species extracted from the incoming proton is the same one that fragments into the observed hadron.
At NLO, this simple pattern is enlarged by the coefficient functions $c^h_{\alpha\beta}$, which permit the PDF and FF flavor indices to differ and therefore generate both diagonal and mixed partonic channels.
They collect the real and virtual radiative corrections together with the collinear-subtraction contributions.
After PDF and FF renormalization, the infrared and ultraviolet singularities cancel, leaving a finite NLO emission function.

In our phenomenological analysis, the same perturbative emission kernel is convoluted either with FFs specific to ${\cal H}_c$ production or with those describing the light-hadron comparison channel ${\cal H}_l$.
The distinction between the two identified-hadron final states therefore resides entirely in the corresponding collinear fragmentation input, while the short-distance structure of the emission function is common to both channels.

The discussion developed in this section highlights the basic architecture of our hybrid high-energy and collinear factorization framework.
The BFKL Green's function controls the resummation of logarithmically enhanced contributions in the high-energy limit, while the process-dependent emission functions provide the interface with collinear dynamics through PDFs and FFs.
Their combination therefore links universal high-energy evolution with the specific structure of the tagged final states.

Having specified the BFKL Green's function and the two process-dependent emission functions, we finally set the characteristic renormalization and factorization scales entering the different high-energy building blocks.
The hadron-emission sector is naturally controlled by $|\vec p_{T_h}|$, whereas the Higgs-emission block is characterized by the transverse mass $m_{T,H}=\sqrt{m_H^2+|\vec p_{T_H}|^2}$.
We therefore adopt a multiscale prescription, setting $\mu_{R_h}=\mu_{F_h}=C_\mu |\vec p_{T_h}|$ for the hadron sector and $\mu_{R_H}=\mu_{F_H}=C_\mu m_{T,H}$ for the Higgs one.
For the renormalization scale entering the BFKL evolution that connects the two emission regions, we choose the geometric mean of the corresponding hard scales, $\mu_{R_c}=\sqrt{\mu_{R_h}\mu_{R_H}}=C_\mu\sqrt{|\vec p_{T_h}|m_{T,H}}$.
The natural configuration corresponds to $C_\mu=1$, while variations of $C_\mu$ are used to estimate hard-factor missing higher-order uncertainties (H-MHOUs), as detailed in Sec.~\ref{ssec:uncertainty}.

\subsection{Collinear ingredients}
\label{ssec:collinear_ingredients}

The fragmentation of a parton into a singly charmed hadron, ${\cal H}_c$, is described through the NLO collinear FF set {\HCFF}, which we have assembled as a private set of {\LHAPDF} grids~\cite{Buckley:2014ana}.
These grids were originally constructed in the context of our studies on quarkonium fragmentation~\cite{Celiberto:2025euy} and can be made available upon reasonable request.
The {\HCFF} parametrization combines the {\tt KKKS08} NLO FFs for $D$-meson production~\cite{Kniehl:2005de,Kneesch:2007ey} with the {\tt KKSS19} NLO determination for $\Lambda_c$ baryons~\cite{Kniehl:2020szu}.
Throughout our analysis, charmed-hadron production is understood inclusively with respect to charge-conjugate states.

The two FF determinations entering {\HCFF} were developed within the same fitting framework and therefore rely on a common theoretical and phenomenological setup.
We refer the reader to the original analyses for a detailed account of their extraction strategy.
Here we only recall that these ZM-VFNS parametrizations employ heavy-flavor inputs at the starting scale inspired by Bowler~\cite{Bowler:1981sb}, Peterson~\cite{Peterson:1982ak}, or powerlike~\cite{Kartvelishvili:1985ac} functional forms, while the lighter partonic components are dynamically generated through DGLAP evolution.

For comparison with light-flavored final states, we employ the {\tt NNFF1.0} NLO collinear FFs~\cite{Bertone:2017tyb}.
Our light-hadron sample is constructed by summing over pion, kaon, and proton species, again inclusively with respect to their charge-conjugate states.
The corresponding FF grids are accessed through {\LHAPDF}.

The collinear structure of the incoming protons is modeled with the {\tt NNPDF4.0} NLO PDF determination~\cite{NNPDF:2021uiq,NNPDF:2021njg}, likewise interfaced through {\LHAPDF}.

\subsection{Emergence of the natural stability}
\label{ssec:natural_stability}

Natural stability denotes a perturbative pattern in which the inclusion of next-to-leading high-energy corrections produces only a moderate departure from the corresponding LL baseline.
This stability of the perturbative series is typically accompanied by a contained sensitivity to hard-factor missing-higher-order uncertainties (H-MHOUs), as probed through renormalization- and factorization-scale variations around their natural values.
The two features are closely related but should not be identified with each other: the reduced scale dependence emerges as a consequence of the same dynamical balance that limits the size of the NLL corrections to the physical observable.
Natural stability should therefore be regarded as an emergent property of the complete hadronic prediction, resulting from the interplay among the running coupling, collinear evolution, process-dependent emission functions, and BFKL dynamics.

Two complementary mechanisms have been identified so far.
We first discuss the one associated with heavy-flavor fragmentation and then the distinct stabilization induced by the intrinsically hard scale of Higgs production.

The fragmentation-driven mechanism can be traced back to the structure of the identified-hadron emission function in Eq.~\eqref{LOHEF}.
In particular, gluon fragmentation can remain numerically important even when a heavy-quark FF is the largest individual component, because it is weighted by both the gluon PDF and the color factor $C_A/C_F$.

The first clear evidence emerged from high-energy $\Lambda_c$ production~\cite{Celiberto:2021dzy}.
There, the smoothly evolving gluon FF of the charmed baryon, combined with the increasing gluon density over the relevant longitudinal region, was found to compensate part of the suppression generated by the running of $\alpha_s$ under scale variations.
A complementary test based on a flavor-independent and scale-independent toy FF produced a similarly mild response~\cite{Celiberto:2021dzy}, supporting the interpretation in terms of the structure of the full fragmentation convolution.

The bottom sector provided an independent confirmation.
For non-charmed $b$-hadrons, whose gluon FF exhibits a more pronounced growth with the factorization scale, an even stronger stabilization pattern was observed~\cite{Celiberto:2021fdp}.
The same qualitative behavior was subsequently found for charmed $b$-hadrons~\cite{Celiberto:2022keu,Celiberto:2024omj}.
Further evidence came from vector~\cite{Celiberto:2022dyf} and pseudoscalar~\cite{Celiberto:2025euy} quarkonia, whose starting-scale FFs are constructed within nonrelativistic QCD.
The persistence of the effect across FF sets obtained with substantially different methodologies suggests that natural stability reflects a more general property of heavy-flavor fragmentation once embedded in the high-energy hadronic convolution.

This favorable perturbative behavior has also opened a phenomenological portal toward rare and exotic hadronic matter in the high-energy regime.
Heavy-flavor fragmentation has consequently been exploited to investigate increasingly complex bound states, including doubly and fully heavy tetraquarks~\cite{Celiberto:2023rzw,Celiberto:2024mab,Celiberto:2024beg,Celiberto:2025dfe,Celiberto:2025ziy,Celiberto:2025vra,Celiberto:2026kks}, pentaquark configurations~\cite{Celiberto:2025ipt,Celiberto:2026rdk}, and rare heavy baryons~\cite{Celiberto:2025ogy,Celiberto:2026qiz}.

The Higgs sector provides a second and conceptually distinct source of stabilization.
The first strong indication arose in Higgs-plus-jet production~\cite{Celiberto:2020tmb}, where the large Higgs transverse mass $m_{T,H}$ drives the natural scales toward a comparatively hard region and thereby reduces the sensitivity of the resummed prediction to their variation.
A partial manifestation of the same hard-scale mechanism was also observed in heavy-light dijet systems~\cite{Bolognino:2021mrc}.
Its interplay with heavy-flavor fragmentation was subsequently exposed in Higgs-plus-charm production~\cite{Celiberto:2022zdg}, where the two mechanisms act simultaneously but originate from different sectors of the factorized cross section.

A dedicated investigation of the physical origin of natural stability is currently in progress~\cite{Celiberto:2026_nst}.
Preliminary analyses indicate that the fragmentation mechanism cannot be reduced either to a generic damping of large Mellin frequencies or to the sign of the scale evolution of a single FF.
Fragmentation reshapes the longitudinal momentum fractions, partonic transverse momenta, flavor composition, and spectral regions sampled by the convolution.
Moreover, at NLO the explicit collinear and renormalization logarithms compensate the leading scale dependence of the evolved inputs and running coupling in all channels.
The phenomenologically relevant differences therefore reside in the finite terms, channel composition, and their convolution with the BFKL evolution.

The same ongoing study also indicates that the Higgs-driven mechanism persists after inclusion of the complete NLO forward-Higgs emission function.
As a preliminary diagnostic, an auxiliary Higgs-plus-jet calculation employing the NLO Higgs impact factor of Ref.~\cite{Celiberto:2022fgx} together with the NLO jet vertex was performed at $\sqrt{s}=14$~TeV and $\DY=y_H-y_J=4.5$, with $20<|\vec p_{T_H}|/{\rm GeV}<60$, $20<|\vec p_{T_J}|/{\rm GeV}<60$, $|y_H|<2.5$, and $|y_J|<4.7$.
We employ the NLL/NLO representation, without the ``$+$'' qualifier.
For the correlated scale variation $C_\mu=1/2,1,2$, used to estimate H-MHOUs, one finds
\begin{equation}
\begin{split}
\label{eq:natural_stability_higgs_check}
C_0^{\rm NLL/NLO}
\,&=\,
\{0.453,\,0.419,\,0.364\}\ {\rm pb}\;,
\\
C_0^{\rm LL/LO}
\,&=\,
\{0.697,\,0.420,\,0.269\}\ {\rm pb}\;.
\end{split}
\end{equation}
Relative to the central prediction, the corresponding excursion is approximately $-13\%$ to $+8\%$ at NLL/NLO, compared with about $-36\%$ to $+66\%$ at LL/LO.
This provides preliminary evidence that the reduced scale sensitivity survives when the finite NLO Higgs-vertex contribution is retained.

Natural stability should nevertheless not be identified with uniformly small higher-order corrections.
Sizeable contributions of opposite sign can arise from the Green's function and the process-dependent emission functions, with the stabilization emerging only after their combination in the physical observable.
Its strength can therefore depend on the observable and on the explored phase-space region.

This interpretation is particularly relevant to Higgs-plus-charm production, where the hard scale supplied by the Higgs boson and the favorable collinear dynamics of heavy-flavor fragmentation coexist within the same factorized cross section.
The phenomenological analysis of Sec.~\ref{sec:pheno} will test whether the stability previously observed at partial-NLL accuracy persists once the complete NLO correction to the Higgs emission function is included.

\subsection{Uncertainty assessment}
\label{ssec:uncertainty}

A realistic phenomenological analysis requires a controlled assessment of uncertainties affecting the theoretical predictions.
In the present study, we distinguish between genuine uncertainty sources, which are propagated to the final error bands, and complementary theory variations, which are instead used as diagnostics of the stability of the high-energy description.

\begin{itemize}

\item[(i)]
\textit{Perturbative H-MHOUs.}
The dominant perturbative uncertainty is associated with the residual dependence on renormalization and factorization scales.
Following the multiscale prescription introduced in Sec.~\ref{ssec:HE_ingredients}, all characteristic scales are varied coherently through the common parameter $C_\mu$.
Starting from the natural configuration $C_\mu=1$, we probe the range $1/2 \leq C_\mu \leq 2$.
This correlated variation simultaneously affects the Higgs and hadron emission sectors, as well as the renormalization scale entering the BFKL Green's function, and provides our estimate of missing higher-order uncertainties.
\\[0.15cm]
\item[(ii)]
\textit{Hadron FFs.}
No FF uncertainty is propagated in the present analysis.
The charmed-hadron input combines the {\tt KKKS08} and {\tt KKSS19} NLO determinations, for which no uncertainty replicas or Hessian error sets are presently available.
A first systematic study of fragmentation uncertainties in this sector has recently been initiated within the Hymn Collaboration~\cite{Epele:2026dia}, but no public uncertainty-enabled release is yet available.
Uncertainty information does exist for the light-hadron {\tt NNFF1.0} parametrization~\cite{Bertone:2017tyb}; however, for consistency between the charmed- and light-hadron channels, we employ only the corresponding central FF members in both cases.
This choice is also supported by previous high-energy fragmentation studies, where FF uncertainties were found not to modify the qualitative perturbative pattern of the resummed series~\cite{Celiberto:2022rfj,Celiberto:2025ziy,Celiberto:2026kks}.
\\[0.15cm]
\item[(iii)]
\textit{Proton PDFs.}
PDF uncertainties are likewise not propagated.
We use the central member of the {\tt NNPDF4.0} NLO set throughout the analysis.
Previous phenomenological studies of heavy-flavor and identified-hadron production indicate that PDF-induced variations are generally subleading with respect to the dominant perturbative scale dependence in the kinematic regime considered here.
Since our main objective is to assess the stability of the high-energy expansion, including PDF replicas would therefore add numerical complexity without materially affecting the conclusions drawn from the perturbative comparison.
\\[0.15cm]
\item[(iv)]
\textit{Phase-space numerical integration.}
A further contribution originates from the multidimensional integrations over the final-state phase space and over the Mellin variable $\nu$.
These integrations are performed through the native routines of {\Jethad}, with numerical errors systematically controlled at the sub-percent level.
Residual uncertainties from the longitudinal convolutions entering the PDF-FF sector are numerically smaller and found to be negligible in comparison.
\\[0.15cm]
\item[(v)]
\textit{RG improvement of the BFKL kernel.}
As an additional stability check, we have performed preliminary tests based on collinear, or RG, improvements of the NLL BFKL kernel~\cite{Salam:1998tj,Ciafaloni:2003kd,Ciafaloni:2003rd,SabioVera:2005tiv}.
Such prescriptions reorganize the Mellin-space pole structure of the kernel so as to improve compatibility with collinear DGLAP dynamics.
These tests show that, throughout the kinematic ranges explored here, the numerical dependence on the RG improvement is negligible.
Having explicitly verified this stability, we do not propagate the corresponding variation as an independent uncertainty, but retain it as a complementary validation of the robustness of the resummed description.
\\[0.15cm]
\item[(vi)]
\textit{Beyond-NLO terms from the ``$+$'' representation.}
Our default predictions are formulated in the ``$+$'' representation, in which the product of the two NLO emission functions is kept unexpanded and therefore retains formally beyond-NLO contributions.
To assess their numerical weight, we compare the ``$+$'' representation with its corresponding pure-NLO counterpart in the transverse-momentum analysis of Sec.~\ref{ssec:pT}.
This observable is particularly suited to such a test, since scanning a broad Higgs transverse-momentum interval allows one to resolve the kinematic dependence of these formally higher-order terms over a wide range of hard scales.
\\[6pt]
We do not fold this difference into the uncertainty bands.
It reflects a change of perturbative representation rather than an independent source of theoretical uncertainty, and its effect is intrinsically correlated with the missing higher-order contributions already probed through scale variations.
Treating it as an additional error and combining it in quadrature with H-MHOUs would therefore risk double counting the same underlying perturbative ambiguity.
For this reason, the ``$+$'' versus pure-NLO comparison is displayed separately as a dedicated diagnostic.
\\[0.15cm]
\item[(vii)]
\textit{Heavy-quark-mass and Born-improvement effects.}
A second dedicated study is performed for the transverse-momentum rates of Sec.~\ref{ssec:pT}, where sensitivity to the heavy-quark structure of the Higgs emission function is expected to be maximal.
We compare the reference setup with finite $m_t$ and $m_b$ supplemented by Born improvement against alternative treatments involving the large-$m_t$ limit, the neglect of finite-$m_b$ effects, and the removal of the Born-improvement prescription.
These variations are not interpreted as statistical or perturbative uncertainties and are therefore not included in the error bands.
Rather, they quantify the sensitivity of the Higgs transverse-momentum spectrum to specific approximations entering the treatment of heavy-quark loops.

\end{itemize}

Accordingly, the uncertainty bands shown throughout our phenomenological analysis include only H-MHOUs and the numerical integration error.
The two contributions are treated as independent and combined in quadrature at the observable level.
All other variations discussed above are displayed separately and serve as dedicated tests of perturbative, resummation, or heavy-mass systematics.

\section{Higgs-plus-hadron studies at the LHC}
\label{sec:pheno}

We begin in Sec.~\ref{ssec:kinematic_ranges} by specifying the fiducial phase-space selections adopted throughout the analysis.
Rapidity-interval rates are then presented in Sec.~\ref{ssec:DY}, while Sec.~\ref{ssec:Rnm} focuses on azimuthal-correlation moments and their ratios.
Normalized azimuthal distributions are discussed in Sec.~\ref{ssec:phi}.
Finally, Sec.~\ref{ssec:pT} is devoted to transverse-momentum spectra, including comparisons among different high-energy representations and dedicated studies of finite top- and bottom-quark mass effects.

The phenomenological analysis is carried out with the \textsc{Python}+\textsc{Fortran} implementation of the multimodular {\Jethad} interface~\cite{Celiberto:2020wpk,Celiberto:2022rfj,Celiberto:2023fzz,Celiberto:2024mrq,Celiberto:2024swu,Celiberto:2025csa,Celiberto:2026ooh}, which provides the numerical backbone for all predictions discussed below.

\subsection{Fiducial-range selection}
\label{ssec:kinematic_ranges}

Our phenomenological analysis is performed for proton-proton collisions at the $\sqrt{s}=14$ TeV LHC and is built upon two transverse-momentum selections and two rapidity acceptances, whose combinations define four distinct fiducial configurations.

Concerning the rapidity coverage, we distinguish between a standard selection, in which both the Higgs boson and the identified hadron are reconstructed within the central barrel region, $|y_H|<2.5$ and $|y_h|<2.5$, and an extended one, where the Higgs remains central while the hadron acceptance is enlarged up to $|y_h|<4.7$.
The former reflects the acceptance typically exploited in present CMS and ATLAS analyses.
The latter should instead be regarded as a theory-oriented benchmark.
Indeed, although jets can be reconstructed also in the endcap region over comparably large rapidity intervals, as in the CMS analysis of Ref.~\cite{Khachatryan:2016udy}, particle identification for identified hadrons becomes significantly more challenging there.

For transverse momenta, we consider a symmetric configuration, where the Higgs boson and the hadron are selected within the same moderate-momentum window, and an asymmetric one, where the Higgs transverse momentum spans a substantially broader range.
Altogether, the four fiducial selections read
\begin{itemize}

\item[(i)] \textit{Symmetric transverse momenta, standard rapidity range:}
\[
\text{\makebox[0.60\linewidth][l]{$20~{\rm GeV}<|\vec p_{T_H}|<60~{\rm GeV},$}}
\qquad |y_H|<2.5,
\]
\[
\text{\makebox[0.60\linewidth][l]{$20~{\rm GeV}<|\vec p_{T_h}|<60~{\rm GeV},$}}
\qquad |y_h|<2.5.
\]

\item[(ii)] \textit{Asymmetric transverse momenta, standard rapidity range:}
\[
\text{\makebox[0.60\linewidth][l]{$10~{\rm GeV}<|\vec p_{T_H}|<2m_t,$}}
\qquad |y_H|<2.5,
\]
\[
\text{\makebox[0.60\linewidth][l]{$20~{\rm GeV}<|\vec p_{T_h}|<60~{\rm GeV},$}}
\qquad |y_h|<2.5.
\]

\item[(iii)] \textit{Symmetric transverse momenta, extended rapidity range:}
\[
\text{\makebox[0.60\linewidth][l]{$20~{\rm GeV}<|\vec p_{T_H}|<60~{\rm GeV},$}}
\qquad |y_H|<2.5,
\]
\[
\text{\makebox[0.60\linewidth][l]{$20~{\rm GeV}<|\vec p_{T_h}|<60~{\rm GeV},$}}
\qquad |y_h|<4.7.
\]

\item[(iv)] \textit{Asymmetric transverse momenta, extended rapidity range:}
\[
\text{\makebox[0.60\linewidth][l]{$10~{\rm GeV}<|\vec p_{T_H}|<2m_t,$}}
\qquad |y_H|<2.5,
\]
\[
\text{\makebox[0.60\linewidth][l]{$20~{\rm GeV}<|\vec p_{T_h}|<60~{\rm GeV},$}}
\qquad |y_h|<4.7.
\]

\end{itemize}
The adopted transverse-momentum cuts are sufficiently hard to ensure, at the same time, the validity of the semi-hard regime and the applicability of the ZM-VFNS description for charmed-hadron fragmentation.

The upper bound $|\vec p_{T_H}|=2m_t$ in the asymmetric selection is chosen to access the kinematic region where finite-heavy-quark-mass effects start becoming increasingly relevant.
Their impact will be investigated more directly through the Higgs transverse-momentum spectra presented in Sec.~\ref{ssec:pT}.

For simplicity, the complete set of four fiducial configurations is employed only for the rapidity-interval distributions discussed in Sec.~\ref{ssec:DY}.
For the remaining observables we work with reduced subsets of these selections, tailored to the quantity under investigation.
In particular, the transverse-momentum rates of Sec.~\ref{ssec:pT} are integrated over the hadron transverse momentum while remaining differential in the Higgs transverse momentum.

\subsection{Rapidity-interval rates}
\label{ssec:DY}

We start our phenomenological study from the cross section differential in the rapidity separation between the Higgs boson and the tagged hadron,
$\DY \equiv y_H-y_h$, as in Eq.~\ref{Delta_Y}.
This observable provides a direct handle on the evolution of the semi-hard reaction with increasing longitudinal distance between the two tagged final-state objects.
For a given perturbative accuracy, the corresponding rapidity-interval rate is defined as
\begin{align}
\label{DY_distribution}
&\frac{\drv \sigma^{\rm [level]}(\DY,s)}{\drv \DY}
=
\!\!\int_{y_H^{\rm min}}^{y_H^{\rm max}}
\!\!\!\!\drv y_H
\int_{y_h^{\rm min}}^{y_h^{\rm max}}
\!\!\!\!\drv y_h\,
\delta\!\left(\DY-y_H+y_h\right)
\nonumber
\\
&\times\,
\int_{|\vec p_{T_H}|^{\rm min}}^{|\vec p_{T_H}|^{\rm max}}
\!\!\drv |\vec p_{T_H}|
\int_{|\vec p_{T_h}|^{\rm min}}^{|\vec p_{T_h}|^{\rm max}}
\!\!\drv |\vec p_{T_h}|\,
{\cal C}_{0}^{\rm [level]}\; .
\end{align}
Here, the integrated coefficient ${\cal C}_{0}$ represents the azimuthally averaged cross section, while the label ${\rm [level]}$ specifies the resummation accuracy adopted in the calculation.

All integration limits in Eq.~\eqref{DY_distribution} are fixed by the fiducial selections of Sec.~\ref{ssec:kinematic_ranges}.
Unlike the observables considered later, rapidity-interval rates are evaluated in the complete set of four configurations defined there.
This allows us to disentangle the impact of symmetric versus asymmetric transverse-momentum selections from that of standard versus extended hadron rapidity coverage, and to assess how these choices affect the $\DY$ dependence of the resummed predictions.

\begin{figure*}[!t]
\centering

   \hspace{0.00cm}
   \includegraphics[scale=0.395,clip]{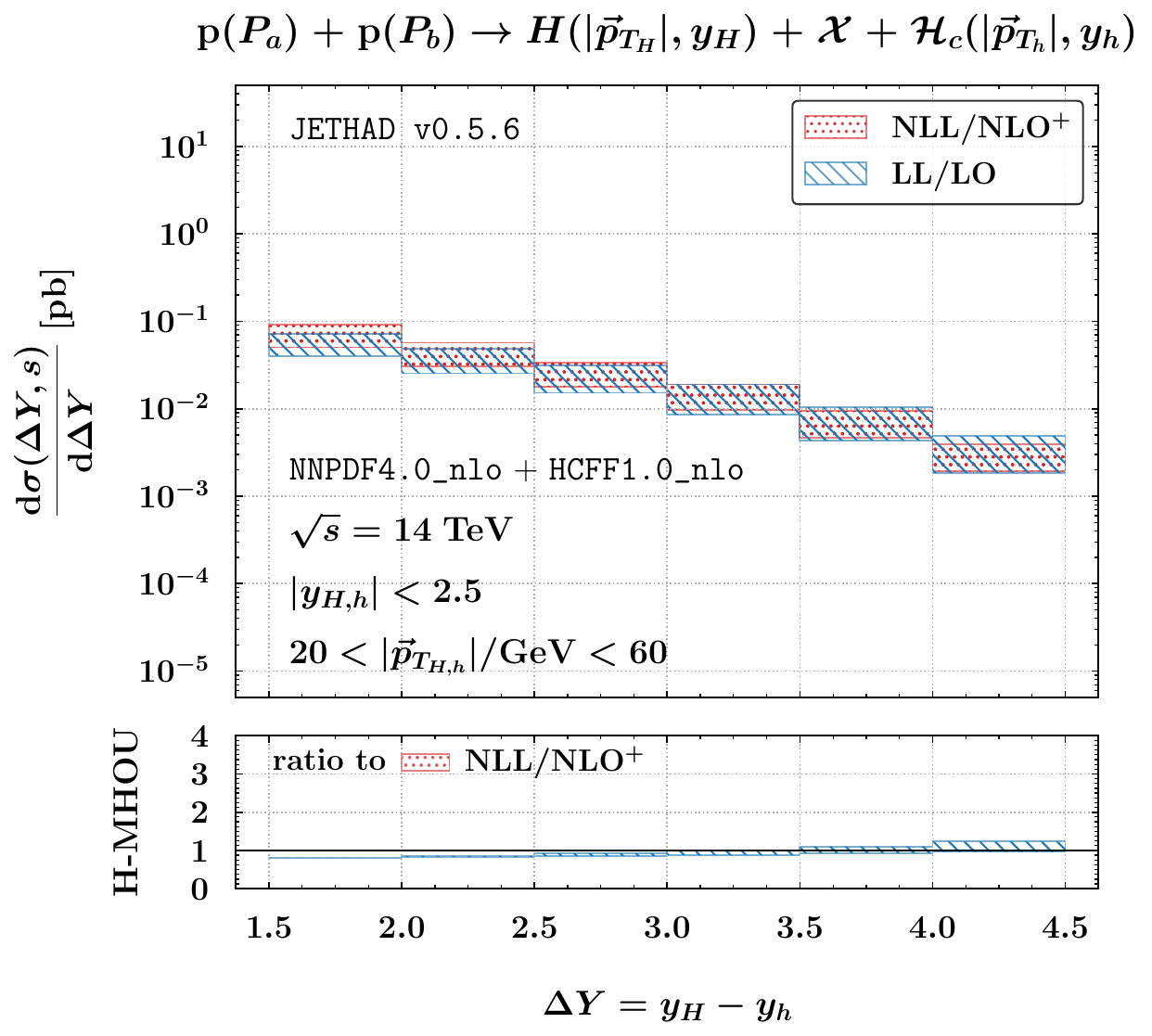}
   \hspace{-0.00cm}
   \includegraphics[scale=0.395,clip]{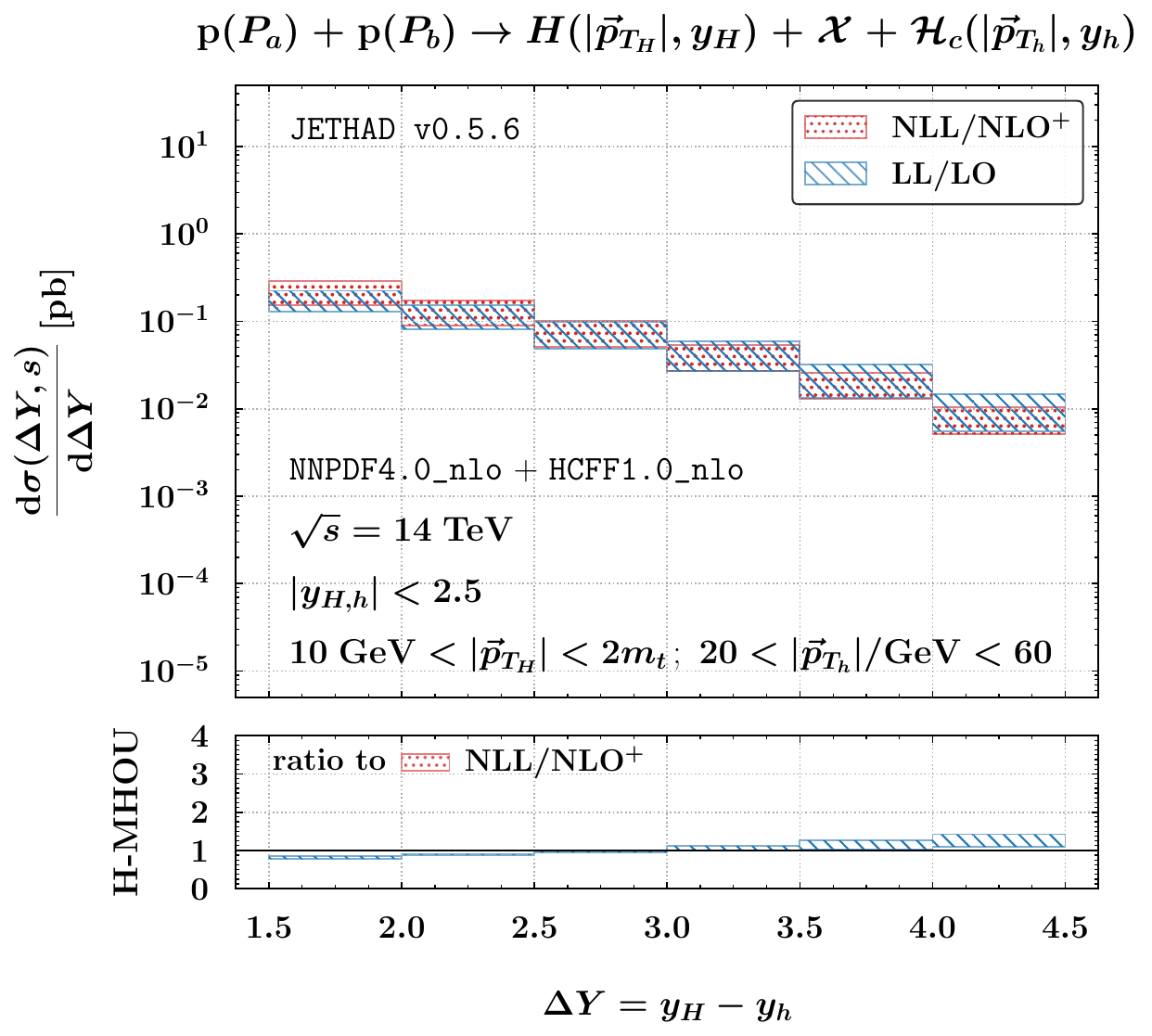}

   \vspace{0.50cm}

   \includegraphics[scale=0.395,clip]{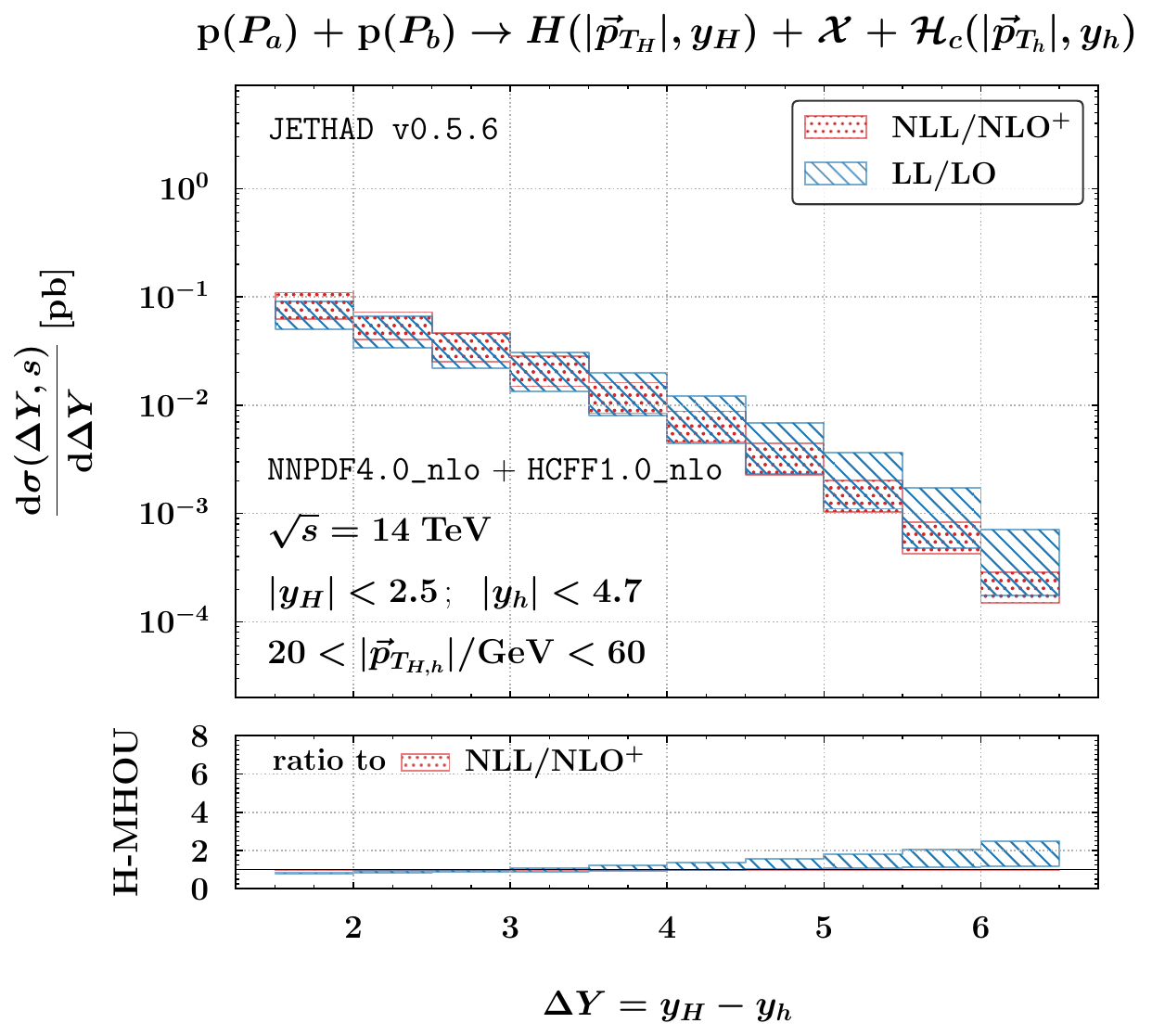}
   \hspace{-0.00cm}
   \includegraphics[scale=0.395,clip]{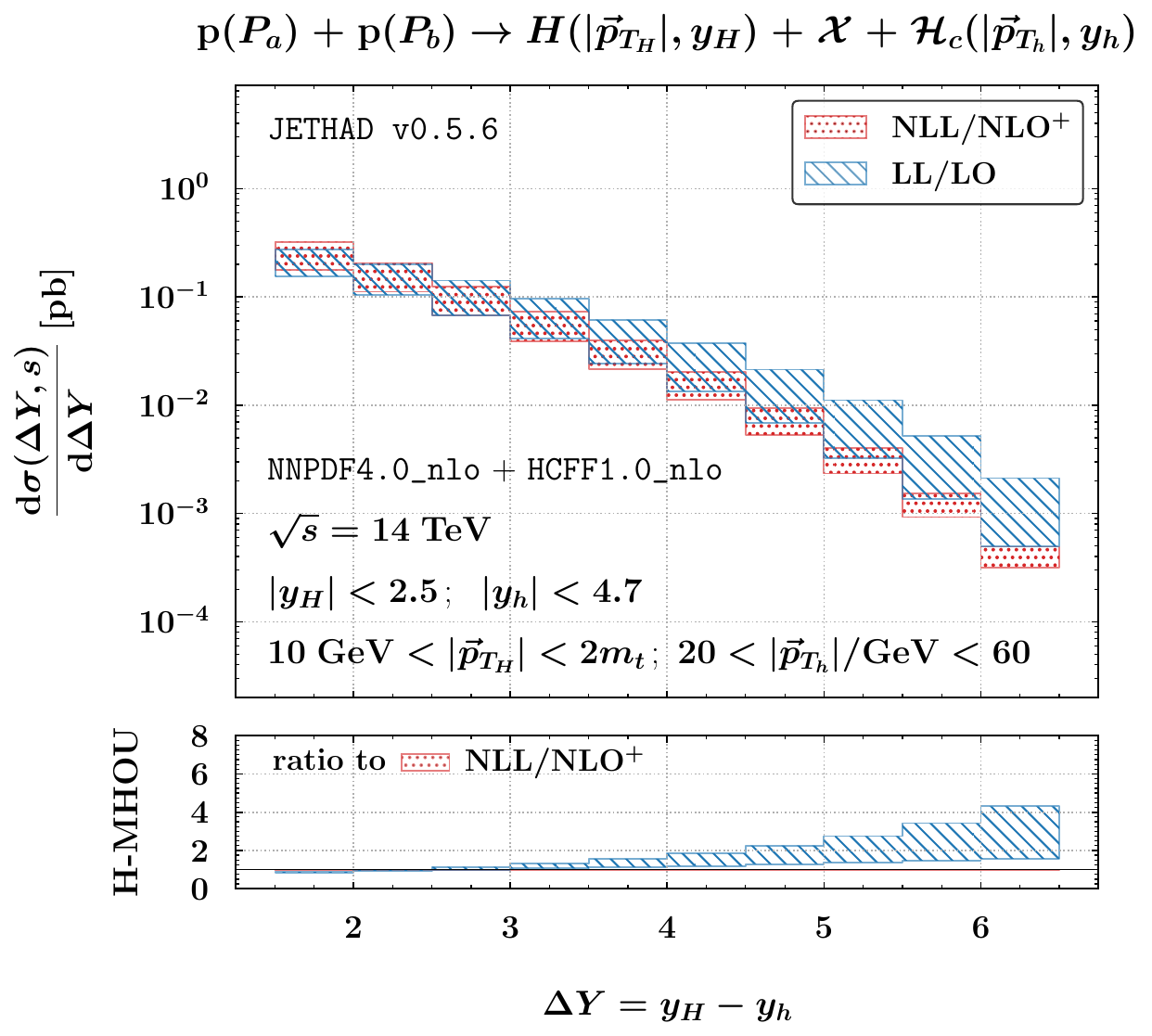}

\caption{Rapidity-interval rates for Higgs-plus-charmed-hadron production at the $\sqrt{s}=14$ TeV LHC, for symmetric (left column) and asymmetric (right column) transverse-momentum configurations, and for standard (upper row) and extended (lower row) rapidity cuts. 
Ancillary panels below the primary plots show the ratio of $\LL$  to $\NLLp$ predictions. 
Uncertainty bands capture the net effect of H-MHOUs and phase-space numerical integration.}
\label{fig:I_0cH}
\end{figure*}

\begin{figure*}[!t]
\centering

   \hspace{0.00cm}
   \includegraphics[scale=0.395,clip]{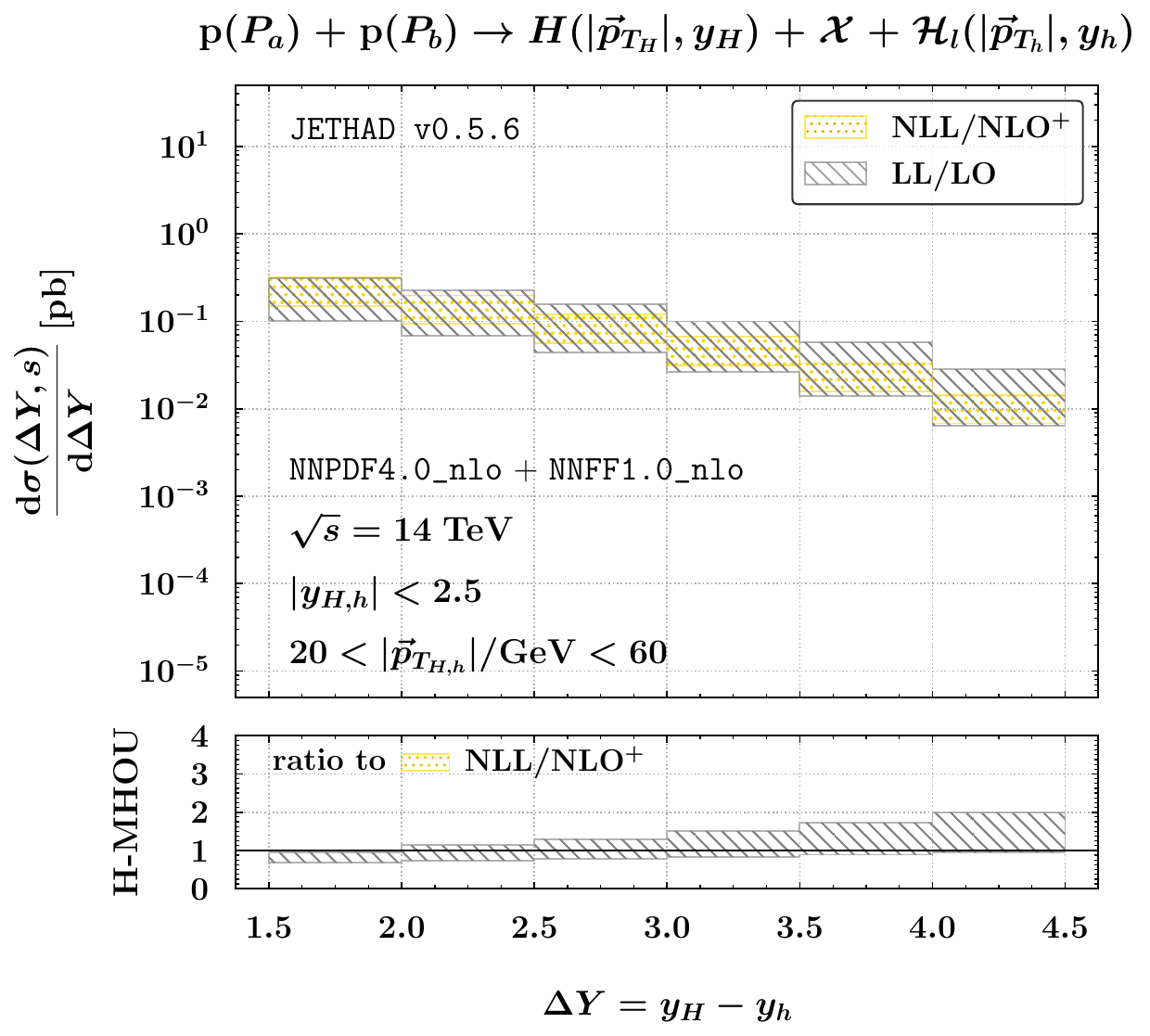}
   \hspace{-0.00cm}
   \includegraphics[scale=0.395,clip]{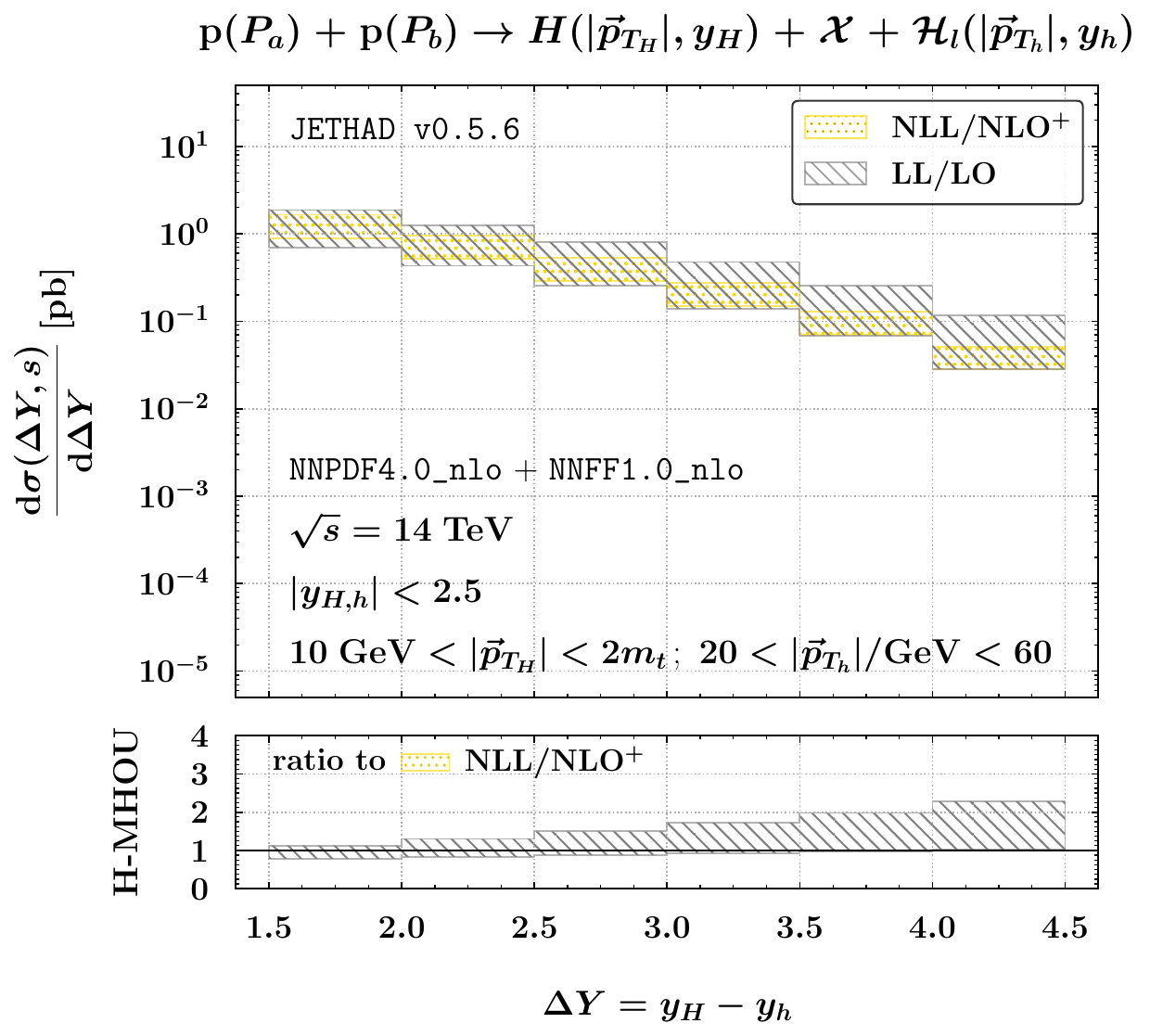}

   \vspace{0.50cm}

   \includegraphics[scale=0.395,clip]{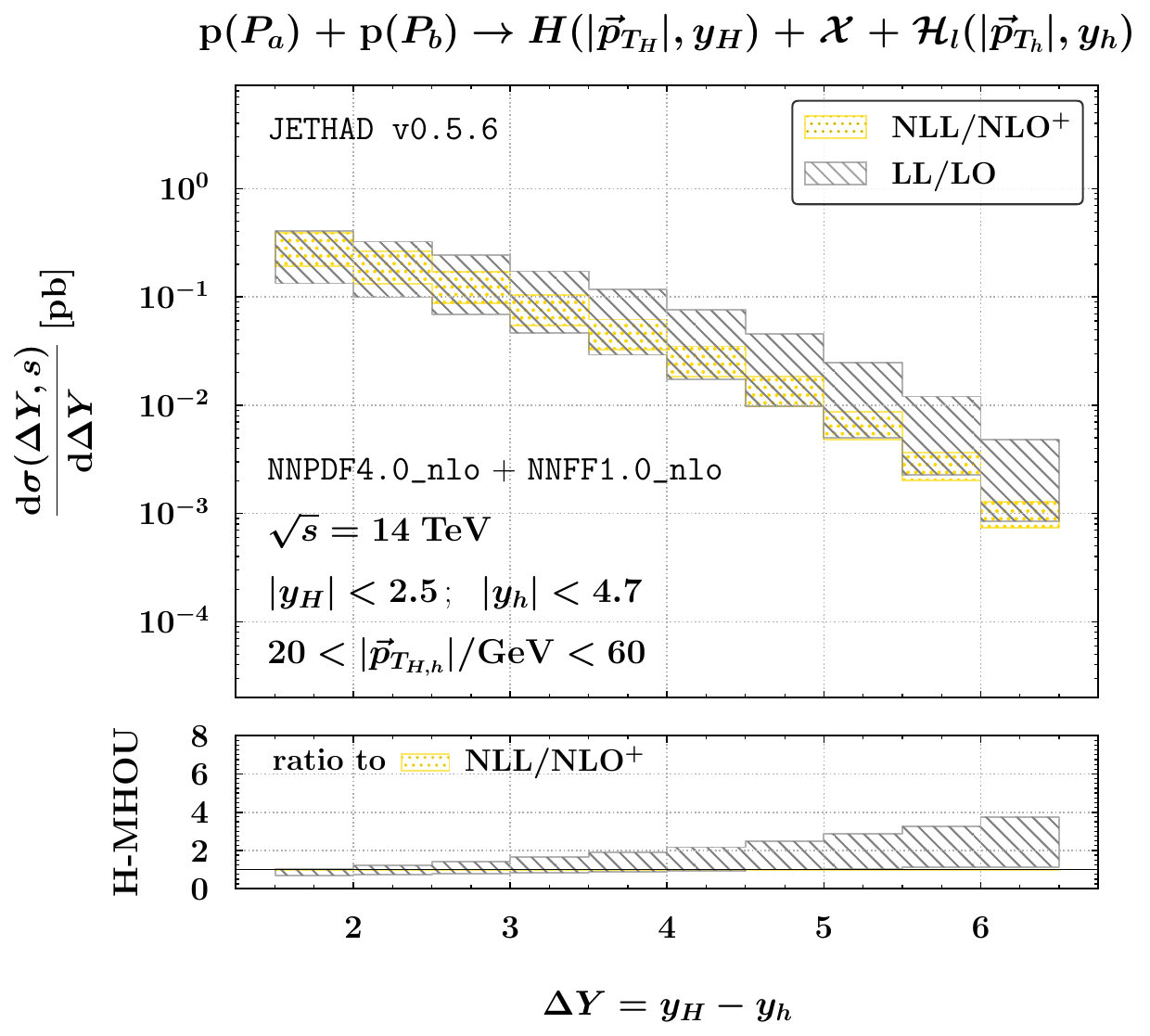}
   \hspace{-0.00cm}
   \includegraphics[scale=0.395,clip]{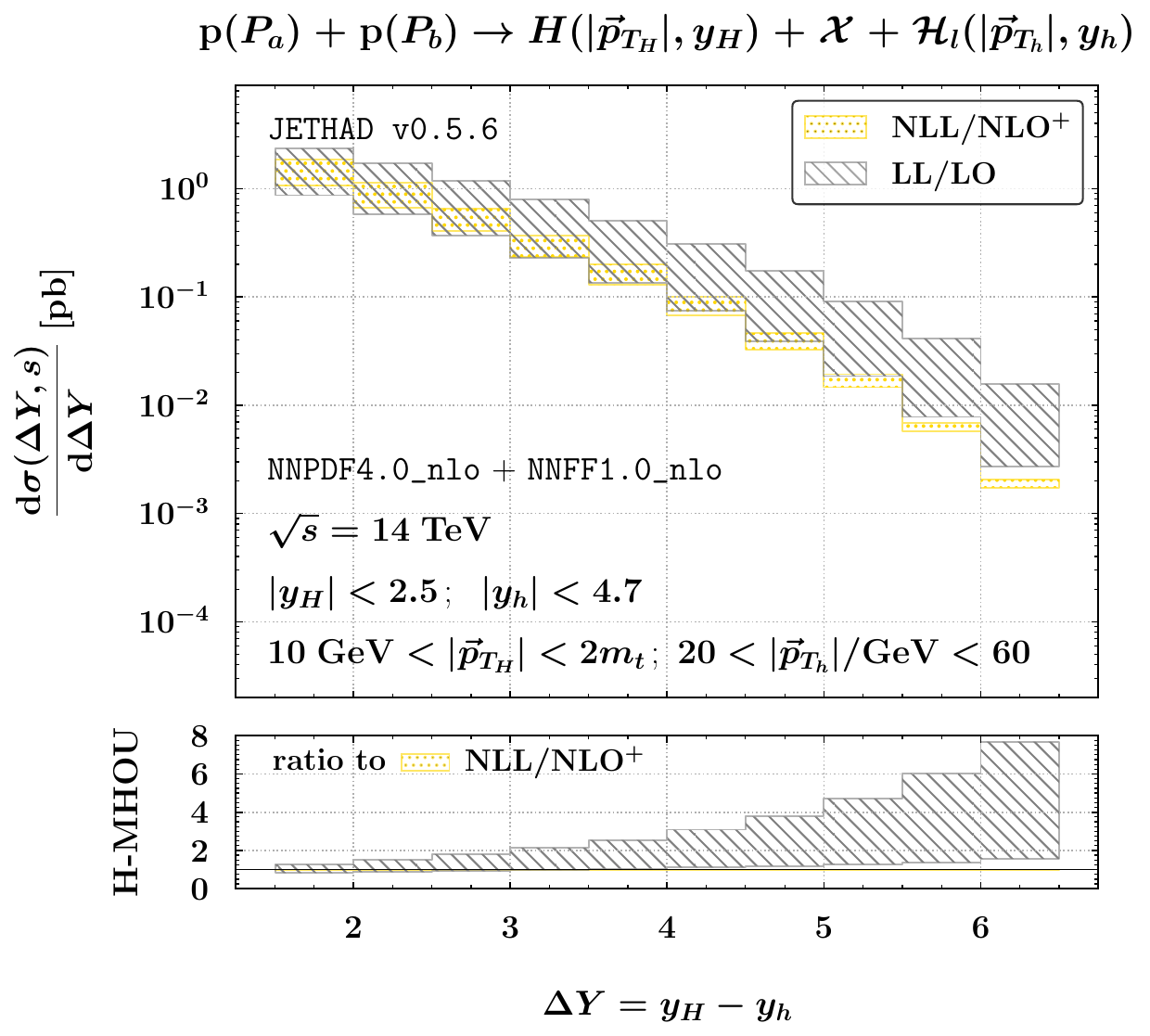}

\caption{Rapidity-interval rates for Higgs-plus-light-hadron production at the $\sqrt{s}=14$ TeV LHC, for symmetric (left column) and asymmetric (right column) transverse-momentum configurations, and for standard (upper row) and extended (lower row) rapidity cuts. 
Ancillary panels below the primary plots show the ratio of $\LL$  to $\NLLp$ predictions. 
Uncertainty bands capture the net effect of H-MHOUs and phase-space numerical integration.}
\label{fig:I_0h}
\end{figure*}

Figures~\ref{fig:I_0cH} and~\ref{fig:I_0h} present the rapidity-interval rates for Higgs production in association with, respectively, a singly charmed hadron and a light-flavored hadron.
For each final state, the four panels correspond to the complete set of fiducial configurations introduced in Sec.~\ref{ssec:kinematic_ranges}, with symmetric and asymmetric transverse-momentum selections displayed in the left and right columns, and standard and extended rapidity ranges in the upper and lower rows.
To facilitate future comparisons with experimentally binned measurements, predictions are reported as averages of $\drv\sigma/\drv\DY$ over finite $\DY$ intervals of width $\delta\DY=0.5$ rapidity units.
The ancillary panels show the $\LL$ versus $\NLLp$ ratio, thus exposing the relative impact of next-to-leading high-energy corrections across the explored rapidity range.

A common feature of all panels in Figs.~\ref{fig:I_0cH} and~\ref{fig:I_0h} is the pronounced decrease of the rapidity-interval rate as $\DY$ grows.
This behavior results from the competition between the increase of the underlying partonic cross section driven by high-energy evolution and the simultaneous suppression induced by the longitudinal convolutions with collinear PDFs and FFs.
At hadron level, the latter effect prevails, producing the characteristic falling pattern of the $\DY$ spectrum.

The Higgs-plus-charmed-hadron channel exhibits a remarkable perturbative stability.
In the standard rapidity range, $\NLLp$ predictions remain close to the corresponding $\LL$ results over the whole $\DY$ interval.
Moreover, the $\NLLp$ uncertainty bands are systematically narrower and, over a large fraction of the explored phase space, almost entirely nested within the $\LL$ ones.
This simultaneous reduction of the shift induced by next-to-leading corrections and of the sensitivity to H-MHOUs provides a particularly clear manifestation of perturbative stability at natural scales.
The agreement is strongest for the symmetric transverse-momentum selection, while a somewhat larger separation develops in the asymmetric configuration.
Extending the hadron rapidity coverage enhances the sensitivity to higher-order corrections, especially toward the largest values of $\DY$, yet the overall $\LL$ versus $\NLLp$ pattern remains comparatively stable.

A different picture emerges in the light-hadron comparison channel.
Although the same qualitative decrease with $\DY$ is observed, the separation between $\LL$ and $\NLLp$ predictions becomes systematically more pronounced as the rapidity interval increases.
This effect is amplified when moving from standard to extended rapidity coverage and is most evident for the asymmetric transverse-momentum selection, where the $\LL$ versus $\NLLp$ ratio departs markedly from unity in the large-$\DY$ region.
The contrast with the charmed-hadron channel therefore provides a direct phenomenological manifestation of the stabilizing role associated with heavy-flavor fragmentation.

This observation is consistent with the natural-stability mechanism identified in earlier heavy-flavor studies, where smoothly behaved gluon FFs were found to reduce the sensitivity of high-energy distributions to higher-order corrections.
In the present reaction, such a fragmentation-driven stabilization acts together with the intrinsically large hard scale supplied by the Higgs transverse mass.
The comparison between Figs.~\ref{fig:I_0cH} and~\ref{fig:I_0h} thus suggests that Higgs-plus-charm production combines two favorable ingredients for perturbatively stable high-energy phenomenology: the hard scale associated with Higgs emission and the collinear dynamics of heavy-flavor fragmentation.

The remarkable stability observed at the level of the $\DY$ rates, both under the inclusion of next-to-leading high-energy corrections and against H-MHOUs, provides only a first test of the perturbative behavior of the process.
A more differential assessment will be carried out in Sec.~\ref{ssec:Rnm} through the azimuthal-correlation moments and in Sec.~\ref{ssec:phi} through the normalized azimuthal distributions.
These observables separately probe the nonzero azimuthal harmonics and can therefore reveal higher-order effects that remain partly hidden in the azimuthally averaged coefficient $C_0$.

We do not display $\HENLOp$ predictions for the rapidity-interval rates.
Previous studies of semi-hard reactions have shown that $\DY$-differential cross sections generally possess a limited discriminating power between high-energy-resummed and high-energy fixed-order descriptions~\cite{Celiberto:2022keu}, and our preliminary tests confirm the same pattern for the present process.
A dedicated comparison between $\NLLp$ and $\HENLOp$ predictions is therefore deferred to the transverse-momentum rates of Sec.~\ref{ssec:pT}, whose broader scan of the hard kinematics provides a more sensitive probe of the genuine impact of the all-order high-energy resummation.

\subsection{Azimuthal-correlation moments}
\label{ssec:Rnm}

\begin{figure*}[!t]
\centering

   \hspace{0.00cm}
   \includegraphics[scale=0.395,clip]{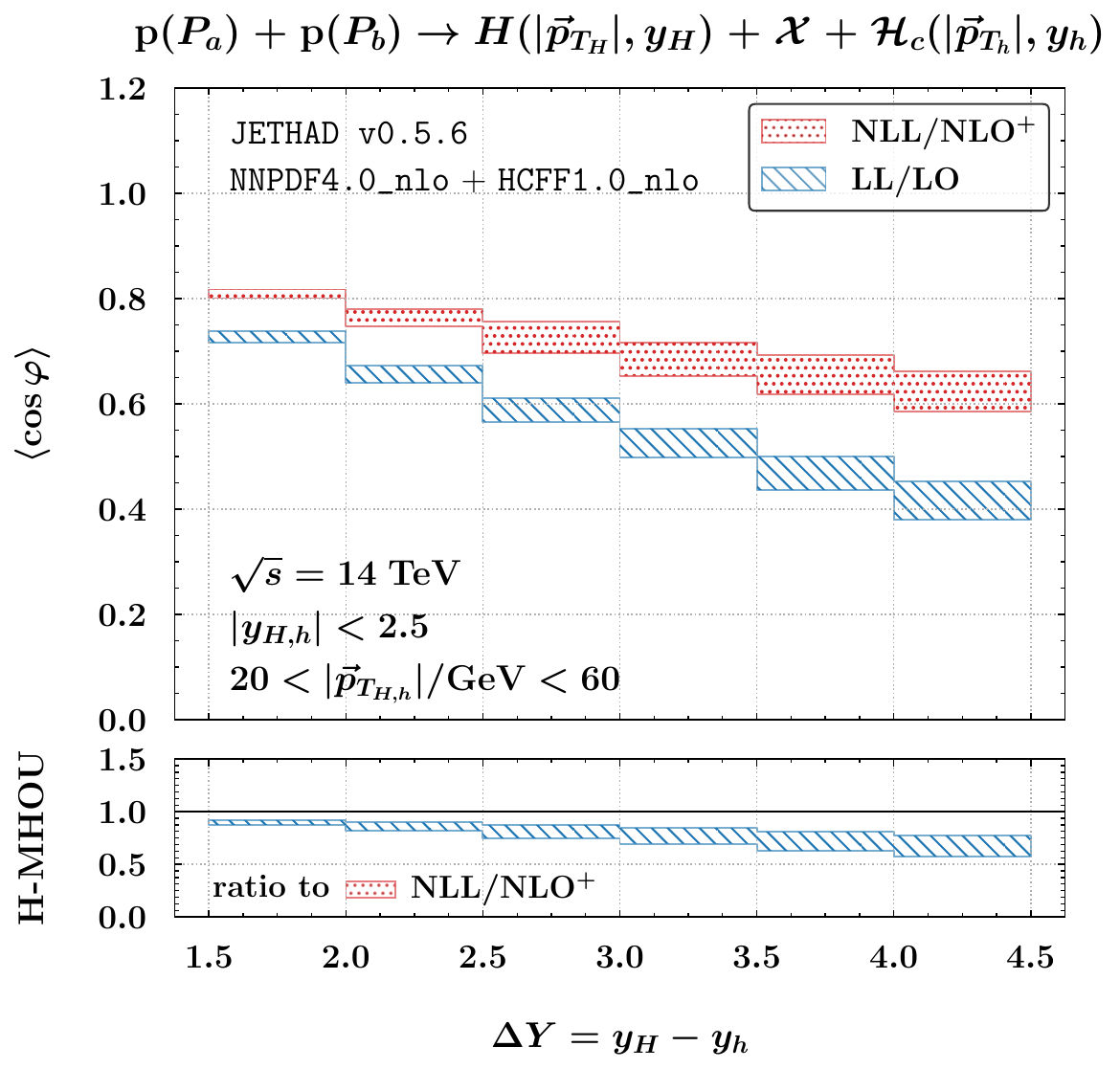}
   \hspace{-0.00cm}
   \includegraphics[scale=0.395,clip]{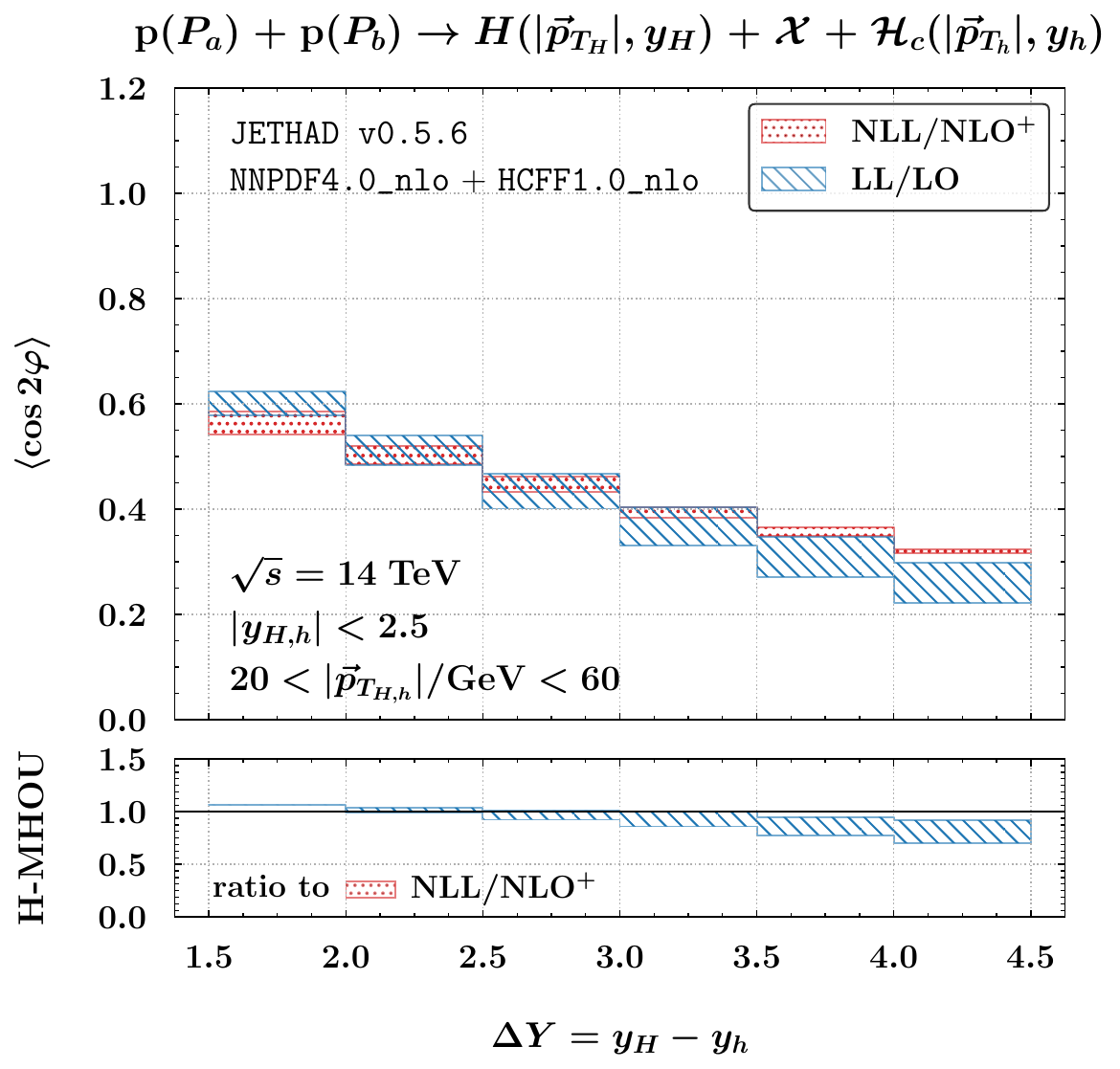}

   \vspace{0.50cm}

   \hspace{0.00cm}
   \includegraphics[scale=0.395,clip]{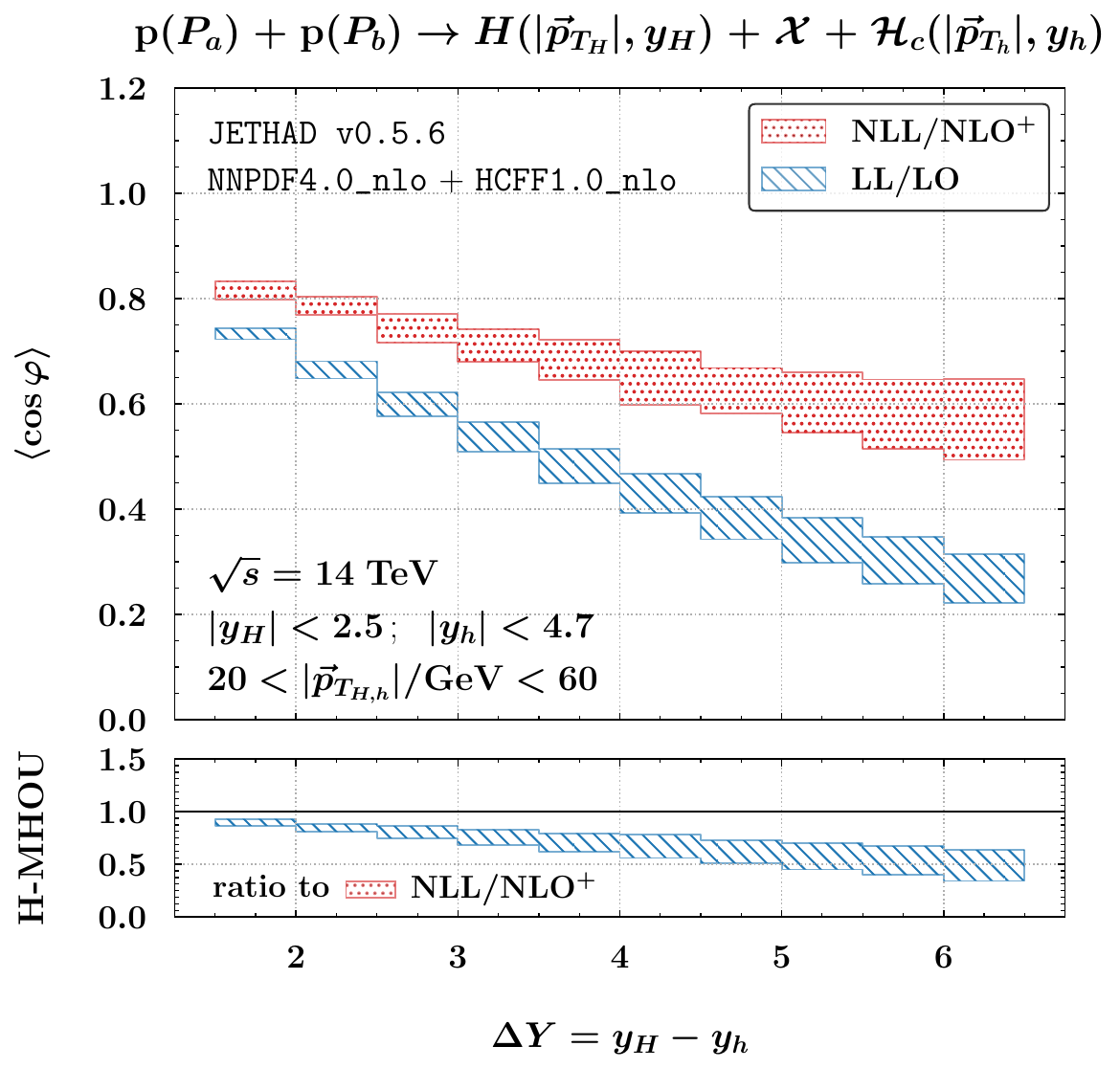}
   \hspace{-0.00cm}
   \includegraphics[scale=0.395,clip]{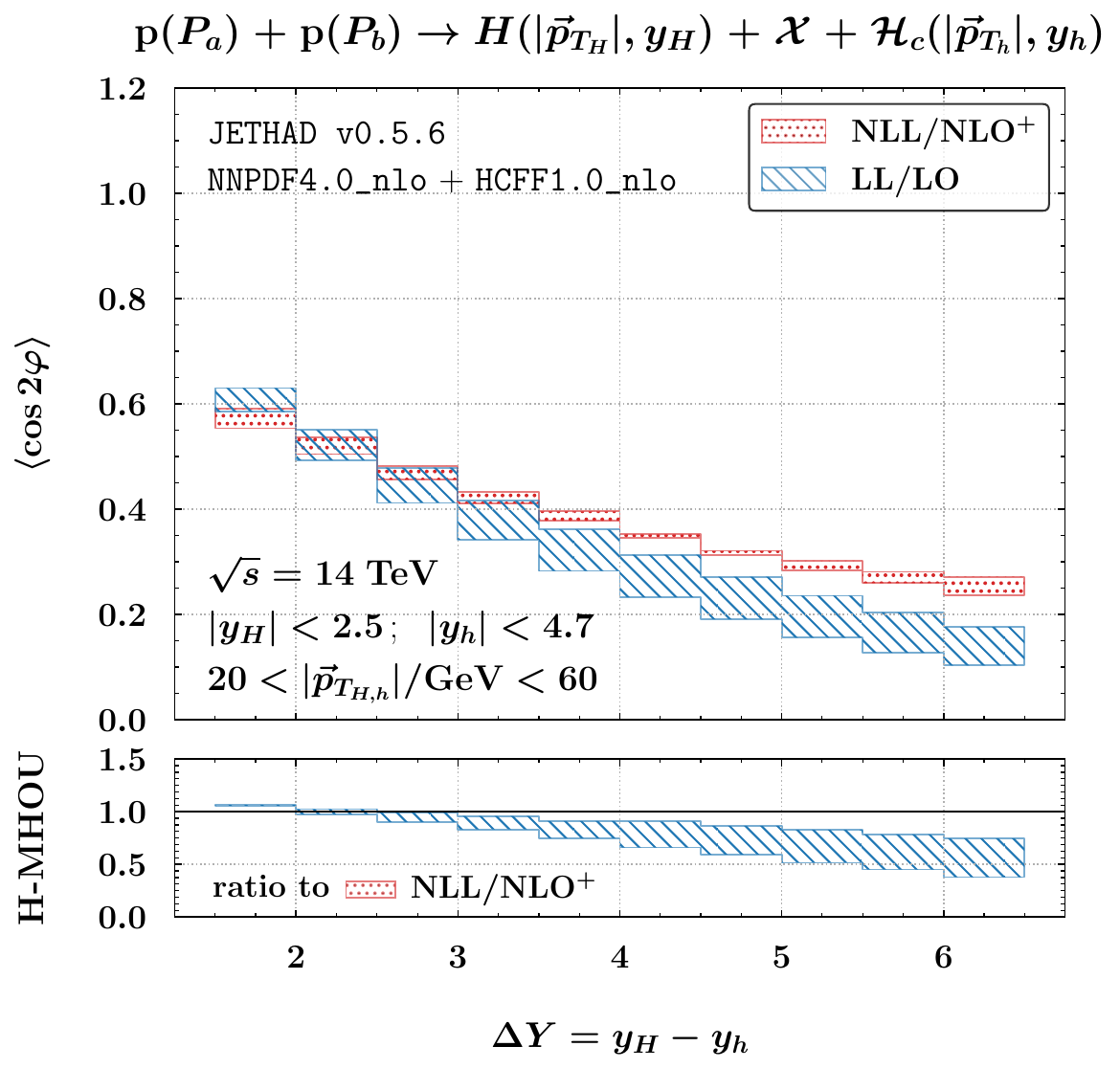}

\caption{Azimuthal-correlation moments $R_{10}=\langle\cos\varphi\rangle$ (left column) and $R_{20}=\langle\cos2\varphi\rangle$ (right column) for Higgs-plus-charmed-hadron production at the $\sqrt{s}=14$ TeV LHC, in the symmetric transverse-momentum configuration and for the standard (upper row) and extended (lower row) rapidity ranges of the charmed hadron. 
Ancillary panels show the $\LL$ versus $\NLLp$ ratio.
Uncertainty bands capture the net effect of H-MHOUs and phase-space numerical integration.}
\label{fig:I-R10-R20_0cH}
\end{figure*}

\begin{figure*}[!t]
\centering

   \hspace{0.00cm}
   \includegraphics[scale=0.395,clip]{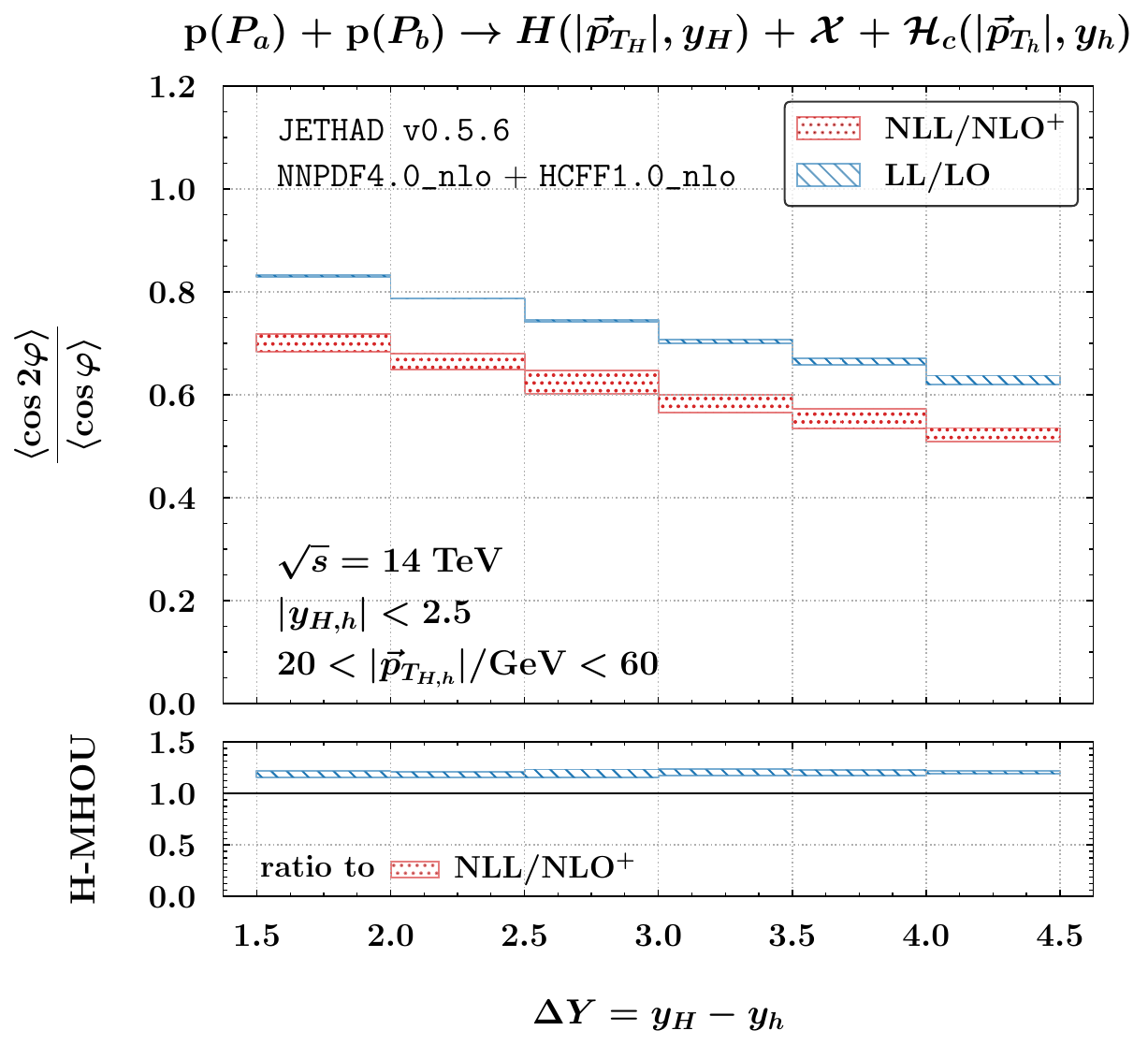}
   \hspace{-0.00cm}
   \includegraphics[scale=0.395,clip]{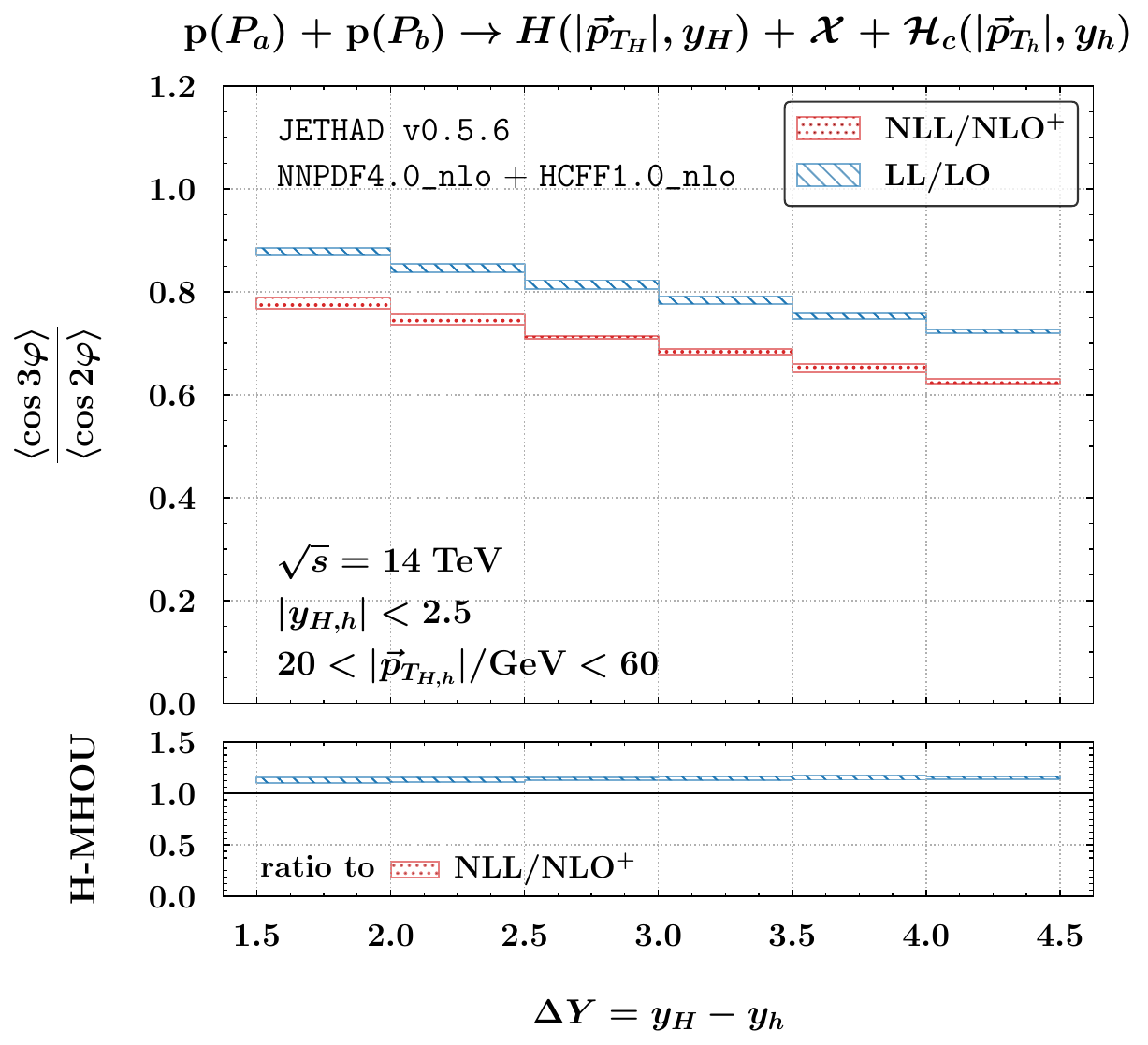}

   \vspace{0.50cm}

   \hspace{0.00cm}
   \includegraphics[scale=0.395,clip]{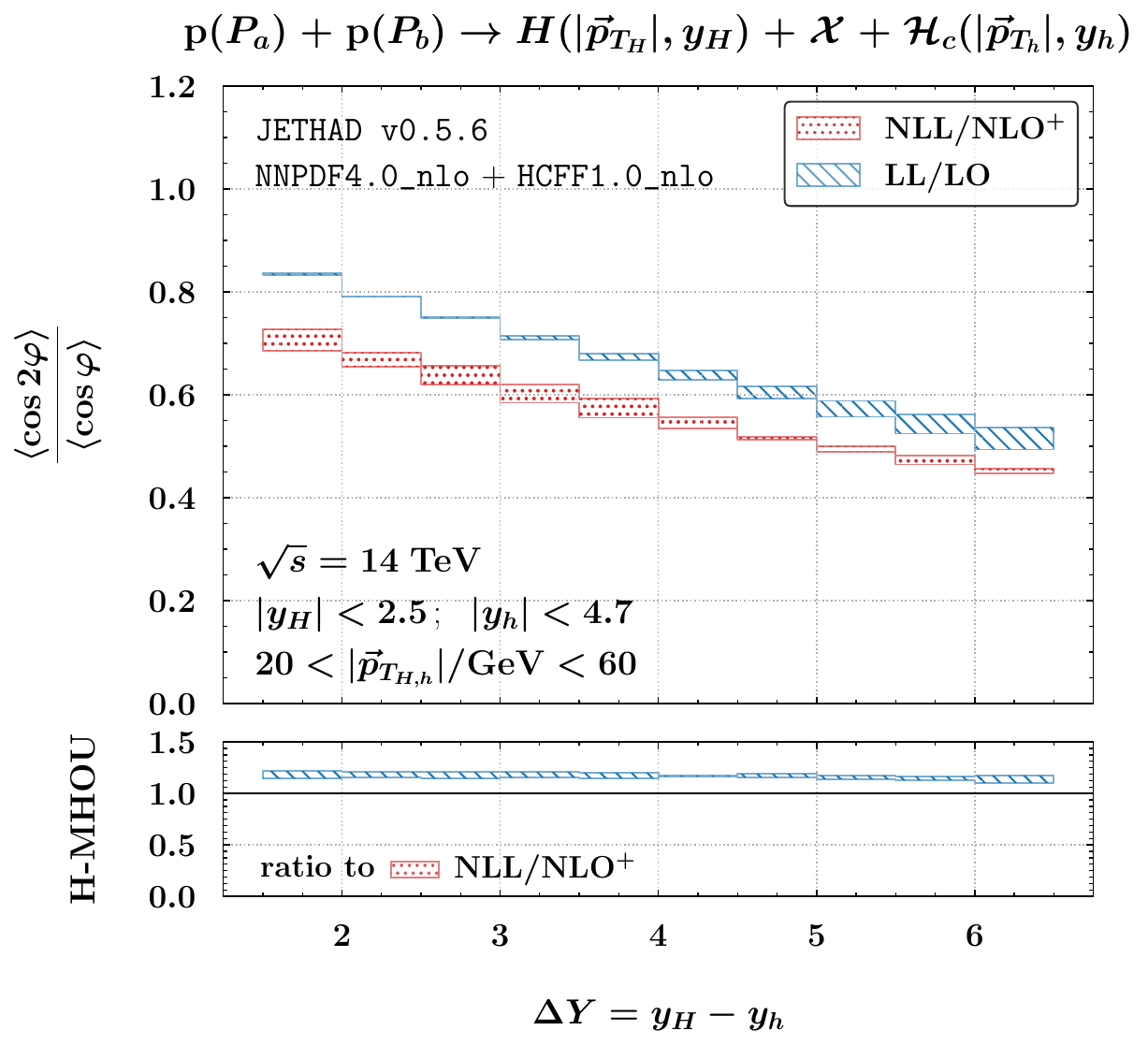}
   \hspace{-0.00cm}
   \includegraphics[scale=0.395,clip]{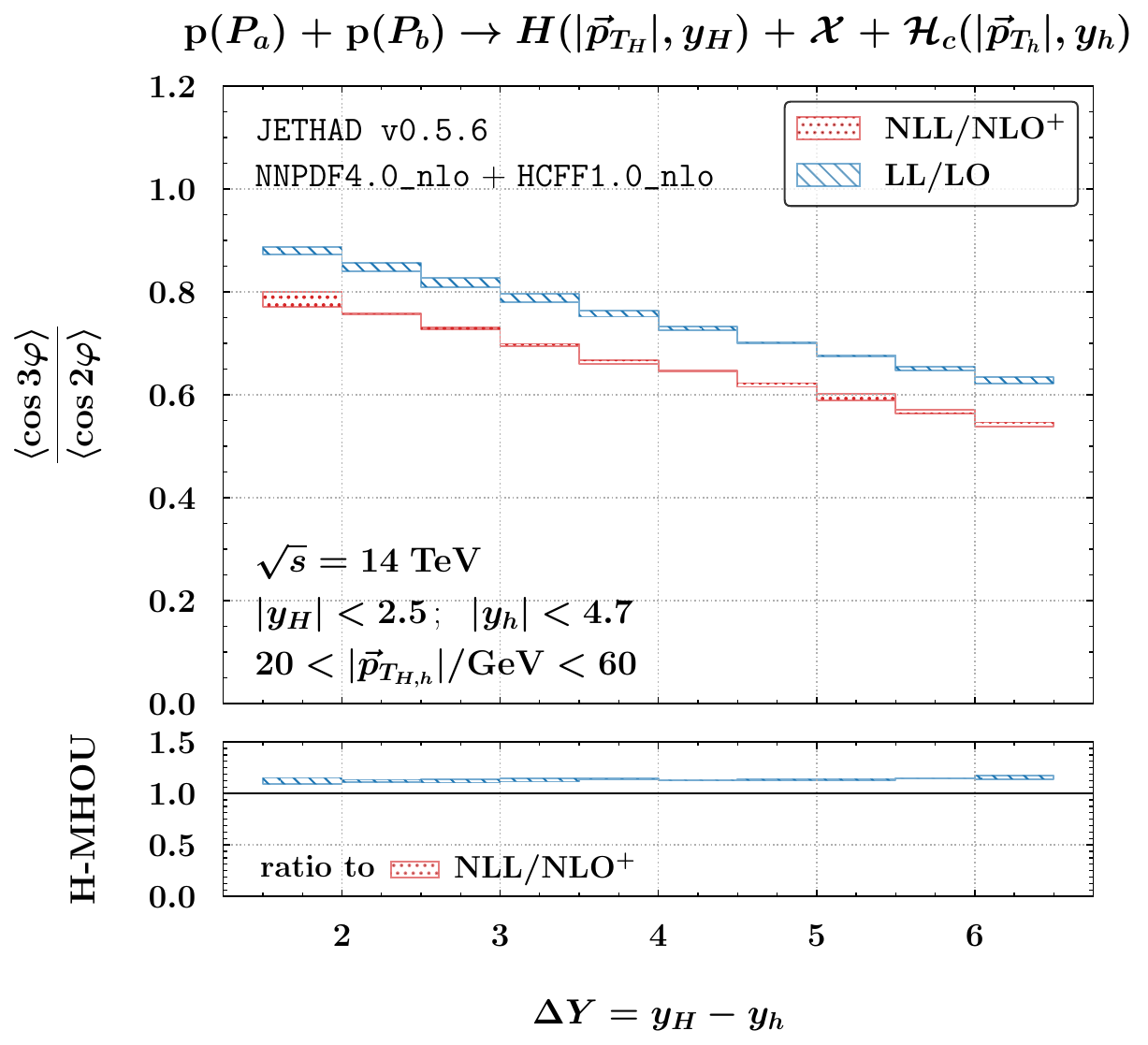}

\caption{Azimuthal-correlation moments $R_{21}=\langle\cos2\varphi\rangle/\langle\cos\varphi\rangle$ (left column) and $R_{32}=\langle\cos3\varphi\rangle/\langle\cos2\varphi\rangle$ (right column) for Higgs-plus-charmed-hadron production at the $\sqrt{s}=14$ TeV LHC, in the symmetric transverse-momentum configuration and for the standard (upper row) and extended (lower row) rapidity ranges of the charmed hadron. 
Ancillary panels show the $\LL$ versus $\NLLp$ ratio.
Uncertainty bands capture the net effect of H-MHOUs and phase-space numerical integration.}
\label{fig:I-R21-R32_0cH}
\end{figure*}

Having established the perturbative stability of the azimuthally averaged rate, we now probe the angular structure of the Higgs-hadron system.
Azimuthal-correlation moments provide a more selective test of high-energy dynamics, since they resolve the individual angular harmonics generated by the BFKL evolution and quantify the progressive decorrelation induced by unresolved radiation in the rapidity interval between the two tagged objects.

For each harmonic $n\geq1$, the phase-space integrated coefficient $C_n^{\rm [level]}(\DY,s)$ is obtained in complete analogy with Eq.~\eqref{DY_distribution}, by performing the same integrations at fixed $\DY$ and replacing ${\cal C}_0^{\rm [level]}$ with the corresponding angular coefficient ${\cal C}_n^{\rm [level]}$.

Azimuthal-correlation moments are then built by normalizing these coefficients to the azimuthally averaged contribution,
\begin{equation}
\begin{split}
\label{Rnm}
R_{n0}^{\rm [level]}(\DY,s)
&=
\frac{C_n^{\rm [level]}(\DY,s)}
     {C_0^{\rm [level]}(\DY,s)}
\equiv
\left\langle \cos(n\varphi)\right\rangle \; ,
\\[0.10cm]
R_{nm}^{\rm [level]}(\DY,s)
&=
\frac{C_n^{\rm [level]}(\DY,s)}
     {C_m^{\rm [level]}(\DY,s)}
\equiv
\frac{\left\langle \cos(n\varphi)\right\rangle}
     {\left\langle \cos(m\varphi)\right\rangle} \; ,
\end{split}
\end{equation}
with $n,m\geq1$~\cite{Vera:2006un,Vera:2007kn}.
The $R_{n0}^{\rm [level]}$ moments directly quantify the degree of azimuthal correlation, while ratios among nonzero harmonics provide complementary information in which the normalization to the total rate is removed.

In our analysis we focus on $R_{10}$, $R_{20}$, $R_{21}$, and $R_{32}$.
Unlike the rapidity-interval rates of Sec.~\ref{ssec:DY}, these observables are studied only in the symmetric transverse-momentum configuration, considering both the standard and extended rapidity ranges introduced in Sec.~\ref{ssec:kinematic_ranges}.
This restriction is made solely for simplicity and conciseness, while retaining the two rapidity configurations needed to inspect the evolution of the angular pattern over different $\DY$ ranges.
As for the rapidity-interval rates, all results are evaluated over finite $\DY$ bins of width $\delta\DY=0.5$.

Figures~\ref{fig:I-R10-R20_0cH} and~\ref{fig:I-R21-R32_0cH} display the resulting azimuthal moments.
The former collects $R_{10}$ (left column) and $R_{20}$ (right column), whereas the latter shows the ratios among nonzero harmonics, $R_{21}$ (left column) and $R_{32}$ (right column).
In both figures, the upper and lower rows correspond to the standard and extended rapidity selections, respectively.
Ancillary panels report the $\LL$ versus $\NLLp$ ratio, providing a direct view of the size and $\DY$ dependence of next-to-leading high-energy corrections.

All four observables decrease as the Higgs-hadron rapidity separation grows.
This is the characteristic signature of high-energy azimuthal decorrelation: enlarging $\DY$ opens an increasingly wider phase space for unresolved emissions between the two tagged objects, thereby progressively washing out their nearly back-to-back configuration.
Such a pattern is a well-established feature of semi-hard reactions and was already observed in earlier Higgs-plus-jet and Higgs-plus-charm studies.

The $R_{10}$ and $R_{20}$ moments provide the more stringent test of the perturbative pattern discussed in Sec.~\ref{ssec:DY}, since both are normalized to the azimuthally averaged coefficient $C_0$.
Their $\DY$ dependence confirms that resolving the nonzero harmonics exposes a stronger sensitivity to higher-order corrections than the one visible in the total rapidity-interval rate.
This is expected on general grounds: azimuthal correlations are particularly sensitive to the radiation generated along the high-energy ladder, and therefore probe aspects of the BFKL dynamics that can remain partially hidden after azimuthal averaging.

A complementary picture is provided by $R_{21}$ and $R_{32}$.
These observables do not involve the $n=0$ component, which is known to be especially sensitive both to high-energy evolution and to the sources of perturbative instability associated with the collinear sector~\cite{Vera:2006un,Vera:2007kn}; see also the discussion in Sec.~\ref{ssec:natural_stability}.
Ratios among nonzero harmonics are consequently expected to display an enhanced perturbative robustness, since correlated effects affecting the individual coefficients partly cancel in the ratio.

This expectation is clearly reflected in Fig.~\ref{fig:I-R21-R32_0cH}.
Both $R_{21}$ and $R_{32}$ exhibit very narrow H-MHOU bands throughout the explored rapidity domain, for both standard and extended selections.
The separation between $\LL$ and $\NLLp$ central predictions remains smooth and moderate, generally at the ten-percent level and somewhat more pronounced for $R_{21}$ than for $R_{32}$.
Thus, removing the $n=0$ component strongly reduces the scale sensitivity without erasing the genuine effect of next-to-leading high-energy corrections.

Extending the hadron rapidity coverage preserves the same qualitative hierarchy while allowing the decorrelation pattern to be followed over a substantially larger $\DY$ interval.
The moments continue to decrease smoothly, with no indication of an enhancement of perturbative instabilities at the largest rapidity separations considered.
The combined picture from Figs.~\ref{fig:I-R10-R20_0cH} and~\ref{fig:I-R21-R32_0cH} therefore strengthens the evidence obtained from the rapidity-interval rates: the Higgs-plus-charm channel remains perturbatively well controlled also when its individual azimuthal modes are resolved.

\subsection{Normalized azimuthal distributions}
\label{ssec:phi}

A more differential view of the Higgs-hadron azimuthal dynamics is provided by normalized distributions in the relative angle
$\varphi=\varphi_H-\varphi_h-\pi$.
At a given perturbative level, they are reconstructed from the full set of azimuthal harmonics according to
\begin{equation}
\label{angular_distribution}
\frac{\drv \sigma^{\rm [level]}(\varphi,s)}
     {\sigma^{\rm [level]} \, \drv \varphi}
=
\frac{1}{2\pi}
+
\frac{1}{\pi}
\sum_{n=1}^{\infty}
\left\langle\cos(n\varphi)\right\rangle
\cos(n\varphi)
\; ,
\end{equation}
where
$\langle\cos(n\varphi)\rangle
=C_n^{\rm [level]}/C_0^{\rm [level]}$
are the correlation moments introduced in Sec.~\ref{ssec:Rnm}.
The coefficients $C_n^{\rm [level]}$ entering Eq.~\eqref{angular_distribution} are obtained through the same phase-space integrations discussed above, with the additional integration over a selected finite $\DY$ bin.

Normalized azimuthal distributions, also referred to as angular multiplicities, combine information from all Fourier harmonics into a single observable and therefore offer a complementary perspective on high-energy azimuthal decorrelation.
Their use in semi-hard reactions was originally explored in the context of light-dijet production~\cite{Marquet:2007xx} and has subsequently proved particularly useful for phenomenological studies of resummed high-energy dynamics~\cite{Celiberto:2022gji}.
Their differential dependence on $\varphi$ also makes them particularly suitable for direct comparisons with experimentally reconstructed angular spectra.

For this observable we make use of the asymmetric transverse-momentum configuration introduced in Sec.~\ref{ssec:kinematic_ranges}.
This choice is adopted for brevity and, for the qualitative study of the angular pattern, entails no loss of generality.
Both the standard and extended rapidity selections are considered.
For the former we study the two intervals
$1<\DY<2$ and $3<\DY<4$, whereas the extended selection additionally gives access to the larger-separation bin $5<\DY<6$.
Since Eq.~\eqref{angular_distribution} is symmetric under $\varphi\to-\varphi$, predictions are displayed over $0\leq\varphi\leq\pi$ and averaged over uniform angular bins of width $\delta\varphi=\pi/14$.

\begin{figure*}[!t]
\centering

   \hspace{0.00cm}
   \includegraphics[scale=0.395,clip]{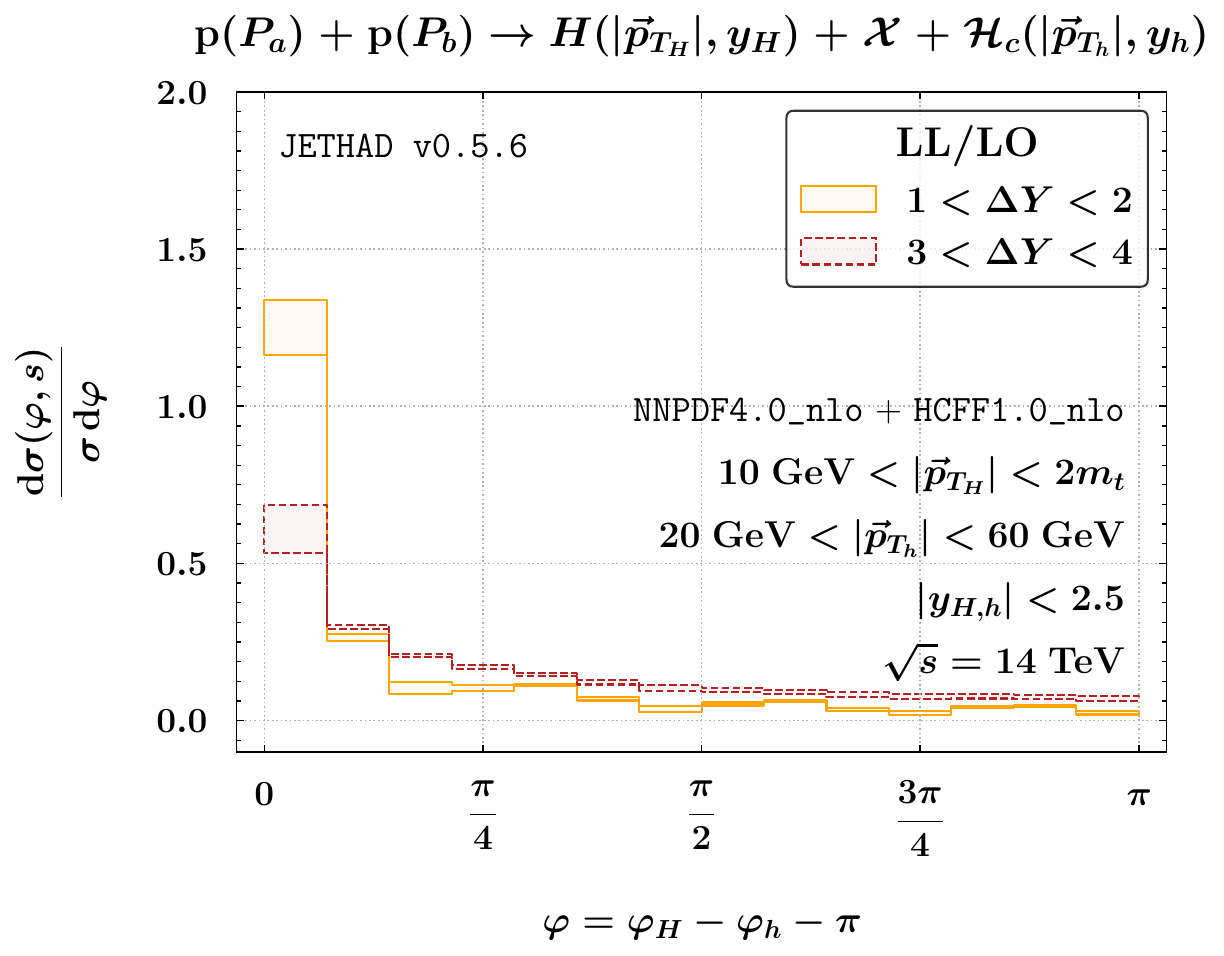}
   \hspace{-0.00cm}
   \includegraphics[scale=0.395,clip]{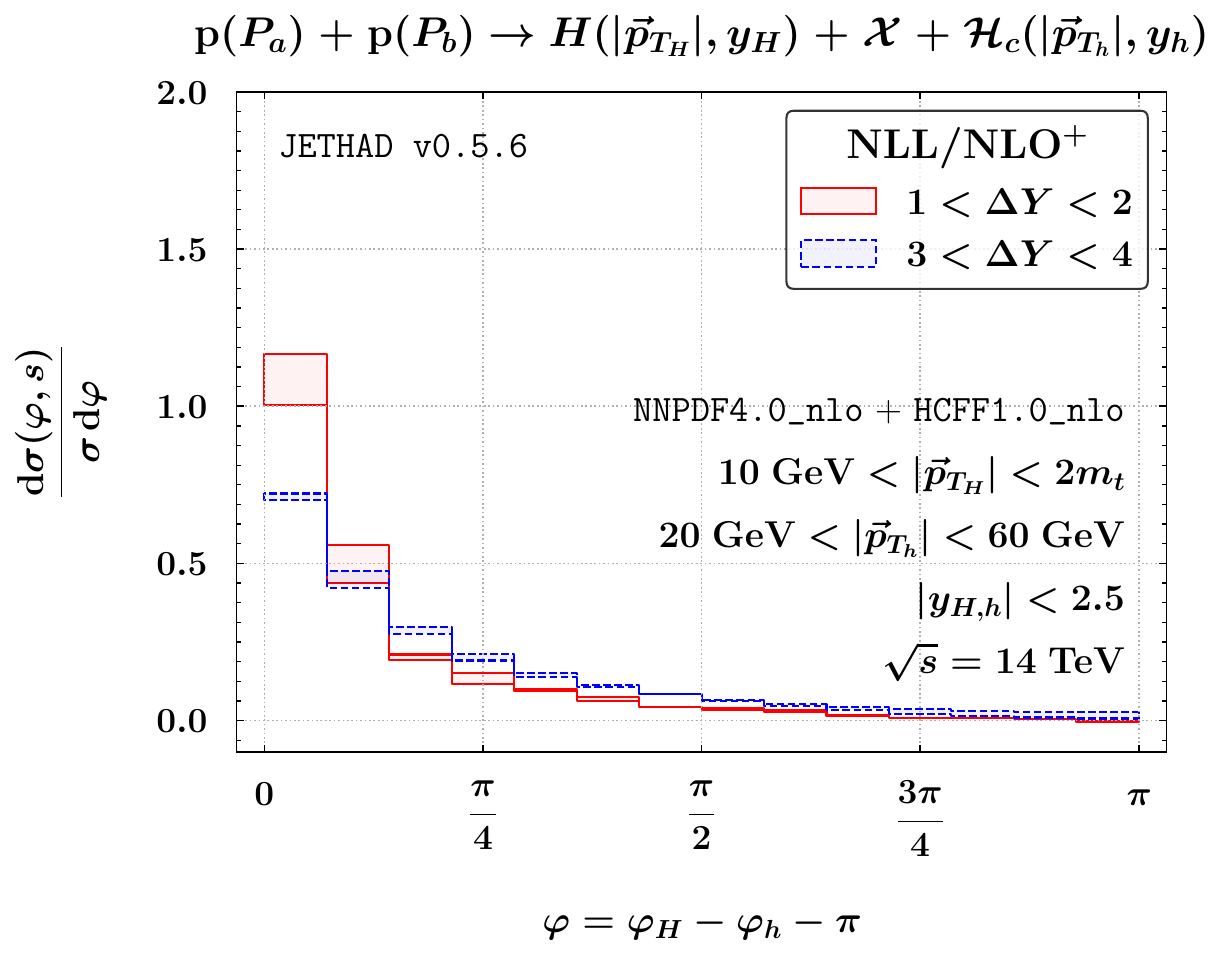}

   \vspace{0.50cm}

   \includegraphics[scale=0.395,clip]{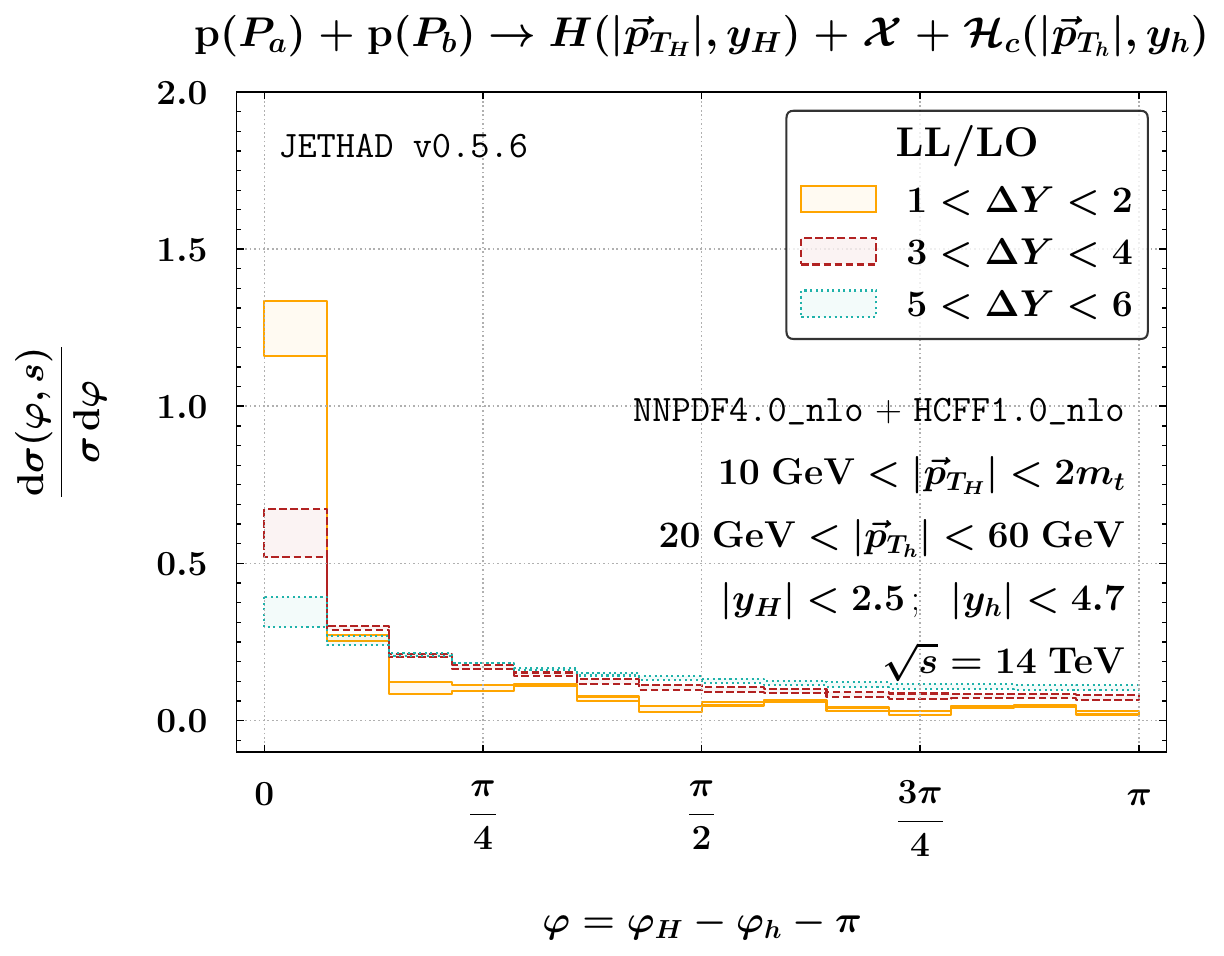}
   \hspace{-0.00cm}
   \includegraphics[scale=0.395,clip]{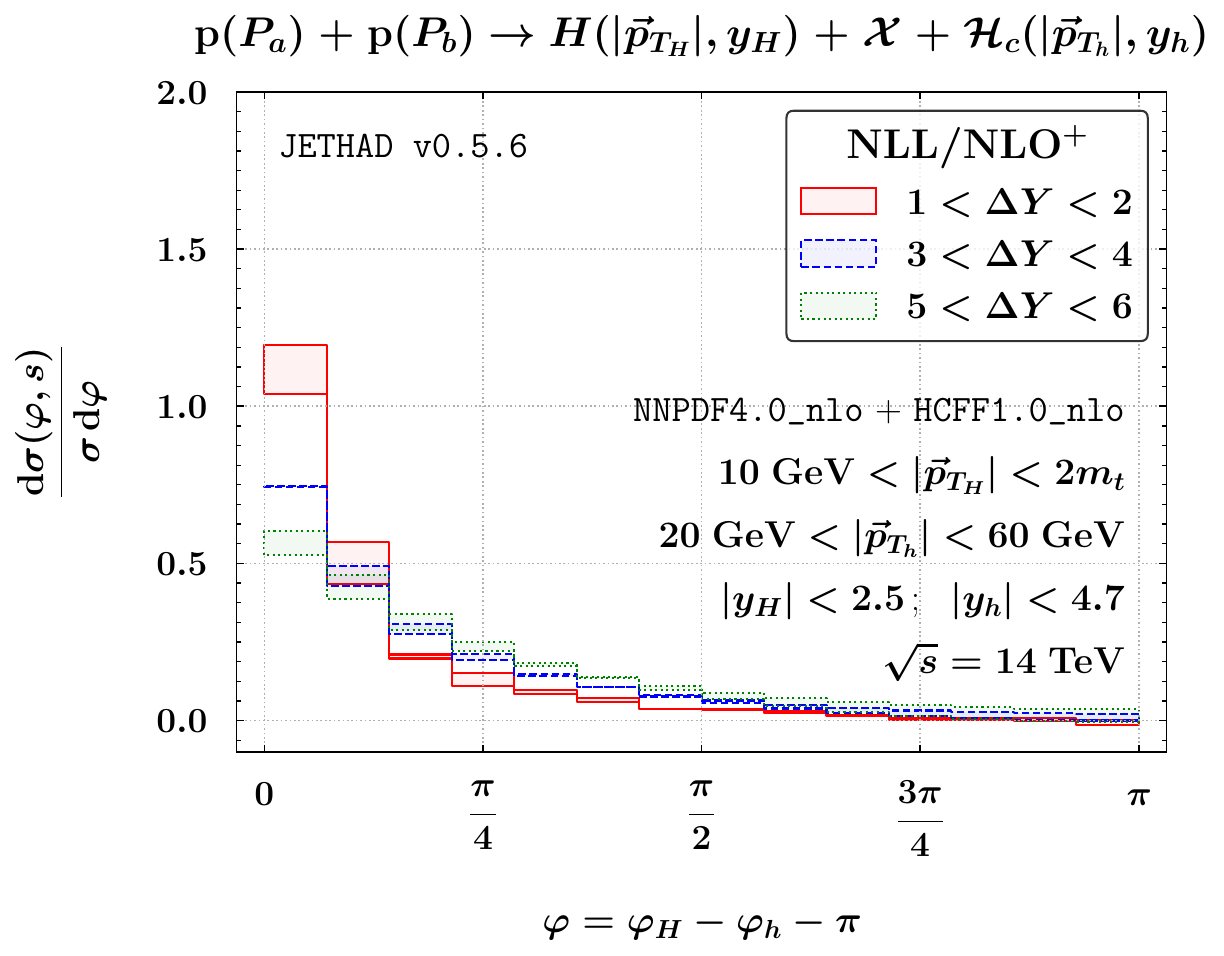}

\caption{Normalized azimuthal distributions at $\LL$ (left column) and $\NLLp$ (right column) for Higgs-plus-charmed-hadron production at the $\sqrt{s}=14$ TeV LHC, in the asymmetric transverse-momentum configuration and for standard (upper row) and extended (lower row) rapidity cuts.
Different curves correspond to selected rapidity-interval bins.
Uncertainty bands capture the net effect of H-MHOUs and phase-space numerical integration.}
\label{fig:I-phib}
\end{figure*}

Figure~\ref{fig:I-phib} collects the resulting distributions.
The left and right columns correspond, respectively, to $\LL$ and $\NLLp$ accuracy, while the upper and lower rows refer to the standard and extended rapidity selections.
Within each panel, the different curves correspond to increasing Higgs-hadron rapidity intervals, thus providing a direct visualization of the evolution of the azimuthal pattern with $\DY$.

All distributions exhibit a pronounced enhancement around $\varphi=0$, corresponding to configurations in which the Higgs boson and the charmed hadron are emitted nearly back to back.
This angular correlation is strongest in the lowest rapidity interval, $1<\DY<2$.
As the rapidity separation increases, the height of the near-back-to-back peak progressively decreases and weight is redistributed toward larger values of $\varphi$, leading to a broader angular profile.

This behavior provides a direct representation of the decorrelation mechanism already inferred from the moments of Sec.~\ref{ssec:Rnm}.
A larger $\DY$ opens additional phase space for unresolved radiation along the high-energy ladder.
The increasing activity in the intermediate rapidity region consequently weakens the correlation between the two tagged transverse momenta.
The extended selection makes this evolution especially transparent, since the additional $5<\DY<6$ interval exhibits the least pronounced peak and the broadest distribution among the configurations considered.

The comparison between the two columns shows that the qualitative decorrelation pattern is preserved when passing from $\LL$ to $\NLLp$ accuracy, while higher-order corrections modify the detailed redistribution among angular modes.
In particular, at the larger rapidity separations the $\NLLp$ prediction retains a more pronounced near-back-to-back structure than its $\LL$ counterpart, consistently with the stronger decorrelation generally generated by the LL series in semi-hard reactions.
At smaller $\DY$, by contrast, the higher-order correction to the peak is more moderate and can act in the opposite direction.
Thus, the effect of the NLL dynamics is not reducible to an overall normalization change, but genuinely reshapes the angular spectrum.

The uncertainty bands remain well controlled across the full angular range.
This is particularly significant for a distribution reconstructed from the complete tower of azimuthal harmonics: the perturbative stability observed separately in the $\DY$ rates and in the correlation moments survives after their information is recombined into a fully differential angular observable.
Figure~\ref{fig:I-phib} therefore provides a compact confirmation that the natural stability of Higgs-plus-charm production extends to the overall azimuthal structure of the final state.

\subsection{Transverse-momentum rates}
\label{ssec:pT}

Rapidity-interval and azimuthal observables provide direct access to the high-energy dynamics associated with large separations in rapidity, but they do not exhaust the information encoded in the transverse kinematics of the tagged final state.
Transverse-momentum spectra offer a complementary perspective, allowing us to test the stability of the high-energy description across different hard scales and to identify kinematic regions where logarithmic structures other than those resummed in the high-energy limit may become relevant.

Different corners of phase space generally call for dedicated resummation frameworks.
At low transverse momentum, large Sudakov logarithms require the use of transverse-momentum resummation techniques~\cite{Catani:2000vq,Bozzi:2005wk,Catani:2010pd,Duhr:2022yyp}, whereas configurations involving markedly different hard scales can enhance collinear DGLAP-type logarithms.
Toward the upper end of the spectrum, threshold-enhanced contributions may also become increasingly relevant.
These logarithmic structures are distinct from those resummed in the high-energy limit and would require dedicated or joint resummation strategies.
A unified treatment of these effects together with high-energy logarithms lies beyond the scope of the present analysis.

Transverse-momentum resummation has been extensively developed for a broad variety of collider observables. 
Representative examples include photon-pair production~\cite{Cieri:2015rqa,Alioli:2020qrd,Becher:2020ugp,Neumann:2021zkb}, Higgs~\cite{Ferrera:2016prr} and $W^\pm$ production~\cite{Ju:2021lah}, as well as associated channels such as Higgs plus jet~\cite{Monni:2019yyr,Buonocore:2021akg} and $Z$ plus photon~\cite{Wiesemann:2020gbm}.
Resummation accuracy up to third order has been reached for Drell--Yan and Higgs observables~\cite{Ebert:2020dfc,Re:2021con,Chen:2022cgv,Neumann:2022lft,Bizon:2017rah,Billis:2021ecs,Caola:2022ayt}.
Of particular relevance to the present study, transverse logarithms in Higgs-plus-jet production were jointly resummed in Ref.~\cite{Monni:2019yyr} within the {\RadISH} framework~\cite{Bizon:2017rah}.
Related transverse-momentum observables have also been investigated for leptonic $W^+W^-$ production at the LHC~\cite{Kallweit:2020gva}.

Within hybrid factorization, transverse-momentum spectra have already proved useful in several semi-hard reactions, encompassing Higgs-plus-jet~\cite{Celiberto:2020tmb}, heavy-flavor~\cite{Bolognino:2021mrc,Celiberto:2021fdp}, and exotic-matter production~\cite{Celiberto:2024mab,Celiberto:2024beg,Celiberto:2025ipt}.
Here, we consider rates differential
in the Higgs transverse momentum, $|\vec p_{T_H}|$, over the asymmetric window
$10~{\rm GeV}<|\vec p_{T_H}|<2m_t$, while the charmed-hadron transverse momentum is integrated over
$20<|\vec p_{T_h}|/{\rm GeV}<60$.
Predictions are evaluated for both the standard and extended rapidity acceptances introduced in Sec.~\ref{ssec:kinematic_ranges}.

The broad Higgs transverse-momentum scan is especially informative.
Besides testing the high-energy description over a wide range of hard scales, it reaches the region where heavy-quark-mass effects in the Higgs emission vertex can progressively become visible.
The spectra also provide a particularly suitable observable for comparing the $\NLLp$ resummed signal with its $\LL$ limit and with the high-energy fixed-order approximation $\HENLOp$, thereby exposing effects that were less discriminating in the rapidity-interval rates of Sec.~\ref{ssec:DY}.

The Higgs transverse-momentum rate is defined as
\begin{align}
\label{TM_distribution}
&\frac{\drv \sigma^{\rm [level]}(|\vec p_{T_H}|,s)}
     {\drv |\vec p_{T_H}|}
=
\int_{\DY^{\rm min}}^{\DY^{\rm max}}
\!\!\drv \DY
\int_{y_H^{\rm min}}^{y_H^{\rm max}}
\!\!\drv y_H
\int_{y_h^{\rm min}}^{y_h^{\rm max}}
\!\!\drv y_h
\nonumber
\\
&\times\,
\delta\!\left(\DY-y_H+y_h\right)
\int_{|\vec p_{T_h}|^{\rm min}}^{|\vec p_{T_h}|^{\rm max}}
\!\!\drv |\vec p_{T_h}|\,
{\cal C}_{0}^{\rm [level]}
\; .
\end{align}
For closer contact with experimentally binned spectra, predictions are evaluated as averages over uniform Higgs transverse-momentum bins of width
$\delta |\vec p_{T_H}|=10$~GeV.

\begin{figure*}[!t]
\centering

   \hspace{0.00cm}
   \includegraphics[scale=0.390,clip]{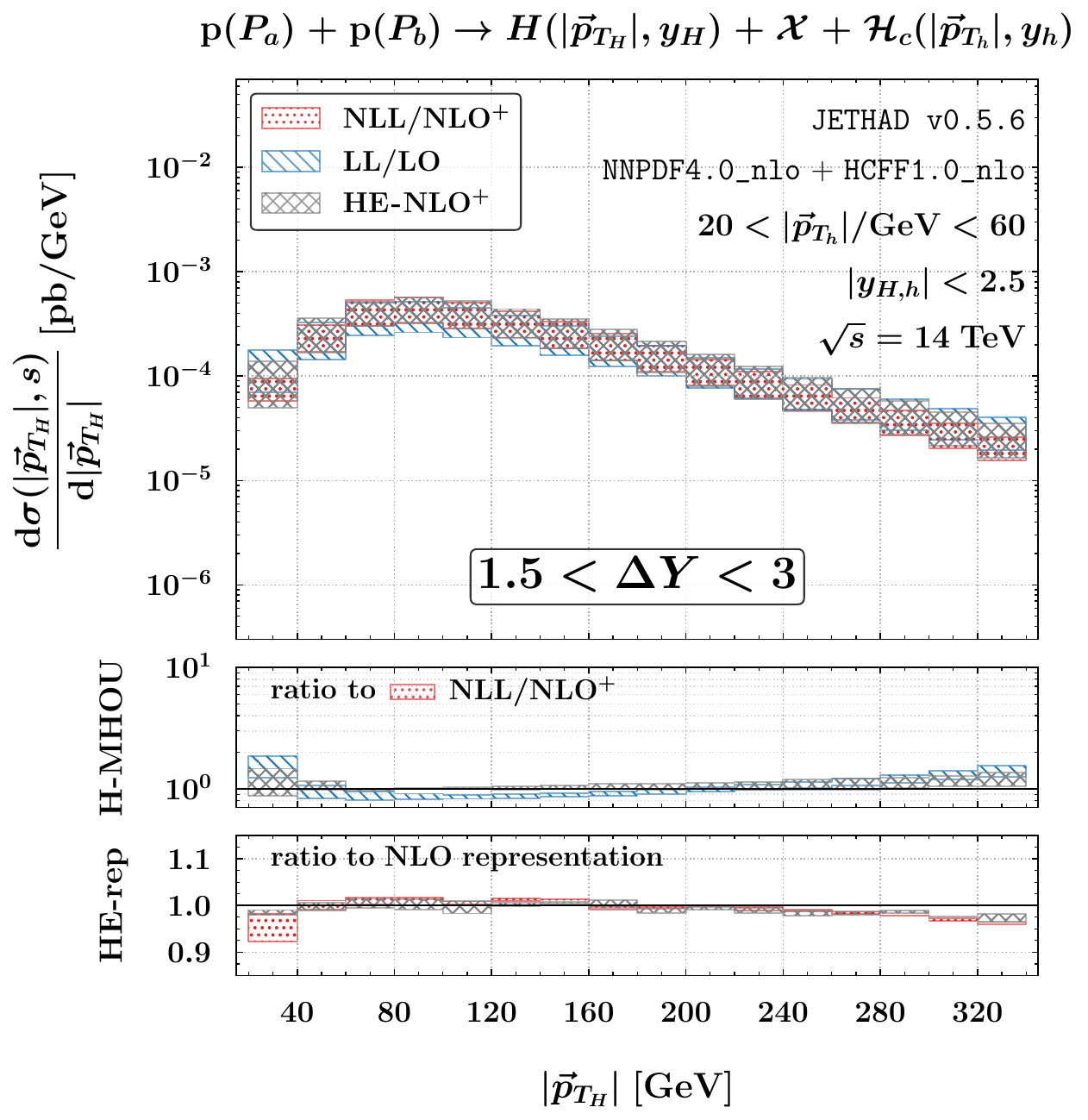}
   \hspace{-0.00cm}
   \includegraphics[scale=0.390,clip]{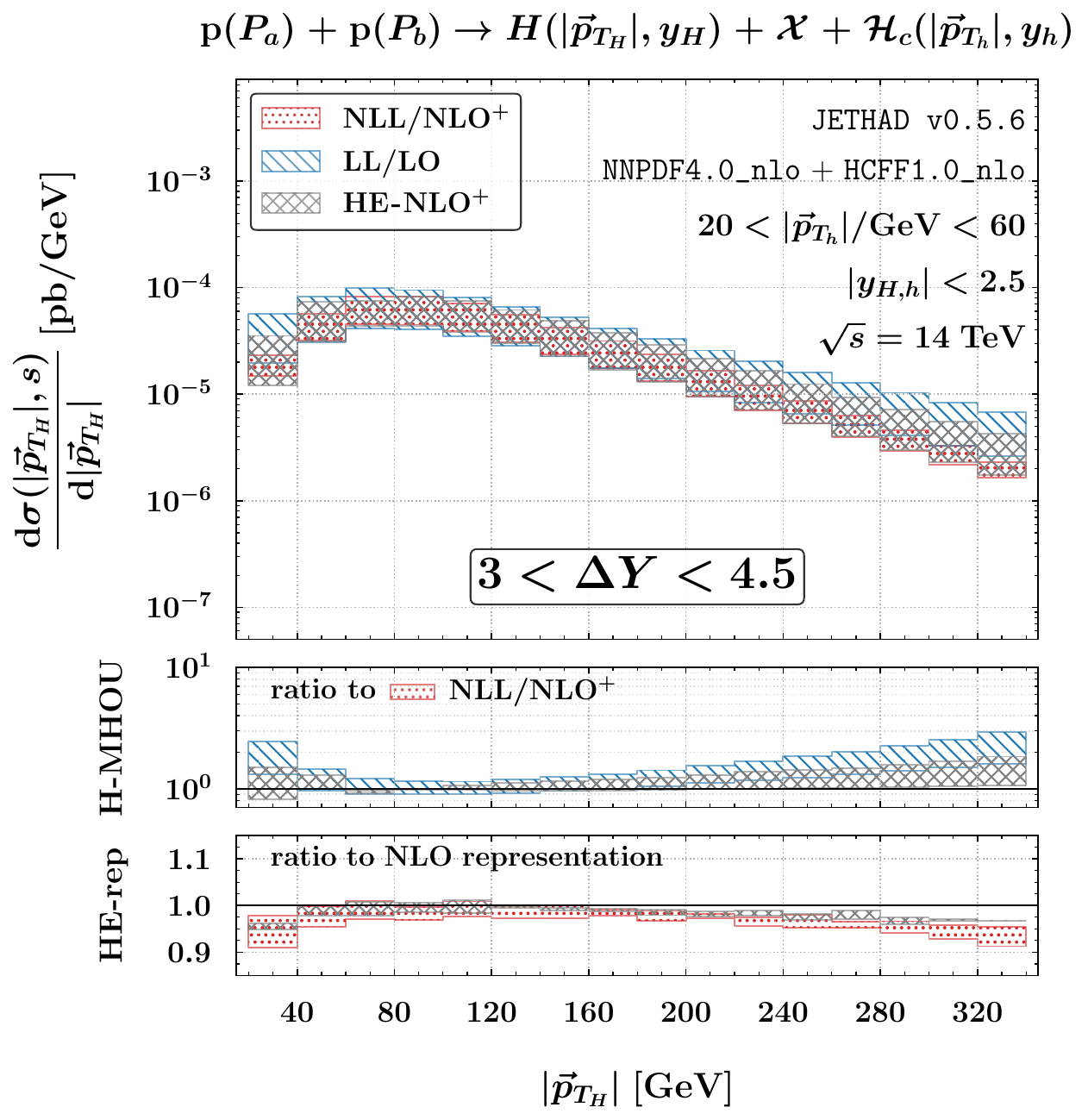}

   \hspace{0.00cm}
   \includegraphics[scale=0.390,clip]{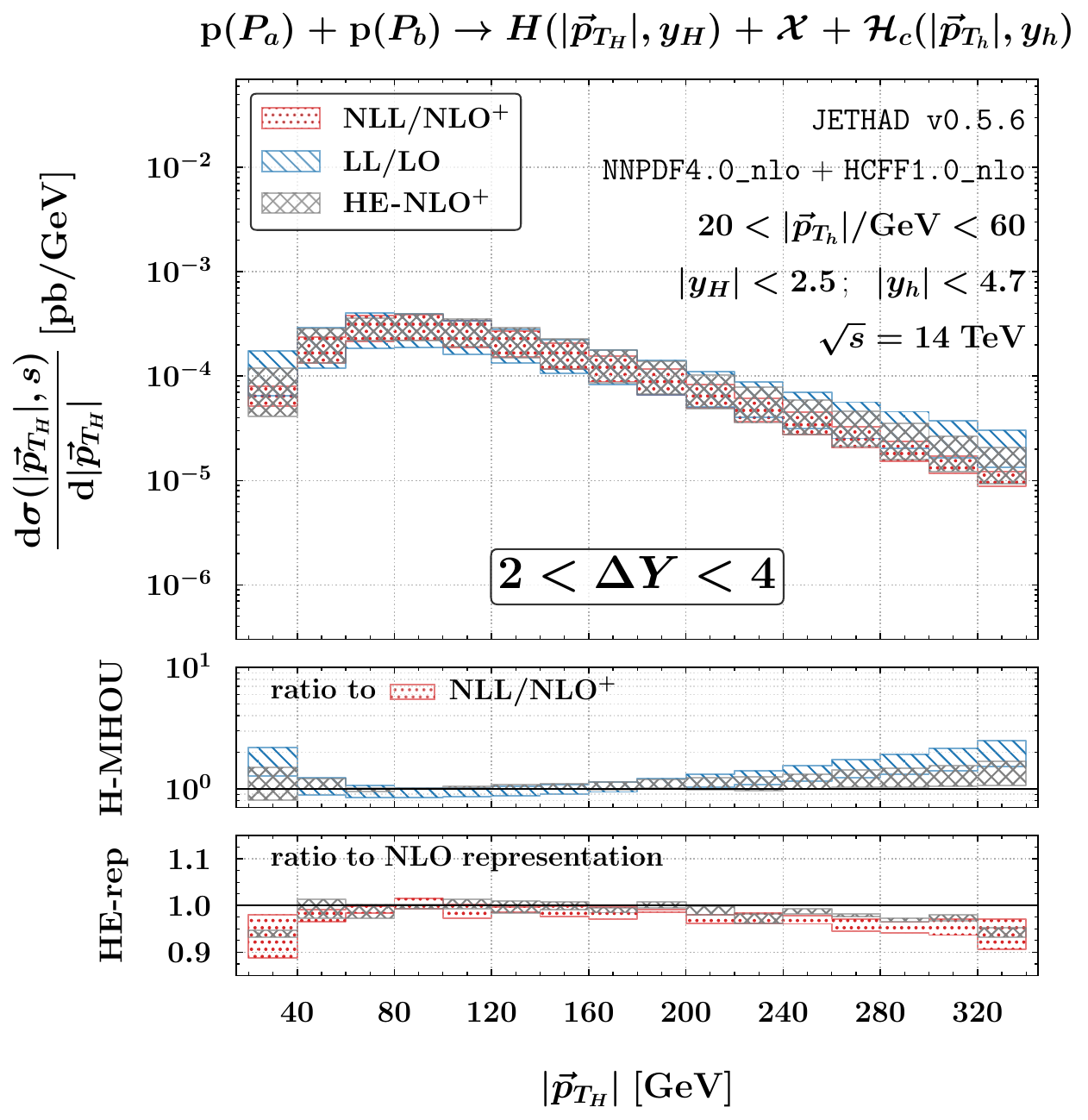}
   \hspace{-0.00cm}
   \includegraphics[scale=0.390,clip]{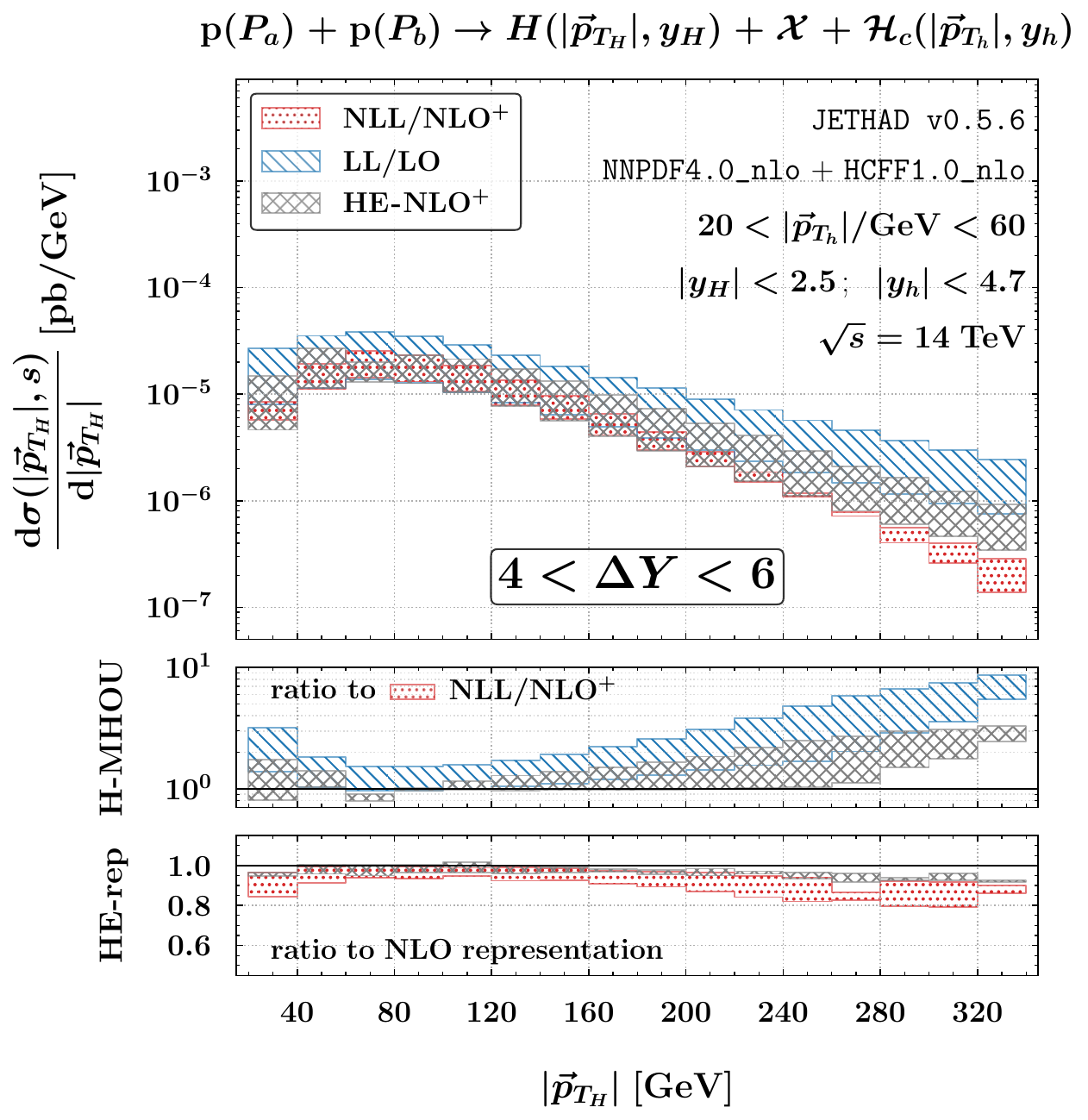}

\caption{Transverse-momentum rates for Higgs-plus-charmed-hadron production at the $\sqrt{s}=14$ TeV LHC, for the standard (upper row) and extended (lower row) rapidity range of the charmed hadron, and for moderate (left column) and large (right column) $\DY$ intervals.
For the standard selection, these correspond to $1.5<\DY<3$ and $3<\DY<4.5$, whereas the extended setup probes $2<\DY<4$ and $4<\DY<6$.
The first ancillary panels show the $\LL$ and $\HENLOp$ predictions normalized to the $\NLLp$ result.
The second ancillary panels show the ratios of the ``$+$'' representations, retaining terms beyond NLO, to their corresponding pure-NLO representations.
Uncertainty bands capture the net effect of H-MHOUs and phase-space numerical integration.}
\label{fig:I-k1b}
\end{figure*}

Figure~\ref{fig:I-k1b} collects our transverse-momentum rates.
The upper row corresponds to the standard rapidity selection, whereas the lower row refers to the extended one.
For each acceptance we consider two separated rapidity regimes.
The left column probes moderate separations, namely
$1.5<\DY<3$ for the standard selection and
$2<\DY<4$ for the extended one.
The right column instead explores larger rapidity intervals,
$3<\DY<4.5$ and $4<\DY<6$, respectively.

The first ancillary panel associated with each spectrum displays the $\LL$ and $\HENLOp$ predictions normalized to the $\NLLp$ result.
The comparison with $\HENLOp$ is particularly informative, since it isolates the effect of promoting the high-energy expansion from its NLO truncation to the corresponding all-order resummed description.
The second ancillary panel addresses a different issue.
As anticipated in Sec.~\ref{ssec:uncertainty}, we compare the default ``$+$'' representations with their counterparts in which the product of the two NLO emission functions is consistently truncated to pure NLO.
This directly quantifies the impact of the formally beyond-NLO terms generated by the unexpanded fixed-order product.
Such a representation dependence is displayed separately and is not folded into the uncertainty bands, since it is correlated with the perturbative higher-order ambiguity already probed through H-MHOUs.

All spectra exhibit a broad maximum in the low-to-intermediate $|\vec p_{T_H}|$ region, followed by a steadily decreasing tail toward $|\vec p_{T_H}|\simeq2m_t$.
Increasing the rapidity separation reduces the overall normalization and simultaneously enhances the discriminating power among the different perturbative descriptions.
This is particularly evident in the extended, large-$\DY$ configuration, where the hard tail spans several orders of magnitude and the separation among $\LL$, $\NLLp$, and $\HENLOp$ becomes most pronounced.

Around the broad maximum, where the Higgs and hadron transverse scales remain comparatively close, the three predictions exhibit their strongest mutual overlap.
The $\HENLOp$ versus $\NLLp$ ratio stays close to unity over a sizeable portion of this region, indicating that the all-order high-energy evolution produces only moderate corrections with respect to its fixed-order expansion.
This is also the domain where the $\NLLp$ uncertainty bands remain particularly narrow, consistently with the natural-stability pattern observed in the previous subsections.

The situation changes progressively along the high-$|\vec p_{T_H}|$ tail.
The $\LL$ versus $\NLLp$ ratio increasingly departs from unity, showing that next-to-leading corrections become essential once a sizeable imbalance develops between the Higgs and hadron transverse scales.
At the same time, the $\HENLOp$ versus $\NLLp$ ratio acquires a marked transverse-momentum dependence.
Its departure from unity grows with both $|\vec p_{T_H}|$ and $\DY$, becoming especially sizeable in the $4<\DY<6$ extended configuration.
Thus, unlike the rapidity-interval rates, the transverse-momentum spectrum provides a clear lever arm to resolve the genuinely all-order component of the high-energy resummation from its fixed-order high-energy truncation.

The pattern found here also illustrates the strongly process-dependent character of this comparison.
The relative ordering of $\HENLOp$ and $\NLLp$ cannot be inferred from high-energy power counting alone, since it results from the interplay among the kernel evolution and the NLO corrections entering the two process-dependent emission functions.
In the present Higgs-plus-charm channel, the high-energy fixed-order approximation becomes increasingly larger than the resummed result in the hardest and most rapidity-separated configurations.
This behavior therefore identifies the large-$|\vec p_{T_H}|$, large-$\DY$ domain as the region with the highest sensitivity to the resummation mechanism.

The lower ancillary panels provide an independent check of the perturbative representation.
Across most of the spectrum, the ratios between ``$+$'' and pure-NLO representations remain close to unity, showing that the formally beyond-NLO terms generated by the product of the two NLO emission functions are numerically controlled.
Their impact becomes somewhat more visible toward the edges of the transverse-momentum range and when the rapidity interval is enlarged, but remains substantially smaller than the separation observed between the different high-energy accuracies.
This supports our default use of the ``$+$'' representation while confirming the choice, discussed in Sec.~\ref{ssec:uncertainty},
of treating the corresponding variation as a dedicated theory diagnostic rather than as an additional independent uncertainty.

Finally, H-MHOUs remain under good control over most of the explored spectrum.
Their relative impact grows in the hardest and largest-$\DY$ configurations, where the various perturbative descriptions also become more separated.
We refrain, however, from interpreting this behavior as a breakdown of the high-energy expansion.
Rather, the broad transverse-momentum scan progressively approaches a multiscale regime in which high-energy dynamics coexist with other logarithmically enhanced structures not resummed within the present formalism.
The transverse-momentum rates therefore represent the most sensitive observable of our analysis for delineating the domain of natural high-energy stability and its transition toward more intricate QCD dynamics.

\begin{figure*}[!t]
\centering

   \hspace{0.00cm}
   \includegraphics[scale=0.390,clip]{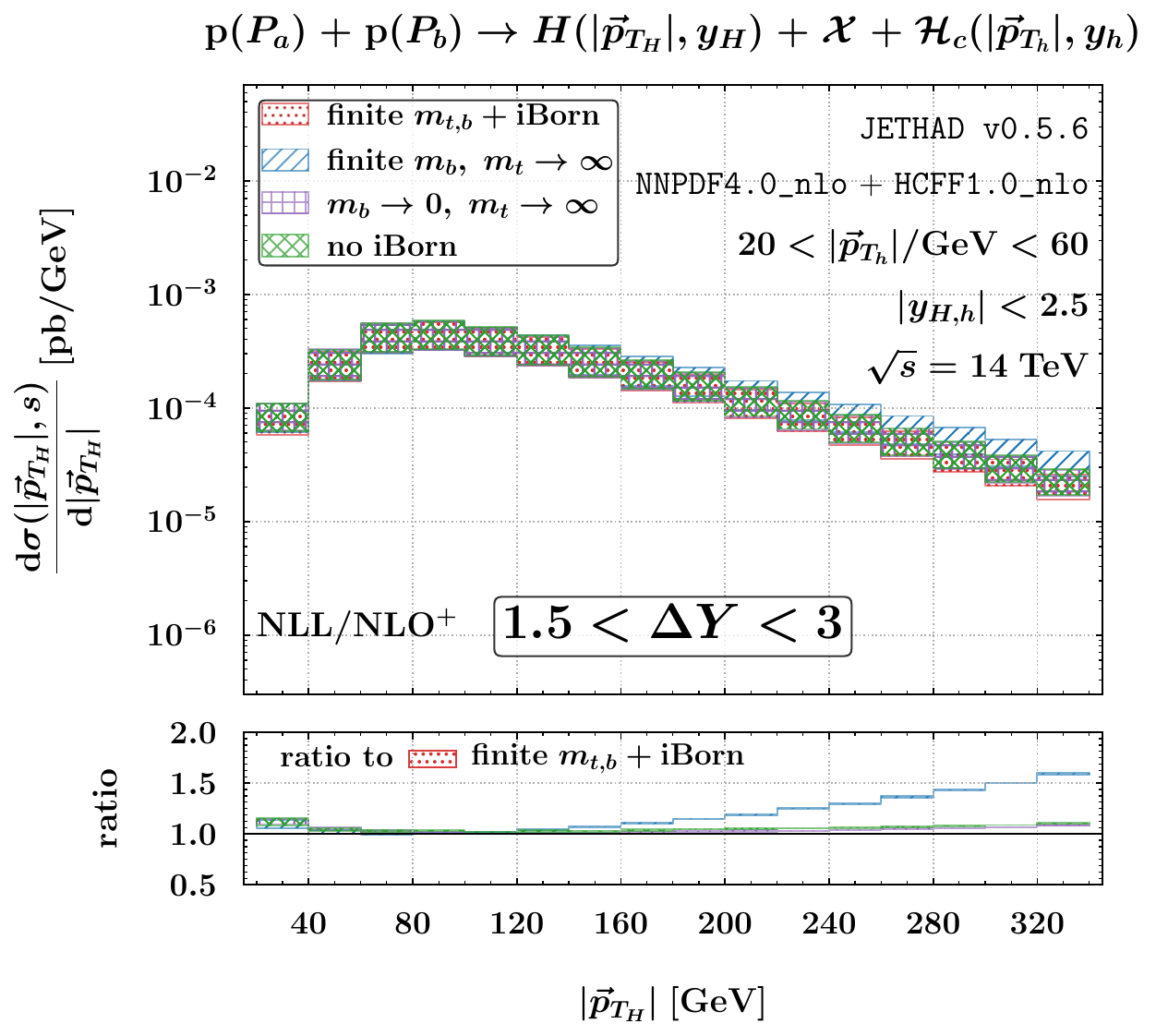}
   \hspace{-0.00cm}
   \includegraphics[scale=0.390,clip]{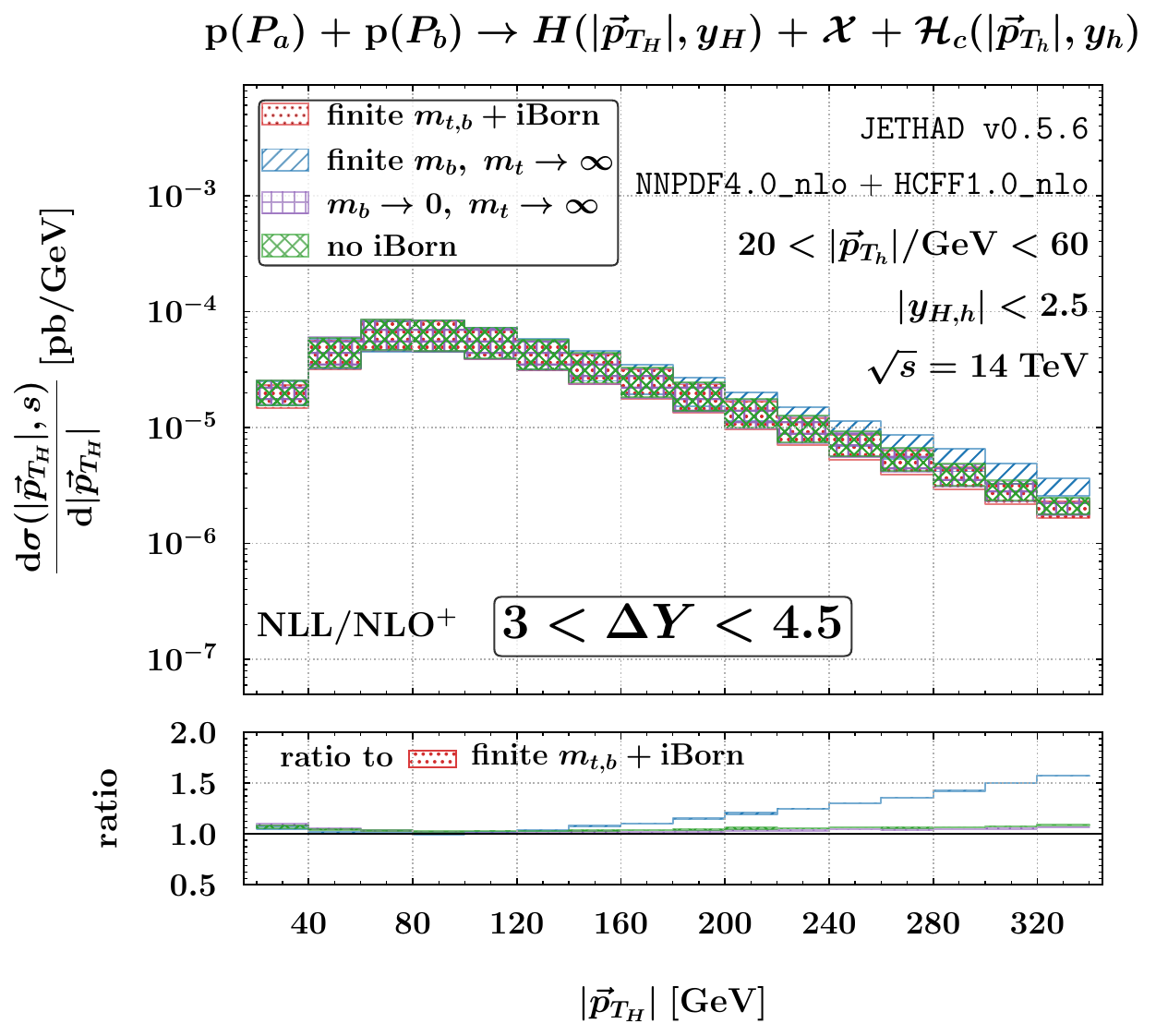}

   \hspace{0.00cm}
   \includegraphics[scale=0.390,clip]{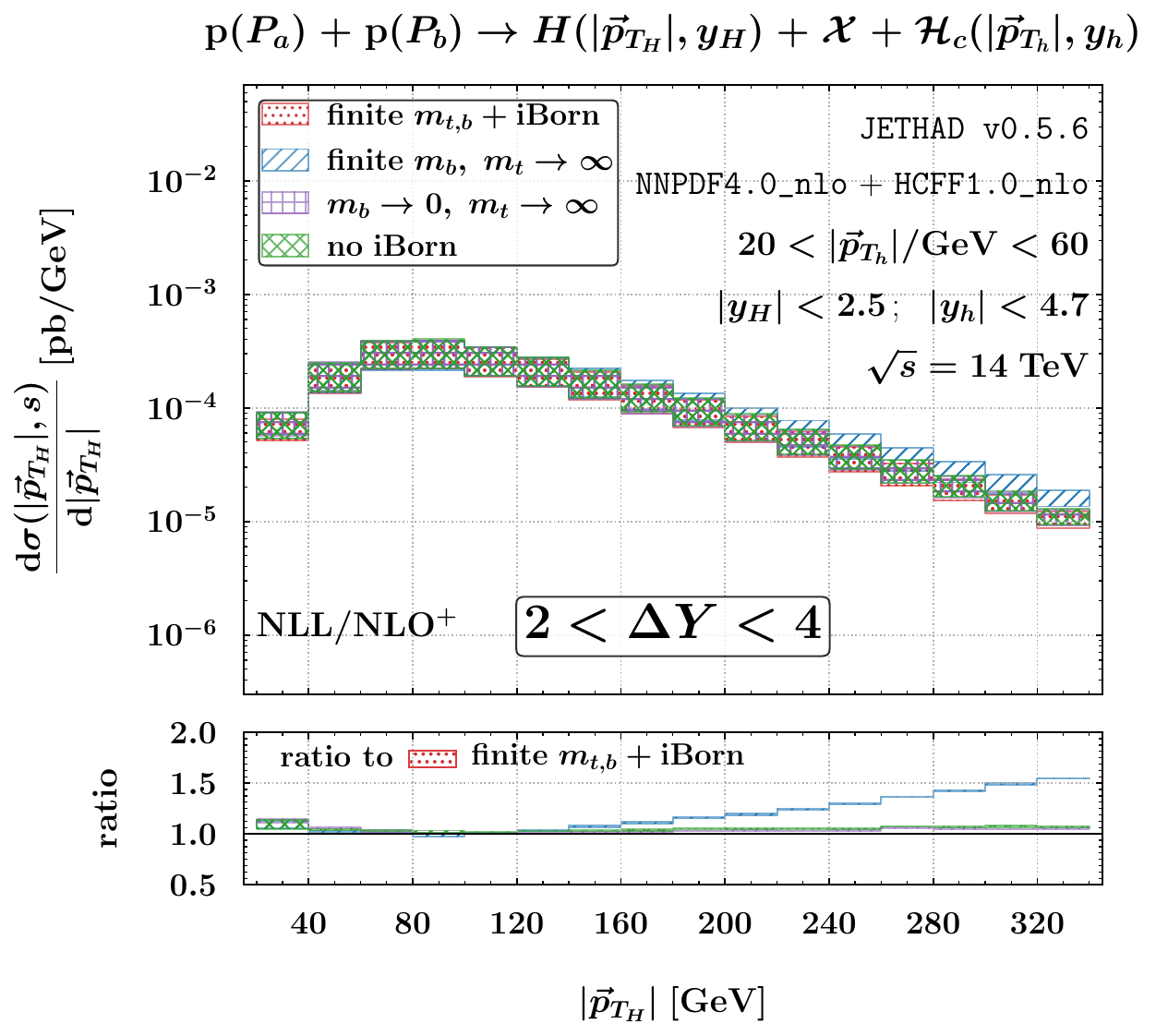}
   \hspace{-0.00cm}
   \includegraphics[scale=0.390,clip]{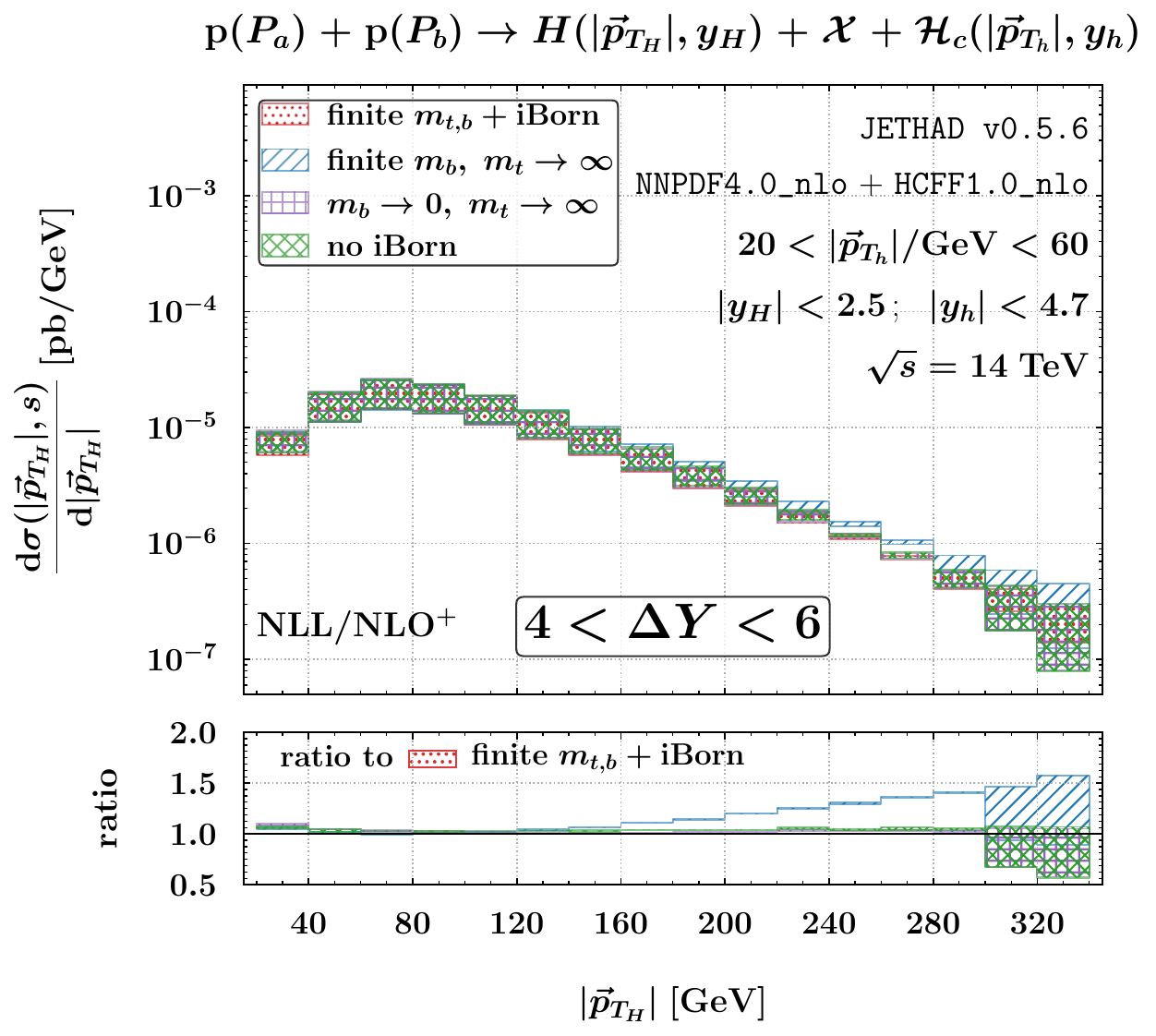}

\caption{Transverse-momentum rates for Higgs-plus-charmed-hadron production at the $\sqrt{s}=14$ TeV LHC, at $\NLLp$ accuracy and for different treatments of top- and bottom-quark mass effects.
The upper and lower rows correspond to the standard and extended rapidity ranges of the charmed hadron, respectively, while the left and right columns refer to moderate and large $\DY$ intervals.
For the standard selection, these correspond to $1.5<\DY<3$ and $3<\DY<4.5$, whereas the extended setup probes $2<\DY<4$ and $4<\DY<6$.
Ancillary panels show the ratios to the finite-$m_{t,b}$ prediction including Born-improvement effects.
Uncertainty bands capture the net effect of H-MHOUs and phase-space numerical integration.}
\label{fig:I-k1b-spec}
\end{figure*}

The broad transverse-momentum lever arm considered here also enables a dedicated study of heavy-quark-mass effects in the Higgs emission function.
This is one of the main motivations for extending the spectrum up to $|\vec p_{T_H}|=2m_t$: while finite-mass corrections are expected to remain comparatively mild when the Higgs transverse momentum is well below the top-quark scale, their relevance progressively increases as the hard transverse scale approaches and eventually becomes comparable with $m_t$.

An earlier Higgs-plus-jet analysis performed at partial $\NLL$ accuracy already provided first evidence of this behavior~\cite{Celiberto:2020tmb}.
There, predictions obtained in the large-$m_t$ approximation differed only mildly from the finite-top-mass result in the region of comparable Higgs and jet transverse momenta, while the sensitivity to finite-$m_t$ effects became more visible along the hard Higgs transverse-momentum tail.
The present $\NLLp$ Higgs-plus-hadron study substantially extends that analysis by combining the complete NLO Higgs emission function in the large-$m_t$ effective theory with finite top- and bottom-quark information incorporated at Born level through the improvement prescription introduced above.

Figure~\ref{fig:I-k1b-spec} is devoted to disentangling these effects.
Our reference prediction retains finite $m_t$ and $m_b$ dependence and includes Born improvement.
It is compared with complementary approximations designed to isolate, in turn, the impact of the finite top mass, of the bottom-quark contribution, and of the Born-improvement procedure.
The ratios displayed in the ancillary panels are normalized to the reference finite-$m_{t,b}$ Born-improved result, making the onset and transverse-momentum dependence of each effect directly visible.

The inspection of Fig.~\ref{fig:I-k1b-spec} reveals a clear hierarchy among the different effects entering the heavy-quark treatment of the Higgs emission function.
The dominant dependence is associated with the finite top-quark mass.
At low and intermediate $|\vec p_{T_H}|$, the large-$m_t$ approximation remains close to the finite-mass reference, consistently with the picture already obtained in the partial-NLL Higgs-plus-jet analysis of Ref.~\cite{Celiberto:2020tmb}.
As the Higgs transverse momentum increases, however, the two descriptions progressively separate.
The large-$m_t$ prediction systematically overshoots the finite-$m_t$ result, with deviations growing to several tens of percent and approaching the ${\cal O}(50\%)$ level toward the upper end of the explored spectrum.

This behavior makes explicit why extending the transverse-momentum scan up to $|\vec p_{T_H}|=2m_t$ is essential.
When the Higgs transverse momentum is well below the top-quark scale, the heavy-top approximation provides a reliable description of the loop-induced Higgs vertex.
As $|\vec p_{T_H}|$ approaches the characteristic top-mass scale, finite-mass information becomes increasingly relevant and the large-$m_t$ approximation progressively loses accuracy.
The present full-$\NLLp$ analysis therefore sharpens the qualitative indication already found in Higgs-plus-jet production~\cite{Celiberto:2020tmb}, where finite-$m_t$ effects were observed to become more visible along the hard Higgs transverse-momentum tail.

By contrast, the impact of the bottom-quark contribution is markedly smaller.
The two large-$m_t$ predictions obtained by retaining finite $m_b$ or by taking the $m_b\to0$ limit remain nearly indistinguishable throughout the investigated range.
Thus, at the level of accuracy and within the kinematics considered here, bottom-mass effects are strongly subleading with respect to the finite-top-mass correction.

The Born-improvement prescription produces a smaller, but clearly structured, effect.
Its removal leads to deviations of about $10\%$ in the first transverse-momentum bin, whereas the two predictions become nearly indistinguishable throughout most of the intermediate region.
A sensitivity of comparable size reappears in the hard tail, approximately above $|\vec p_{T_H}|\simeq250$~GeV.
Thus, although Born-improvement effects remain substantially smaller than the finite-$m_t$ corrections, they exhibit a nontrivial transverse-momentum dependence and become relevant precisely toward the edges of the explored spectrum.

Another noteworthy feature is the persistence of this hierarchy across the different rapidity configurations.
The onset and growth of finite-$m_t$ effects remain qualitatively similar when moving from moderate to large $\DY$ and from standard to extended hadron rapidity coverage.
This indicates that the dominant distortion is primarily tied to the local hard structure of the Higgs emission function, rather than to the long-range high-energy evolution connecting the two tagged objects.
Only in the hardest bins and at the largest rapidity separations do the enlarged H-MHOUs reduce the resolving power among the numerically smaller mass prescriptions.

Taken together, Figs.~\ref{fig:I-k1b} and~\ref{fig:I-k1b-spec} show that the Higgs transverse-momentum spectrum simultaneously probes two complementary layers of QCD dynamics.
Its dependence on $\DY$ and on the perturbative representation resolves the interplay between all-order high-energy resummation and its high-energy fixed-order limit, whereas the extension toward $|\vec p_{T_H}|\sim m_t$ directly exposes the finite-mass structure of the Higgs-production vertex.
Transverse-momentum rates therefore provide the most differential testing ground of our analysis, combining sensitivity to high-energy resummation, higher-order effects, and heavy-quark dynamics within a single observable.

\section{Paving the way towards precision}
\label{sec:conclusions}

We investigated Higgs-plus-charmed-hadron production at the $\sqrt{s}=14$~TeV LHC within hybrid high-energy and collinear factorization at $\NLLp$ accuracy, using light-flavored hadrons as a comparison channel.
For the first time in this class of phenomenological studies, we implemented the complete NLO correction to the singly off-shell Higgs emission function together with NLO identified-hadron production~\cite{Hentschinski:2020tbi,Celiberto:2022fgx}.
This completed a key step toward a genuinely next-to-leading description of Higgs-associated observables in the high-energy regime and allowed us to assess their perturbative behavior under controlled variations of the natural hard scales.

Our analysis revealed a remarkable stability of Higgs-plus-charm observables under both NLL corrections and H-MHOUs.
Rapidity-interval rates showed a particularly mild $\LL$ versus $\NLLp$ dependence, with the charmed-hadron channel exhibiting a more stable pattern than its light-flavored counterpart.
Azimuthal-correlation moments and normalized angular distributions confirmed that this behavior persisted when the individual Fourier modes of the high-energy evolution were resolved.
These results strengthened the interpretation of natural stability as the combined outcome of the hard scale supplied by Higgs production and the stabilizing collinear dynamics associated with heavy-flavor fragmentation.

Transverse-momentum rates provided the most differential probe of this picture.
They resolved the transition from the naturally stable region, where the transverse scales remained comparable, to configurations characterized by a sizeable momentum imbalance and enhanced sensitivity to additional logarithmic structures.
The comparison of $\NLLp$ predictions with their $\LL$ and $\HENLOp$ counterparts exposed the growing relevance of all-order high-energy dynamics across the hard tail, while the study of different perturbative representations showed that formally beyond-NLO terms remained under control.
The same spectra also resolved the finite-mass structure of the Higgs vertex: finite-$m_t$ corrections became increasingly important as $|\vec p_{T_H}|$ approached the top-quark scale, whereas bottom-mass and Born-improvement effects remained substantially more moderate.

Systematic matching of high-energy resummation with fixed-order calculations represents the principal next step toward phenomenological precision, with developments already in progress~\cite{Celiberto:2023uuk,Celiberto:2023eba,Celiberto:2023nym}.
Single-forward Higgs production at the LHC represents a complementary asset for high-precision hadron-structure studies, providing direct sensitivity to the multidimensional gluon content of the proton in transverse-momentum space and, through polarization-sensitive extensions, to its gluon-polarization structure, thereby linking high-energy dynamics with gluon transverse-momentum-dependent tomography~\cite{Boer:2011kf,Boer:2014lka,Echevarria:2015uaa,Gutierrez-Reyes:2019rug,Bacchetta:2020vty,Celiberto:2021zww,Bacchetta:2024fci,Anedda:2026cox}.
Electroweak-boson-plus-jet reactions, and in particular $Z$-plus-jet production, also represent a natural extension toward fully reconstructible final states directly connected with precision LHC measurements~\cite{Celiberto:2024mdt,Celiberto:2025edg,Celiberto:2026mjs,Celiberto:2026dwf}.

Taken together, our results broaden the phenomenological scope of high-energy resummation, reinforcing its established role in probing asymptotic QCD dynamics while supporting its use as a quantitatively controlled ingredient for precision collider phenomenology across increasingly realistic and experimentally relevant final states.

\section*{Acknowledgments}
\label{sec:acknowledgments}
\addcontentsline{toc}{section}{\nameref{sec:acknowledgments}}

FGC gratefully acknowledges Dmitry Yu. Ivanov, Mohammed M. A. Mohammed, and Alessandro Papa for insightful discussions during the early stages of studies on Higgs production in hybrid factorization at partial $\NLL$ accuracy~\cite{Celiberto:2020tmb}.
FGC also thanks Gabriele Gatto for preliminary discussions during the first phase of the numerical implementation of the Higgs emission function at NLO.
The authors thank Valerio Bertone, Marco Bonvini, Giuseppe Bozzi, Andreas Maier, Sebastian E. Jaskiewicz, Simone Marzani, Michael Spira, and Federico Silvetti for fruitful discussions on Higgs production in precision QCD.
The authors acknowledge support from the Atracci\'on de Talento Grant n. 2022-T1/TIC-24176 of the Comunidad Aut\'onoma de Madrid, Spain.

\setlength{\bibsep}{0.5em}
\bibliographystyle{apsrev}
\bibliography{references}

\begin{thebibliography}{159}
\expandafter\ifx\csname natexlab\endcsname\relax\def\natexlab#1{#1}\fi
\expandafter\ifx\csname bibnamefont\endcsname\relax
  \def\bibnamefont#1{#1}\fi
\expandafter\ifx\csname bibfnamefont\endcsname\relax
  \def\bibfnamefont#1{#1}\fi
\expandafter\ifx\csname citenamefont\endcsname\relax
  \def\citenamefont#1{#1}\fi
\expandafter\ifx\csname url\endcsname\relax
  \def\url#1{\texttt{#1}}\fi
\expandafter\ifx\csname urlprefix\endcsname\relax\def\urlprefix{URL }\fi
\providecommand{\bibinfo}[2]{#2}
\providecommand{\eprint}[2][]{\url{#2}}

\bibitem[{\citenamefont{Aad et~al.}(2012)}]{ATLAS:2012yve}
\bibinfo{author}{\bibfnamefont{G.}~\bibnamefont{Aad}} \bibnamefont{et~al.} (\bibinfo{collaboration}{ATLAS}), \bibinfo{journal}{Phys. Lett. B} \textbf{\bibinfo{volume}{716}}, \bibinfo{pages}{1} (\bibinfo{year}{2012}), \eprint{1207.7214}.

\bibitem[{\citenamefont{Chatrchyan et~al.}(2012)}]{CMS:2012qbp}
\bibinfo{author}{\bibfnamefont{S.}~\bibnamefont{Chatrchyan}} \bibnamefont{et~al.} (\bibinfo{collaboration}{CMS}), \bibinfo{journal}{Phys. Lett. B} \textbf{\bibinfo{volume}{716}}, \bibinfo{pages}{30} (\bibinfo{year}{2012}), \eprint{1207.7235}.

\bibitem[{\citenamefont{Bargiela et~al.}(2023)\citenamefont{Bargiela, Buccioni, Caola, Devoto, von Manteuffel, and Tancredi}}]{Bargiela:2022dla}
\bibinfo{author}{\bibfnamefont{P.}~\bibnamefont{Bargiela}}, \bibinfo{author}{\bibfnamefont{F.}~\bibnamefont{Buccioni}}, \bibinfo{author}{\bibfnamefont{F.}~\bibnamefont{Caola}}, \bibinfo{author}{\bibfnamefont{F.}~\bibnamefont{Devoto}}, \bibinfo{author}{\bibfnamefont{A.}~\bibnamefont{von Manteuffel}}, \bibnamefont{and} \bibinfo{author}{\bibfnamefont{L.}~\bibnamefont{Tancredi}}, \bibinfo{journal}{Eur. Phys. J. C} \textbf{\bibinfo{volume}{83}}, \bibinfo{pages}{174} (\bibinfo{year}{2023}), \eprint{2212.06287}.

\bibitem[{\citenamefont{Anastasiou et~al.}(2015)\citenamefont{Anastasiou, Duhr, Dulat, Herzog, and Mistlberger}}]{Anastasiou:2015vya}
\bibinfo{author}{\bibfnamefont{C.}~\bibnamefont{Anastasiou}}, \bibinfo{author}{\bibfnamefont{C.}~\bibnamefont{Duhr}}, \bibinfo{author}{\bibfnamefont{F.}~\bibnamefont{Dulat}}, \bibinfo{author}{\bibfnamefont{F.}~\bibnamefont{Herzog}}, \bibnamefont{and} \bibinfo{author}{\bibfnamefont{B.}~\bibnamefont{Mistlberger}}, \bibinfo{journal}{Phys. Rev. Lett.} \textbf{\bibinfo{volume}{114}}, \bibinfo{pages}{212001} (\bibinfo{year}{2015}), \eprint{1503.06056}.

\bibitem[{\citenamefont{Anastasiou et~al.}(2016)\citenamefont{Anastasiou, Duhr, Dulat, Furlan, Gehrmann, Herzog, Lazopoulos, and Mistlberger}}]{Anastasiou:2016cez}
\bibinfo{author}{\bibfnamefont{C.}~\bibnamefont{Anastasiou}}, \bibinfo{author}{\bibfnamefont{C.}~\bibnamefont{Duhr}}, \bibinfo{author}{\bibfnamefont{F.}~\bibnamefont{Dulat}}, \bibinfo{author}{\bibfnamefont{E.}~\bibnamefont{Furlan}}, \bibinfo{author}{\bibfnamefont{T.}~\bibnamefont{Gehrmann}}, \bibinfo{author}{\bibfnamefont{F.}~\bibnamefont{Herzog}}, \bibinfo{author}{\bibfnamefont{A.}~\bibnamefont{Lazopoulos}}, \bibnamefont{and} \bibinfo{author}{\bibfnamefont{B.}~\bibnamefont{Mistlberger}}, \bibinfo{journal}{JHEP} \textbf{\bibinfo{volume}{05}}, \bibinfo{pages}{058} (\bibinfo{year}{2016}), \eprint{1602.00695}.

\bibitem[{\citenamefont{Mistlberger}(2018)}]{Mistlberger:2018etf}
\bibinfo{author}{\bibfnamefont{B.}~\bibnamefont{Mistlberger}}, \bibinfo{journal}{JHEP} \textbf{\bibinfo{volume}{05}}, \bibinfo{pages}{028} (\bibinfo{year}{2018}), \eprint{1802.00833}.

\bibitem[{\citenamefont{Spira et~al.}(1995)\citenamefont{Spira, Djouadi, Graudenz, and Zerwas}}]{Spira:1995rr}
\bibinfo{author}{\bibfnamefont{M.}~\bibnamefont{Spira}}, \bibinfo{author}{\bibfnamefont{A.}~\bibnamefont{Djouadi}}, \bibinfo{author}{\bibfnamefont{D.}~\bibnamefont{Graudenz}}, \bibnamefont{and} \bibinfo{author}{\bibfnamefont{P.~M.} \bibnamefont{Zerwas}}, \bibinfo{journal}{Nucl. Phys. B} \textbf{\bibinfo{volume}{453}}, \bibinfo{pages}{17} (\bibinfo{year}{1995}), \eprint{hep-ph/9504378}.

\bibitem[{\citenamefont{Harlander et~al.}(2010)\citenamefont{Harlander, Mantler, Marzani, and Ozeren}}]{Harlander:2009my}
\bibinfo{author}{\bibfnamefont{R.~V.} \bibnamefont{Harlander}}, \bibinfo{author}{\bibfnamefont{H.}~\bibnamefont{Mantler}}, \bibinfo{author}{\bibfnamefont{S.}~\bibnamefont{Marzani}}, \bibnamefont{and} \bibinfo{author}{\bibfnamefont{K.~J.} \bibnamefont{Ozeren}}, \bibinfo{journal}{Eur. Phys. J. C} \textbf{\bibinfo{volume}{66}}, \bibinfo{pages}{359} (\bibinfo{year}{2010}), \eprint{0912.2104}.

\bibitem[{\citenamefont{Harlander and Ozeren}(2009)}]{Harlander:2009mq}
\bibinfo{author}{\bibfnamefont{R.~V.} \bibnamefont{Harlander}} \bibnamefont{and} \bibinfo{author}{\bibfnamefont{K.~J.} \bibnamefont{Ozeren}}, \bibinfo{journal}{JHEP} \textbf{\bibinfo{volume}{11}}, \bibinfo{pages}{088} (\bibinfo{year}{2009}), \eprint{0909.3420}.

\bibitem[{\citenamefont{Davies et~al.}(2020)\citenamefont{Davies, Herren, and Steinhauser}}]{Davies:2019wmk}
\bibinfo{author}{\bibfnamefont{J.}~\bibnamefont{Davies}}, \bibinfo{author}{\bibfnamefont{F.}~\bibnamefont{Herren}}, \bibnamefont{and} \bibinfo{author}{\bibfnamefont{M.}~\bibnamefont{Steinhauser}}, \bibinfo{journal}{Phys. Rev. Lett.} \textbf{\bibinfo{volume}{124}}, \bibinfo{pages}{112002} (\bibinfo{year}{2020}), \eprint{1911.10214}.

\bibitem[{\citenamefont{Czakon et~al.}(2021)\citenamefont{Czakon, Harlander, Klappert, and Niggetiedt}}]{Czakon:2021yub}
\bibinfo{author}{\bibfnamefont{M.}~\bibnamefont{Czakon}}, \bibinfo{author}{\bibfnamefont{R.~V.} \bibnamefont{Harlander}}, \bibinfo{author}{\bibfnamefont{J.}~\bibnamefont{Klappert}}, \bibnamefont{and} \bibinfo{author}{\bibfnamefont{M.}~\bibnamefont{Niggetiedt}}, \bibinfo{journal}{Phys. Rev. Lett.} \textbf{\bibinfo{volume}{127}}, \bibinfo{pages}{162002} (\bibinfo{year}{2021}), \bibinfo{note}{[Erratum: Phys.Rev.Lett. 131, 179901 (2023)]}, \eprint{2105.04436}.

\bibitem[{\citenamefont{Catani et~al.}(2001)\citenamefont{Catani, de~Florian, and Grazzini}}]{Catani:2000vq}
\bibinfo{author}{\bibfnamefont{S.}~\bibnamefont{Catani}}, \bibinfo{author}{\bibfnamefont{D.}~\bibnamefont{de~Florian}}, \bibnamefont{and} \bibinfo{author}{\bibfnamefont{M.}~\bibnamefont{Grazzini}}, \bibinfo{journal}{Nucl. Phys. B} \textbf{\bibinfo{volume}{596}}, \bibinfo{pages}{299} (\bibinfo{year}{2001}), \eprint{hep-ph/0008184}.

\bibitem[{\citenamefont{Bozzi et~al.}(2006)\citenamefont{Bozzi, Catani, de~Florian, and Grazzini}}]{Bozzi:2005wk}
\bibinfo{author}{\bibfnamefont{G.}~\bibnamefont{Bozzi}}, \bibinfo{author}{\bibfnamefont{S.}~\bibnamefont{Catani}}, \bibinfo{author}{\bibfnamefont{D.}~\bibnamefont{de~Florian}}, \bibnamefont{and} \bibinfo{author}{\bibfnamefont{M.}~\bibnamefont{Grazzini}}, \bibinfo{journal}{Nucl. Phys. B} \textbf{\bibinfo{volume}{737}}, \bibinfo{pages}{73} (\bibinfo{year}{2006}), \eprint{hep-ph/0508068}.

\bibitem[{\citenamefont{Catani and Grazzini}(2011)}]{Catani:2010pd}
\bibinfo{author}{\bibfnamefont{S.}~\bibnamefont{Catani}} \bibnamefont{and} \bibinfo{author}{\bibfnamefont{M.}~\bibnamefont{Grazzini}}, \bibinfo{journal}{Nucl. Phys. B} \textbf{\bibinfo{volume}{845}}, \bibinfo{pages}{297} (\bibinfo{year}{2011}), \eprint{1011.3918}.

\bibitem[{\citenamefont{Catani et~al.}(2014)\citenamefont{Catani, Cieri, de~Florian, Ferrera, and Grazzini}}]{Catani:2013tia}
\bibinfo{author}{\bibfnamefont{S.}~\bibnamefont{Catani}}, \bibinfo{author}{\bibfnamefont{L.}~\bibnamefont{Cieri}}, \bibinfo{author}{\bibfnamefont{D.}~\bibnamefont{de~Florian}}, \bibinfo{author}{\bibfnamefont{G.}~\bibnamefont{Ferrera}}, \bibnamefont{and} \bibinfo{author}{\bibfnamefont{M.}~\bibnamefont{Grazzini}}, \bibinfo{journal}{Nucl. Phys. B} \textbf{\bibinfo{volume}{881}}, \bibinfo{pages}{414} (\bibinfo{year}{2014}), \eprint{1311.1654}.

\bibitem[{\citenamefont{Duhr et~al.}(2022)\citenamefont{Duhr, Mistlberger, and Vita}}]{Duhr:2022yyp}
\bibinfo{author}{\bibfnamefont{C.}~\bibnamefont{Duhr}}, \bibinfo{author}{\bibfnamefont{B.}~\bibnamefont{Mistlberger}}, \bibnamefont{and} \bibinfo{author}{\bibfnamefont{G.}~\bibnamefont{Vita}}, \bibinfo{journal}{Phys. Rev. Lett.} \textbf{\bibinfo{volume}{129}}, \bibinfo{pages}{162001} (\bibinfo{year}{2022}), \eprint{2205.02242}.

\bibitem[{\citenamefont{Sterman}(1987)}]{Sterman:1986aj}
\bibinfo{author}{\bibfnamefont{G.~F.} \bibnamefont{Sterman}}, \bibinfo{journal}{Nucl. Phys. B} \textbf{\bibinfo{volume}{281}}, \bibinfo{pages}{310} (\bibinfo{year}{1987}).

\bibitem[{\citenamefont{Catani et~al.}(1996)\citenamefont{Catani, Mangano, Nason, and Trentadue}}]{Catani:1996yz}
\bibinfo{author}{\bibfnamefont{S.}~\bibnamefont{Catani}}, \bibinfo{author}{\bibfnamefont{M.~L.} \bibnamefont{Mangano}}, \bibinfo{author}{\bibfnamefont{P.}~\bibnamefont{Nason}}, \bibnamefont{and} \bibinfo{author}{\bibfnamefont{L.}~\bibnamefont{Trentadue}}, \bibinfo{journal}{Nucl. Phys. B} \textbf{\bibinfo{volume}{478}}, \bibinfo{pages}{273} (\bibinfo{year}{1996}), \eprint{hep-ph/9604351}.

\bibitem[{\citenamefont{Bonciani et~al.}(2003)\citenamefont{Bonciani, Catani, Mangano, and Nason}}]{Bonciani:2003nt}
\bibinfo{author}{\bibfnamefont{R.}~\bibnamefont{Bonciani}}, \bibinfo{author}{\bibfnamefont{S.}~\bibnamefont{Catani}}, \bibinfo{author}{\bibfnamefont{M.~L.} \bibnamefont{Mangano}}, \bibnamefont{and} \bibinfo{author}{\bibfnamefont{P.}~\bibnamefont{Nason}}, \bibinfo{journal}{Phys. Lett. B} \textbf{\bibinfo{volume}{575}}, \bibinfo{pages}{268} (\bibinfo{year}{2003}), \eprint{hep-ph/0307035}.

\bibitem[{\citenamefont{de~Florian et~al.}(2006)\citenamefont{de~Florian, Kulesza, and Vogelsang}}]{deFlorian:2005fzc}
\bibinfo{author}{\bibfnamefont{D.}~\bibnamefont{de~Florian}}, \bibinfo{author}{\bibfnamefont{A.}~\bibnamefont{Kulesza}}, \bibnamefont{and} \bibinfo{author}{\bibfnamefont{W.}~\bibnamefont{Vogelsang}}, \bibinfo{journal}{JHEP} \textbf{\bibinfo{volume}{02}}, \bibinfo{pages}{047} (\bibinfo{year}{2006}), \eprint{hep-ph/0511205}.

\bibitem[{\citenamefont{Forte et~al.}(2021)\citenamefont{Forte, Ridolfi, and Rota}}]{Forte:2021wxe}
\bibinfo{author}{\bibfnamefont{S.}~\bibnamefont{Forte}}, \bibinfo{author}{\bibfnamefont{G.}~\bibnamefont{Ridolfi}}, \bibnamefont{and} \bibinfo{author}{\bibfnamefont{S.}~\bibnamefont{Rota}}, \bibinfo{journal}{JHEP} \textbf{\bibinfo{volume}{08}}, \bibinfo{pages}{110} (\bibinfo{year}{2021}), \eprint{2106.11321}.

\bibitem[{\citenamefont{Gribov et~al.}(1983)\citenamefont{Gribov, Levin, and Ryskin}}]{Gribov:1983ivg}
\bibinfo{author}{\bibfnamefont{L.~V.} \bibnamefont{Gribov}}, \bibinfo{author}{\bibfnamefont{E.~M.} \bibnamefont{Levin}}, \bibnamefont{and} \bibinfo{author}{\bibfnamefont{M.~G.} \bibnamefont{Ryskin}}, \bibinfo{journal}{Phys. Rept.} \textbf{\bibinfo{volume}{100}}, \bibinfo{pages}{1} (\bibinfo{year}{1983}).

\bibitem[{\citenamefont{Celiberto}(2017)}]{Celiberto:2017ius}
\bibinfo{author}{\bibfnamefont{F.~G.} \bibnamefont{Celiberto}}, \bibinfo{type}{Phd thesis}, \bibinfo{school}{Universit\`a della Calabria and INFN-Cosenza} (\bibinfo{year}{2017}), \eprint{1707.04315}.

\bibitem[{\citenamefont{Bonvini et~al.}(2012)\citenamefont{Bonvini, Forte, and Ridolfi}}]{Bonvini:2012an}
\bibinfo{author}{\bibfnamefont{M.}~\bibnamefont{Bonvini}}, \bibinfo{author}{\bibfnamefont{S.}~\bibnamefont{Forte}}, \bibnamefont{and} \bibinfo{author}{\bibfnamefont{G.}~\bibnamefont{Ridolfi}}, \bibinfo{journal}{Phys. Rev. Lett.} \textbf{\bibinfo{volume}{109}}, \bibinfo{pages}{102002} (\bibinfo{year}{2012}), \eprint{1204.5473}.

\bibitem[{\citenamefont{Bonvini and Marzani}(2014)}]{Bonvini:2014joa}
\bibinfo{author}{\bibfnamefont{M.}~\bibnamefont{Bonvini}} \bibnamefont{and} \bibinfo{author}{\bibfnamefont{S.}~\bibnamefont{Marzani}}, \bibinfo{journal}{JHEP} \textbf{\bibinfo{volume}{09}}, \bibinfo{pages}{007} (\bibinfo{year}{2014}), \eprint{1405.3654}.

\bibitem[{\citenamefont{Bonvini et~al.}(2015)\citenamefont{Bonvini, Marzani, Rojo, Rottoli, Ubiali, Ball, Bertone, Carrazza, and Hartland}}]{Bonvini:2015ira}
\bibinfo{author}{\bibfnamefont{M.}~\bibnamefont{Bonvini}}, \bibinfo{author}{\bibfnamefont{S.}~\bibnamefont{Marzani}}, \bibinfo{author}{\bibfnamefont{J.}~\bibnamefont{Rojo}}, \bibinfo{author}{\bibfnamefont{L.}~\bibnamefont{Rottoli}}, \bibinfo{author}{\bibfnamefont{M.}~\bibnamefont{Ubiali}}, \bibinfo{author}{\bibfnamefont{R.~D.} \bibnamefont{Ball}}, \bibinfo{author}{\bibfnamefont{V.}~\bibnamefont{Bertone}}, \bibinfo{author}{\bibfnamefont{S.}~\bibnamefont{Carrazza}}, \bibnamefont{and} \bibinfo{author}{\bibfnamefont{N.~P.} \bibnamefont{Hartland}}, \bibinfo{journal}{JHEP} \textbf{\bibinfo{volume}{09}}, \bibinfo{pages}{191} (\bibinfo{year}{2015}), \eprint{1507.01006}.

\bibitem[{\citenamefont{Bonvini et~al.}(2016)\citenamefont{Bonvini, Marzani, Muselli, and Rottoli}}]{Bonvini:2016frm}
\bibinfo{author}{\bibfnamefont{M.}~\bibnamefont{Bonvini}}, \bibinfo{author}{\bibfnamefont{S.}~\bibnamefont{Marzani}}, \bibinfo{author}{\bibfnamefont{C.}~\bibnamefont{Muselli}}, \bibnamefont{and} \bibinfo{author}{\bibfnamefont{L.}~\bibnamefont{Rottoli}}, \bibinfo{journal}{JHEP} \textbf{\bibinfo{volume}{08}}, \bibinfo{pages}{105} (\bibinfo{year}{2016}), \eprint{1603.08000}.

\bibitem[{\citenamefont{Ball and Forte}(1997)}]{Ball:1997vf}
\bibinfo{author}{\bibfnamefont{R.~D.} \bibnamefont{Ball}} \bibnamefont{and} \bibinfo{author}{\bibfnamefont{S.}~\bibnamefont{Forte}}, \bibinfo{journal}{Phys. Lett. B} \textbf{\bibinfo{volume}{405}}, \bibinfo{pages}{317} (\bibinfo{year}{1997}), \eprint{hep-ph/9703417}.

\bibitem[{\citenamefont{Altarelli et~al.}(2006)\citenamefont{Altarelli, Ball, and Forte}}]{Altarelli:2005ni}
\bibinfo{author}{\bibfnamefont{G.}~\bibnamefont{Altarelli}}, \bibinfo{author}{\bibfnamefont{R.~D.} \bibnamefont{Ball}}, \bibnamefont{and} \bibinfo{author}{\bibfnamefont{S.}~\bibnamefont{Forte}}, \bibinfo{journal}{Nucl. Phys. B} \textbf{\bibinfo{volume}{742}}, \bibinfo{pages}{1} (\bibinfo{year}{2006}), \eprint{hep-ph/0512237}.

\bibitem[{\citenamefont{Ball et~al.}(2018)\citenamefont{Ball, Bertone, Bonvini, Marzani, Rojo, and Rottoli}}]{Ball:2017otu}
\bibinfo{author}{\bibfnamefont{R.~D.} \bibnamefont{Ball}}, \bibinfo{author}{\bibfnamefont{V.}~\bibnamefont{Bertone}}, \bibinfo{author}{\bibfnamefont{M.}~\bibnamefont{Bonvini}}, \bibinfo{author}{\bibfnamefont{S.}~\bibnamefont{Marzani}}, \bibinfo{author}{\bibfnamefont{J.}~\bibnamefont{Rojo}}, \bibnamefont{and} \bibinfo{author}{\bibfnamefont{L.}~\bibnamefont{Rottoli}}, \bibinfo{journal}{Eur. Phys. J.} \textbf{\bibinfo{volume}{C78}}, \bibinfo{pages}{321} (\bibinfo{year}{2018}), \eprint{1710.05935}.

\bibitem[{\citenamefont{Bonvini et~al.}(2026)\citenamefont{Bonvini, Frixione, and Stagnitto}}]{Bonvini:2026cxp}
\bibinfo{author}{\bibfnamefont{M.}~\bibnamefont{Bonvini}}, \bibinfo{author}{\bibfnamefont{S.}~\bibnamefont{Frixione}}, \bibnamefont{and} \bibinfo{author}{\bibfnamefont{G.}~\bibnamefont{Stagnitto}} (\bibinfo{year}{2026}), \eprint{2603.02312}.

\bibitem[{\citenamefont{Ball et~al.}(2013)\citenamefont{Ball, Bonvini, Forte, Marzani, and Ridolfi}}]{Ball:2013bra}
\bibinfo{author}{\bibfnamefont{R.~D.} \bibnamefont{Ball}}, \bibinfo{author}{\bibfnamefont{M.}~\bibnamefont{Bonvini}}, \bibinfo{author}{\bibfnamefont{S.}~\bibnamefont{Forte}}, \bibinfo{author}{\bibfnamefont{S.}~\bibnamefont{Marzani}}, \bibnamefont{and} \bibinfo{author}{\bibfnamefont{G.}~\bibnamefont{Ridolfi}}, \bibinfo{journal}{Nucl. Phys. B} \textbf{\bibinfo{volume}{874}}, \bibinfo{pages}{746} (\bibinfo{year}{2013}), \eprint{1303.3590}.

\bibitem[{\citenamefont{Bonvini and Marzani}(2018)}]{Bonvini:2018ixe}
\bibinfo{author}{\bibfnamefont{M.}~\bibnamefont{Bonvini}} \bibnamefont{and} \bibinfo{author}{\bibfnamefont{S.}~\bibnamefont{Marzani}}, \bibinfo{journal}{Phys. Rev. Lett.} \textbf{\bibinfo{volume}{120}}, \bibinfo{pages}{202003} (\bibinfo{year}{2018}), \eprint{1802.07758}.

\bibitem[{\citenamefont{Mangano et~al.}(2017)}]{Mangano:2016jyj}
\bibinfo{author}{\bibfnamefont{M.~L.} \bibnamefont{Mangano}} \bibnamefont{et~al.}, \bibinfo{journal}{CERN Yellow Rep. 3} \textbf{\bibinfo{volume}{3}}, \bibinfo{pages}{1} (\bibinfo{year}{2017}), \eprint{1607.01831}.

\bibitem[{\citenamefont{Benedikt et~al.}(2025{\natexlab{a}})}]{FCC:2025lpp}
\bibinfo{author}{\bibfnamefont{M.}~\bibnamefont{Benedikt}} \bibnamefont{et~al.} (\bibinfo{collaboration}{FCC}), \bibinfo{journal}{Eur. Phys. J. C} \textbf{\bibinfo{volume}{85}}, \bibinfo{pages}{1468} (\bibinfo{year}{2025}{\natexlab{a}}), \eprint{2505.00272}.

\bibitem[{\citenamefont{Benedikt et~al.}(2025{\natexlab{b}})}]{FCC:2025uan}
\bibinfo{author}{\bibfnamefont{M.}~\bibnamefont{Benedikt}} \bibnamefont{et~al.} (\bibinfo{collaboration}{FCC}), \bibinfo{journal}{Eur. Phys. J. ST} \textbf{\bibinfo{volume}{234}}, \bibinfo{pages}{5713} (\bibinfo{year}{2025}{\natexlab{b}}), \eprint{2505.00274}.

\bibitem[{\citenamefont{Benedikt et~al.}(2025{\natexlab{c}})}]{FCC:2025jtd}
\bibinfo{author}{\bibfnamefont{M.}~\bibnamefont{Benedikt}} \bibnamefont{et~al.} (\bibinfo{collaboration}{FCC}), \bibinfo{journal}{Eur. Phys. J. ST} \textbf{\bibinfo{volume}{234}}, \bibinfo{pages}{5113} (\bibinfo{year}{2025}{\natexlab{c}}), \bibinfo{note}{[Erratum: Eur.Phys.J.ST None, (2025)]}, \eprint{2505.00273}.

\bibitem[{\citenamefont{Silvetti and Bonvini}(2023)}]{Silvetti:2022hyc}
\bibinfo{author}{\bibfnamefont{F.}~\bibnamefont{Silvetti}} \bibnamefont{and} \bibinfo{author}{\bibfnamefont{M.}~\bibnamefont{Bonvini}}, \bibinfo{journal}{Eur. Phys. J. C} \textbf{\bibinfo{volume}{83}}, \bibinfo{pages}{267} (\bibinfo{year}{2023}), \eprint{2211.10142}.

\bibitem[{\citenamefont{Fadin et~al.}(1975)\citenamefont{Fadin, Kuraev, and Lipatov}}]{Fadin:1975cb}
\bibinfo{author}{\bibfnamefont{V.~S.} \bibnamefont{Fadin}}, \bibinfo{author}{\bibfnamefont{E.}~\bibnamefont{Kuraev}}, \bibnamefont{and} \bibinfo{author}{\bibfnamefont{L.}~\bibnamefont{Lipatov}}, \bibinfo{journal}{Phys. Lett. B} \textbf{\bibinfo{volume}{60}}, \bibinfo{pages}{50} (\bibinfo{year}{1975}).

\bibitem[{\citenamefont{Kuraev et~al.}(1976)\citenamefont{Kuraev, Lipatov, and Fadin}}]{Kuraev:1976ge}
\bibinfo{author}{\bibfnamefont{E.~A.} \bibnamefont{Kuraev}}, \bibinfo{author}{\bibfnamefont{L.~N.} \bibnamefont{Lipatov}}, \bibnamefont{and} \bibinfo{author}{\bibfnamefont{V.~S.} \bibnamefont{Fadin}}, \bibinfo{journal}{Sov. Phys. JETP} \textbf{\bibinfo{volume}{44}}, \bibinfo{pages}{443} (\bibinfo{year}{1976}).

\bibitem[{\citenamefont{Kuraev et~al.}(1977)\citenamefont{Kuraev, Lipatov, and Fadin}}]{Kuraev:1977fs}
\bibinfo{author}{\bibfnamefont{E.}~\bibnamefont{Kuraev}}, \bibinfo{author}{\bibfnamefont{L.}~\bibnamefont{Lipatov}}, \bibnamefont{and} \bibinfo{author}{\bibfnamefont{V.~S.} \bibnamefont{Fadin}}, \bibinfo{journal}{Sov.\ Phys.\ JETP} \textbf{\bibinfo{volume}{45}}, \bibinfo{pages}{199} (\bibinfo{year}{1977}).

\bibitem[{\citenamefont{Celiberto et~al.}(2021{\natexlab{a}})\citenamefont{Celiberto, Ivanov, Mohammed, and Papa}}]{Celiberto:2020tmb}
\bibinfo{author}{\bibfnamefont{F.~G.} \bibnamefont{Celiberto}}, \bibinfo{author}{\bibfnamefont{D.~{\relax Yu}.} \bibnamefont{Ivanov}}, \bibinfo{author}{\bibfnamefont{M.~M.~A.} \bibnamefont{Mohammed}}, \bibnamefont{and} \bibinfo{author}{\bibfnamefont{A.}~\bibnamefont{Papa}}, \bibinfo{journal}{Eur. Phys. J. C} \textbf{\bibinfo{volume}{81}}, \bibinfo{pages}{293} (\bibinfo{year}{2021}{\natexlab{a}}), \eprint{2008.00501}.

\bibitem[{\citenamefont{Mohammed}(2022)}]{Mohammed:2022gbk}
\bibinfo{author}{\bibfnamefont{M.~M.~A.} \bibnamefont{Mohammed}}, \bibinfo{type}{Phd thesis}, \bibinfo{school}{Universit\`a della Calabria and INFN-Cosenza} (\bibinfo{year}{2022}), \eprint{2204.11606}.

\bibitem[{\citenamefont{Celiberto et~al.}(2021{\natexlab{b}})\citenamefont{Celiberto, Fucilla, Ivanov, and Papa}}]{Celiberto:2021dzy}
\bibinfo{author}{\bibfnamefont{F.~G.} \bibnamefont{Celiberto}}, \bibinfo{author}{\bibfnamefont{M.}~\bibnamefont{Fucilla}}, \bibinfo{author}{\bibfnamefont{D.~{\relax Yu}.} \bibnamefont{Ivanov}}, \bibnamefont{and} \bibinfo{author}{\bibfnamefont{A.}~\bibnamefont{Papa}}, \bibinfo{journal}{Eur. Phys. J. C} \textbf{\bibinfo{volume}{81}}, \bibinfo{pages}{780} (\bibinfo{year}{2021}{\natexlab{b}}), \eprint{2105.06432}.

\bibitem[{\citenamefont{Celiberto et~al.}(2021{\natexlab{c}})\citenamefont{Celiberto, Fucilla, Ivanov, Mohammed, and Papa}}]{Celiberto:2021fdp}
\bibinfo{author}{\bibfnamefont{F.~G.} \bibnamefont{Celiberto}}, \bibinfo{author}{\bibfnamefont{M.}~\bibnamefont{Fucilla}}, \bibinfo{author}{\bibfnamefont{D.~{\relax Yu}.} \bibnamefont{Ivanov}}, \bibinfo{author}{\bibfnamefont{M.~M.~A.} \bibnamefont{Mohammed}}, \bibnamefont{and} \bibinfo{author}{\bibfnamefont{A.}~\bibnamefont{Papa}}, \bibinfo{journal}{Phys. Rev. D} \textbf{\bibinfo{volume}{104}}, \bibinfo{pages}{114007} (\bibinfo{year}{2021}{\natexlab{c}}), \eprint{2109.11875}.

\bibitem[{\citenamefont{Celiberto}(2022{\natexlab{a}})}]{Celiberto:2022rfj}
\bibinfo{author}{\bibfnamefont{F.~G.} \bibnamefont{Celiberto}}, \bibinfo{journal}{Phys. Rev. D} \textbf{\bibinfo{volume}{105}}, \bibinfo{pages}{114008} (\bibinfo{year}{2022}{\natexlab{a}}), \eprint{2204.06497}.

\bibitem[{\citenamefont{Celiberto et~al.}(2022{\natexlab{a}})\citenamefont{Celiberto, Fucilla, Mohammed, and Papa}}]{Celiberto:2022zdg}
\bibinfo{author}{\bibfnamefont{F.~G.} \bibnamefont{Celiberto}}, \bibinfo{author}{\bibfnamefont{M.}~\bibnamefont{Fucilla}}, \bibinfo{author}{\bibfnamefont{M.~M.~A.} \bibnamefont{Mohammed}}, \bibnamefont{and} \bibinfo{author}{\bibfnamefont{A.}~\bibnamefont{Papa}}, \bibinfo{journal}{Phys. Rev. D} \textbf{\bibinfo{volume}{105}}, \bibinfo{pages}{114056} (\bibinfo{year}{2022}{\natexlab{a}}), \eprint{2205.13429}.

\bibitem[{\citenamefont{Celiberto}(2023{\natexlab{a}})}]{Celiberto:2022grc}
\bibinfo{author}{\bibfnamefont{F.~G.} \bibnamefont{Celiberto}}, \bibinfo{journal}{Acta Phys. Polon. Supp.} \textbf{\bibinfo{volume}{16}}, \bibinfo{pages}{41} (\bibinfo{year}{2023}{\natexlab{a}}), \eprint{2211.11780}.

\bibitem[{\citenamefont{Aad et~al.}(2025)}]{ATLAS:2025pxp}
\bibinfo{author}{\bibfnamefont{G.}~\bibnamefont{Aad}} \bibnamefont{et~al.} (\bibinfo{collaboration}{ATLAS}) (\bibinfo{year}{2025}), \eprint{2511.21911}.

\bibitem[{\citenamefont{Giannakopoulou et~al.}(2025)\citenamefont{Giannakopoulou, Meade, and Valli}}]{Giannakopoulou:2024unn}
\bibinfo{author}{\bibfnamefont{A.~S.} \bibnamefont{Giannakopoulou}}, \bibinfo{author}{\bibfnamefont{P.}~\bibnamefont{Meade}}, \bibnamefont{and} \bibinfo{author}{\bibfnamefont{M.}~\bibnamefont{Valli}}, \bibinfo{journal}{JHEP} \textbf{\bibinfo{volume}{02}}, \bibinfo{pages}{067} (\bibinfo{year}{2025}), \eprint{2410.05236}.

\bibitem[{\citenamefont{Kopp and Nardecchia}(2014)}]{Kopp:2014rva}
\bibinfo{author}{\bibfnamefont{J.}~\bibnamefont{Kopp}} \bibnamefont{and} \bibinfo{author}{\bibfnamefont{M.}~\bibnamefont{Nardecchia}}, \bibinfo{journal}{JHEP} \textbf{\bibinfo{volume}{10}}, \bibinfo{pages}{156} (\bibinfo{year}{2014}), \eprint{1406.5303}.

\bibitem[{\citenamefont{Di~Luzio et~al.}(2023)\citenamefont{Di~Luzio, Guerrera, D{\'\i}az, and Rigolin}}]{DiLuzio:2023ndz}
\bibinfo{author}{\bibfnamefont{L.}~\bibnamefont{Di~Luzio}}, \bibinfo{author}{\bibfnamefont{A.~W.~M.} \bibnamefont{Guerrera}}, \bibinfo{author}{\bibfnamefont{X.~P.} \bibnamefont{D{\'\i}az}}, \bibnamefont{and} \bibinfo{author}{\bibfnamefont{S.}~\bibnamefont{Rigolin}}, \bibinfo{journal}{JHEP} \textbf{\bibinfo{volume}{06}}, \bibinfo{pages}{046} (\bibinfo{year}{2023}), \eprint{2304.04643}.

\bibitem[{\citenamefont{Brivio et~al.}(2015)\citenamefont{Brivio, Goertz, and Isidori}}]{Brivio:2015fxa}
\bibinfo{author}{\bibfnamefont{I.}~\bibnamefont{Brivio}}, \bibinfo{author}{\bibfnamefont{F.}~\bibnamefont{Goertz}}, \bibnamefont{and} \bibinfo{author}{\bibfnamefont{G.}~\bibnamefont{Isidori}}, \bibinfo{journal}{Phys. Rev. Lett.} \textbf{\bibinfo{volume}{115}}, \bibinfo{pages}{211801} (\bibinfo{year}{2015}), \eprint{1507.02916}.

\bibitem[{\citenamefont{Nefedov}(2019)}]{Nefedov:2019mrg}
\bibinfo{author}{\bibfnamefont{M.~A.} \bibnamefont{Nefedov}}, \bibinfo{journal}{Nucl. Phys. B} \textbf{\bibinfo{volume}{946}}, \bibinfo{pages}{114715} (\bibinfo{year}{2019}), \eprint{1902.11030}.

\bibitem[{\citenamefont{Hentschinski et~al.}(2021)\citenamefont{Hentschinski, Kutak, and van Hameren}}]{Hentschinski:2020tbi}
\bibinfo{author}{\bibfnamefont{M.}~\bibnamefont{Hentschinski}}, \bibinfo{author}{\bibfnamefont{K.}~\bibnamefont{Kutak}}, \bibnamefont{and} \bibinfo{author}{\bibfnamefont{A.}~\bibnamefont{van Hameren}}, \bibinfo{journal}{Eur. Phys. J. C} \textbf{\bibinfo{volume}{81}}, \bibinfo{pages}{112} (\bibinfo{year}{2021}), \bibinfo{note}{[Erratum: Eur. Phys. J. C 81, 262 (2021)]}, \eprint{2011.03193}.

\bibitem[{\citenamefont{Celiberto et~al.}(2022{\natexlab{b}})\citenamefont{Celiberto, Fucilla, Ivanov, Mohammed, and Papa}}]{Celiberto:2022fgx}
\bibinfo{author}{\bibfnamefont{F.~G.} \bibnamefont{Celiberto}}, \bibinfo{author}{\bibfnamefont{M.}~\bibnamefont{Fucilla}}, \bibinfo{author}{\bibfnamefont{D.~{\relax Yu}.} \bibnamefont{Ivanov}}, \bibinfo{author}{\bibfnamefont{M.~M.~A.} \bibnamefont{Mohammed}}, \bibnamefont{and} \bibinfo{author}{\bibfnamefont{A.}~\bibnamefont{Papa}}, \bibinfo{journal}{JHEP} \textbf{\bibinfo{volume}{08}}, \bibinfo{pages}{092} (\bibinfo{year}{2022}{\natexlab{b}}), \eprint{2205.02681}.

\bibitem[{\citenamefont{Celiberto et~al.}(2024{\natexlab{a}})\citenamefont{Celiberto, Delle~Rose, Fucilla, Gatto, and Papa}}]{Celiberto:2024bbv}
\bibinfo{author}{\bibfnamefont{F.~G.} \bibnamefont{Celiberto}}, \bibinfo{author}{\bibfnamefont{L.}~\bibnamefont{Delle~Rose}}, \bibinfo{author}{\bibfnamefont{M.}~\bibnamefont{Fucilla}}, \bibinfo{author}{\bibfnamefont{G.}~\bibnamefont{Gatto}}, \bibnamefont{and} \bibinfo{author}{\bibfnamefont{A.}~\bibnamefont{Papa}}, \bibinfo{journal}{JHEP} \textbf{\bibinfo{volume}{12}}, \bibinfo{pages}{061} (\bibinfo{year}{2024}{\natexlab{a}}), \eprint{2409.20354}.

\bibitem[{\citenamefont{Del~Duca and Falcioni}(2025)}]{DelDuca:2025vux}
\bibinfo{author}{\bibfnamefont{V.}~\bibnamefont{Del~Duca}} \bibnamefont{and} \bibinfo{author}{\bibfnamefont{G.}~\bibnamefont{Falcioni}}, \bibinfo{journal}{JHEP} \textbf{\bibinfo{volume}{07}}, \bibinfo{pages}{018} (\bibinfo{year}{2025}), \eprint{2504.06184}.

\bibitem[{\citenamefont{Celiberto}(2021{\natexlab{a}})}]{Celiberto:2020wpk}
\bibinfo{author}{\bibfnamefont{F.~G.} \bibnamefont{Celiberto}}, \bibinfo{journal}{Eur. Phys. J. C} \textbf{\bibinfo{volume}{81}}, \bibinfo{pages}{691} (\bibinfo{year}{2021}{\natexlab{a}}), \eprint{2008.07378}.

\bibitem[{\citenamefont{Celiberto}(2023{\natexlab{b}})}]{Celiberto:2023fzz}
\bibinfo{author}{\bibfnamefont{F.~G.} \bibnamefont{Celiberto}}, \bibinfo{journal}{Universe} \textbf{\bibinfo{volume}{9}}, \bibinfo{pages}{324} (\bibinfo{year}{2023}{\natexlab{b}}), \eprint{2305.14295}.

\bibitem[{\citenamefont{Celiberto}(2024{\natexlab{a}})}]{Celiberto:2024mrq}
\bibinfo{author}{\bibfnamefont{F.~G.} \bibnamefont{Celiberto}}, \bibinfo{journal}{Symmetry} \textbf{\bibinfo{volume}{16}}, \bibinfo{pages}{550} (\bibinfo{year}{2024}{\natexlab{a}}), \eprint{2403.15639}.

\bibitem[{\citenamefont{Celiberto}(2024{\natexlab{b}})}]{Celiberto:2024swu}
\bibinfo{author}{\bibfnamefont{F.~G.} \bibnamefont{Celiberto}}, \bibinfo{journal}{Particles} \textbf{\bibinfo{volume}{7}}, \bibinfo{pages}{502} (\bibinfo{year}{2024}{\natexlab{b}}), \eprint{2405.09526}.

\bibitem[{\citenamefont{Celiberto}(2026{\natexlab{a}})}]{Celiberto:2025csa}
\bibinfo{author}{\bibfnamefont{F.~G.} \bibnamefont{Celiberto}}, \bibinfo{journal}{Symmetry} \textbf{\bibinfo{volume}{18}}, \bibinfo{pages}{29} (\bibinfo{year}{2026}{\natexlab{a}}), \eprint{2604.01871}.

\bibitem[{\citenamefont{Celiberto}(2026{\natexlab{b}})}]{Celiberto:2026ooh}
\bibinfo{author}{\bibfnamefont{F.~G.} \bibnamefont{Celiberto}}, \bibinfo{journal}{Particles} \textbf{\bibinfo{volume}{9}}, \bibinfo{pages}{23} (\bibinfo{year}{2026}{\natexlab{b}}), \eprint{2604.13769}.

\bibitem[{\citenamefont{Celiberto et~al.}(2023)\citenamefont{Celiberto, Delle~Rose, Fucilla, Gatto, and Papa}}]{Celiberto:2023uuk}
\bibinfo{author}{\bibfnamefont{F.~G.} \bibnamefont{Celiberto}}, \bibinfo{author}{\bibfnamefont{L.}~\bibnamefont{Delle~Rose}}, \bibinfo{author}{\bibfnamefont{M.}~\bibnamefont{Fucilla}}, \bibinfo{author}{\bibfnamefont{G.}~\bibnamefont{Gatto}}, \bibnamefont{and} \bibinfo{author}{\bibfnamefont{A.}~\bibnamefont{Papa}}, in \emph{\bibinfo{booktitle}{{57th Rencontres de Moriond on QCD and High Energy Interactions}}} (\bibinfo{year}{2023}), \eprint{2305.05052}.

\bibitem[{\citenamefont{Celiberto et~al.}(2024{\natexlab{b}})\citenamefont{Celiberto, Delle~Rose, Fucilla, Gatto, and Papa}}]{Celiberto:2023eba}
\bibinfo{author}{\bibfnamefont{F.~G.} \bibnamefont{Celiberto}}, \bibinfo{author}{\bibfnamefont{L.}~\bibnamefont{Delle~Rose}}, \bibinfo{author}{\bibfnamefont{M.}~\bibnamefont{Fucilla}}, \bibinfo{author}{\bibfnamefont{G.}~\bibnamefont{Gatto}}, \bibnamefont{and} \bibinfo{author}{\bibfnamefont{A.}~\bibnamefont{Papa}}, \bibinfo{journal}{PoS} \textbf{\bibinfo{volume}{RADCOR2023}}, \bibinfo{pages}{069} (\bibinfo{year}{2024}{\natexlab{b}}), \eprint{2309.11573}.

\bibitem[{\citenamefont{Celiberto et~al.}(2024{\natexlab{c}})\citenamefont{Celiberto, Delle~Rose, Fucilla, Gatto, and Papa}}]{Celiberto:2023nym}
\bibinfo{author}{\bibfnamefont{F.~G.} \bibnamefont{Celiberto}}, \bibinfo{author}{\bibfnamefont{L.}~\bibnamefont{Delle~Rose}}, \bibinfo{author}{\bibfnamefont{M.}~\bibnamefont{Fucilla}}, \bibinfo{author}{\bibfnamefont{G.}~\bibnamefont{Gatto}}, \bibnamefont{and} \bibinfo{author}{\bibfnamefont{A.}~\bibnamefont{Papa}}, \bibinfo{journal}{PoS} \textbf{\bibinfo{volume}{EPS-HEP2023}}, \bibinfo{pages}{390} (\bibinfo{year}{2024}{\natexlab{c}}), \eprint{2310.16967}.

\bibitem[{\citenamefont{Mueller and Navelet}(1987)}]{Mueller:1986ey}
\bibinfo{author}{\bibfnamefont{A.~H.} \bibnamefont{Mueller}} \bibnamefont{and} \bibinfo{author}{\bibfnamefont{H.}~\bibnamefont{Navelet}}, \bibinfo{journal}{Nucl. Phys. B} \textbf{\bibinfo{volume}{282}}, \bibinfo{pages}{727} (\bibinfo{year}{1987}).

\bibitem[{\citenamefont{Duclou\'e et~al.}(2013)\citenamefont{Duclou\'e, Szymanowski, and Wallon}}]{Ducloue:2013hia}
\bibinfo{author}{\bibfnamefont{B.}~\bibnamefont{Duclou\'e}}, \bibinfo{author}{\bibfnamefont{L.}~\bibnamefont{Szymanowski}}, \bibnamefont{and} \bibinfo{author}{\bibfnamefont{S.}~\bibnamefont{Wallon}}, \bibinfo{journal}{JHEP} \textbf{\bibinfo{volume}{05}}, \bibinfo{pages}{096} (\bibinfo{year}{2013}), \eprint{1302.7012}.

\bibitem[{\citenamefont{Duclou\'e et~al.}(2014)\citenamefont{Duclou\'e, Szymanowski, and Wallon}}]{Ducloue:2013bva}
\bibinfo{author}{\bibfnamefont{B.}~\bibnamefont{Duclou\'e}}, \bibinfo{author}{\bibfnamefont{L.}~\bibnamefont{Szymanowski}}, \bibnamefont{and} \bibinfo{author}{\bibfnamefont{S.}~\bibnamefont{Wallon}}, \bibinfo{journal}{Phys. Rev. Lett.} \textbf{\bibinfo{volume}{112}}, \bibinfo{pages}{082003} (\bibinfo{year}{2014}), \eprint{1309.3229}.

\bibitem[{\citenamefont{Caporale et~al.}(2014)\citenamefont{Caporale, Ivanov, Murdaca, and Papa}}]{Caporale:2014gpa}
\bibinfo{author}{\bibfnamefont{F.}~\bibnamefont{Caporale}}, \bibinfo{author}{\bibfnamefont{D.~{\relax Yu}.} \bibnamefont{Ivanov}}, \bibinfo{author}{\bibfnamefont{B.}~\bibnamefont{Murdaca}}, \bibnamefont{and} \bibinfo{author}{\bibfnamefont{A.}~\bibnamefont{Papa}}, \bibinfo{journal}{Eur. Phys. J. C} \textbf{\bibinfo{volume}{74}}, \bibinfo{pages}{3084} (\bibinfo{year}{2014}), \bibinfo{note}{[Erratum: Eur.Phys.J.C 75, 535 (2015)]}, \eprint{1407.8431}.

\bibitem[{\citenamefont{Colferai and Niccoli}(2015)}]{Colferai:2015zfa}
\bibinfo{author}{\bibfnamefont{D.}~\bibnamefont{Colferai}} \bibnamefont{and} \bibinfo{author}{\bibfnamefont{A.}~\bibnamefont{Niccoli}}, \bibinfo{journal}{JHEP} \textbf{\bibinfo{volume}{04}}, \bibinfo{pages}{071} (\bibinfo{year}{2015}), \eprint{1501.07442}.

\bibitem[{\citenamefont{Celiberto et~al.}(2015{\natexlab{a}})\citenamefont{Celiberto, Ivanov, Murdaca, and Papa}}]{Celiberto:2015yba}
\bibinfo{author}{\bibfnamefont{F.~G.} \bibnamefont{Celiberto}}, \bibinfo{author}{\bibfnamefont{D.~{\relax Yu}.} \bibnamefont{Ivanov}}, \bibinfo{author}{\bibfnamefont{B.}~\bibnamefont{Murdaca}}, \bibnamefont{and} \bibinfo{author}{\bibfnamefont{A.}~\bibnamefont{Papa}}, \bibinfo{journal}{Eur. Phys. J. C} \textbf{\bibinfo{volume}{75}}, \bibinfo{pages}{292} (\bibinfo{year}{2015}{\natexlab{a}}), \eprint{1504.08233}.

\bibitem[{\citenamefont{Celiberto et~al.}(2016{\natexlab{a}})\citenamefont{Celiberto, Ivanov, Murdaca, and Papa}}]{Celiberto:2016ygs}
\bibinfo{author}{\bibfnamefont{F.~G.} \bibnamefont{Celiberto}}, \bibinfo{author}{\bibfnamefont{D.~{\relax Yu}.} \bibnamefont{Ivanov}}, \bibinfo{author}{\bibfnamefont{B.}~\bibnamefont{Murdaca}}, \bibnamefont{and} \bibinfo{author}{\bibfnamefont{A.}~\bibnamefont{Papa}}, \bibinfo{journal}{Eur. Phys. J. C} \textbf{\bibinfo{volume}{76}}, \bibinfo{pages}{224} (\bibinfo{year}{2016}{\natexlab{a}}), \eprint{1601.07847}.

\bibitem[{\citenamefont{Caporale et~al.}(2018)\citenamefont{Caporale, Celiberto, Chachamis, Gordo~G\'omez, and Sabio~Vera}}]{Caporale:2018qnm}
\bibinfo{author}{\bibfnamefont{F.}~\bibnamefont{Caporale}}, \bibinfo{author}{\bibfnamefont{F.~G.} \bibnamefont{Celiberto}}, \bibinfo{author}{\bibfnamefont{G.}~\bibnamefont{Chachamis}}, \bibinfo{author}{\bibfnamefont{D.}~\bibnamefont{Gordo~G\'omez}}, \bibnamefont{and} \bibinfo{author}{\bibfnamefont{A.}~\bibnamefont{Sabio~Vera}}, \bibinfo{journal}{Nucl. Phys. B} \textbf{\bibinfo{volume}{935}}, \bibinfo{pages}{412} (\bibinfo{year}{2018}), \eprint{1806.06309}.

\bibitem[{\citenamefont{Celiberto and Papa}(2022)}]{Celiberto:2022gji}
\bibinfo{author}{\bibfnamefont{F.~G.} \bibnamefont{Celiberto}} \bibnamefont{and} \bibinfo{author}{\bibfnamefont{A.}~\bibnamefont{Papa}}, \bibinfo{journal}{Phys. Rev. D} \textbf{\bibinfo{volume}{106}}, \bibinfo{pages}{114004} (\bibinfo{year}{2022}), \eprint{2207.05015}.

\bibitem[{\citenamefont{Celiberto et~al.}(2016{\natexlab{b}})\citenamefont{Celiberto, Ivanov, Murdaca, and Papa}}]{Celiberto:2016hae}
\bibinfo{author}{\bibfnamefont{F.~G.} \bibnamefont{Celiberto}}, \bibinfo{author}{\bibfnamefont{D.~{\relax Yu}.} \bibnamefont{Ivanov}}, \bibinfo{author}{\bibfnamefont{B.}~\bibnamefont{Murdaca}}, \bibnamefont{and} \bibinfo{author}{\bibfnamefont{A.}~\bibnamefont{Papa}}, \bibinfo{journal}{Phys. Rev. D} \textbf{\bibinfo{volume}{94}}, \bibinfo{pages}{034013} (\bibinfo{year}{2016}{\natexlab{b}}), \eprint{1604.08013}.

\bibitem[{\citenamefont{Celiberto et~al.}(2017)\citenamefont{Celiberto, Ivanov, Murdaca, and Papa}}]{Celiberto:2017ptm}
\bibinfo{author}{\bibfnamefont{F.~G.} \bibnamefont{Celiberto}}, \bibinfo{author}{\bibfnamefont{D.~{\relax Yu}.} \bibnamefont{Ivanov}}, \bibinfo{author}{\bibfnamefont{B.}~\bibnamefont{Murdaca}}, \bibnamefont{and} \bibinfo{author}{\bibfnamefont{A.}~\bibnamefont{Papa}}, \bibinfo{journal}{Eur. Phys. J. C} \textbf{\bibinfo{volume}{77}}, \bibinfo{pages}{382} (\bibinfo{year}{2017}), \eprint{1701.05077}.

\bibitem[{\citenamefont{Celiberto et~al.}(2020)\citenamefont{Celiberto, Ivanov, and Papa}}]{Celiberto:2020rxb}
\bibinfo{author}{\bibfnamefont{F.~G.} \bibnamefont{Celiberto}}, \bibinfo{author}{\bibfnamefont{D.~{\relax Yu}.} \bibnamefont{Ivanov}}, \bibnamefont{and} \bibinfo{author}{\bibfnamefont{A.}~\bibnamefont{Papa}}, \bibinfo{journal}{Phys. Rev. D} \textbf{\bibinfo{volume}{102}}, \bibinfo{pages}{094019} (\bibinfo{year}{2020}), \eprint{2008.10513}.

\bibitem[{\citenamefont{Bolognino et~al.}(2018)\citenamefont{Bolognino, Celiberto, Ivanov, Mohammed, and Papa}}]{Bolognino:2018oth}
\bibinfo{author}{\bibfnamefont{A.~D.} \bibnamefont{Bolognino}}, \bibinfo{author}{\bibfnamefont{F.~G.} \bibnamefont{Celiberto}}, \bibinfo{author}{\bibfnamefont{D.~{\relax Yu}.} \bibnamefont{Ivanov}}, \bibinfo{author}{\bibfnamefont{M.~M.~A.} \bibnamefont{Mohammed}}, \bibnamefont{and} \bibinfo{author}{\bibfnamefont{A.}~\bibnamefont{Papa}}, \bibinfo{journal}{Eur. Phys. J. C} \textbf{\bibinfo{volume}{78}}, \bibinfo{pages}{772} (\bibinfo{year}{2018}), \eprint{1808.05483}.

\bibitem[{\citenamefont{Celiberto}(2023{\natexlab{c}})}]{Celiberto:2022kxx}
\bibinfo{author}{\bibfnamefont{F.~G.} \bibnamefont{Celiberto}}, \bibinfo{journal}{Eur. Phys. J. C} \textbf{\bibinfo{volume}{83}}, \bibinfo{pages}{332} (\bibinfo{year}{2023}{\natexlab{c}}), \eprint{2208.14577}.

\bibitem[{\citenamefont{Celiberto and Fucilla}(2022)}]{Celiberto:2022dyf}
\bibinfo{author}{\bibfnamefont{F.~G.} \bibnamefont{Celiberto}} \bibnamefont{and} \bibinfo{author}{\bibfnamefont{M.}~\bibnamefont{Fucilla}}, \bibinfo{journal}{Eur. Phys. J. C} \textbf{\bibinfo{volume}{82}}, \bibinfo{pages}{929} (\bibinfo{year}{2022}), \eprint{2202.12227}.

\bibitem[{\citenamefont{Celiberto}(2022{\natexlab{b}})}]{Celiberto:2022keu}
\bibinfo{author}{\bibfnamefont{F.~G.} \bibnamefont{Celiberto}}, \bibinfo{journal}{Phys. Lett. B} \textbf{\bibinfo{volume}{835}}, \bibinfo{pages}{137554} (\bibinfo{year}{2022}{\natexlab{b}}), \eprint{2206.09413}.

\bibitem[{\citenamefont{Celiberto}(2024{\natexlab{c}})}]{Celiberto:2024omj}
\bibinfo{author}{\bibfnamefont{F.~G.} \bibnamefont{Celiberto}}, \bibinfo{journal}{Eur. Phys. J. C} \textbf{\bibinfo{volume}{84}}, \bibinfo{pages}{384} (\bibinfo{year}{2024}{\natexlab{c}}), \eprint{2401.01410}.

\bibitem[{\citenamefont{Celiberto and Lonigro}(2025)}]{Celiberto:2025euy}
\bibinfo{author}{\bibfnamefont{F.~G.} \bibnamefont{Celiberto}} \bibnamefont{and} \bibinfo{author}{\bibfnamefont{F.}~\bibnamefont{Lonigro}}, \bibinfo{journal}{Phys. Rev. D} \textbf{\bibinfo{volume}{112}}, \bibinfo{pages}{114040} (\bibinfo{year}{2025}), \eprint{2510.10593}.

\bibitem[{\citenamefont{Caporale et~al.}(2017)\citenamefont{Caporale, Celiberto, Chachamis, Gordo~G\'omez, and Sabio~Vera}}]{Caporale:2016zkc}
\bibinfo{author}{\bibfnamefont{F.}~\bibnamefont{Caporale}}, \bibinfo{author}{\bibfnamefont{F.~G.} \bibnamefont{Celiberto}}, \bibinfo{author}{\bibfnamefont{G.}~\bibnamefont{Chachamis}}, \bibinfo{author}{\bibfnamefont{D.}~\bibnamefont{Gordo~G\'omez}}, \bibnamefont{and} \bibinfo{author}{\bibfnamefont{A.}~\bibnamefont{Sabio~Vera}}, \bibinfo{journal}{Phys. Rev. D} \textbf{\bibinfo{volume}{95}}, \bibinfo{pages}{074007} (\bibinfo{year}{2017}), \eprint{1612.05428}.

\bibitem[{\citenamefont{Celiberto et~al.}(2018)\citenamefont{Celiberto, Gordo~G\'omez, and Sabio~Vera}}]{Celiberto:2018muu}
\bibinfo{author}{\bibfnamefont{F.~G.} \bibnamefont{Celiberto}}, \bibinfo{author}{\bibfnamefont{D.}~\bibnamefont{Gordo~G\'omez}}, \bibnamefont{and} \bibinfo{author}{\bibfnamefont{A.}~\bibnamefont{Sabio~Vera}}, \bibinfo{journal}{Phys. Lett.} \textbf{\bibinfo{volume}{B786}}, \bibinfo{pages}{201} (\bibinfo{year}{2018}), \eprint{1808.09511}.

\bibitem[{\citenamefont{Golec-Biernat et~al.}(2018)\citenamefont{Golec-Biernat, Motyka, and Stebel}}]{Golec-Biernat:2018kem}
\bibinfo{author}{\bibfnamefont{K.}~\bibnamefont{Golec-Biernat}}, \bibinfo{author}{\bibfnamefont{L.}~\bibnamefont{Motyka}}, \bibnamefont{and} \bibinfo{author}{\bibfnamefont{T.}~\bibnamefont{Stebel}}, \bibinfo{journal}{JHEP} \textbf{\bibinfo{volume}{12}}, \bibinfo{pages}{091} (\bibinfo{year}{2018}), \eprint{1811.04361}.

\bibitem[{\citenamefont{Binosi et~al.}(2009)\citenamefont{Binosi, Collins, Kaufhold, and Theussl}}]{Binosi:2008ig}
\bibinfo{author}{\bibfnamefont{D.}~\bibnamefont{Binosi}}, \bibinfo{author}{\bibfnamefont{J.}~\bibnamefont{Collins}}, \bibinfo{author}{\bibfnamefont{C.}~\bibnamefont{Kaufhold}}, \bibnamefont{and} \bibinfo{author}{\bibfnamefont{L.}~\bibnamefont{Theussl}}, \bibinfo{journal}{Comput. Phys. Commun.} \textbf{\bibinfo{volume}{180}}, \bibinfo{pages}{1709} (\bibinfo{year}{2009}), \eprint{0811.4113}.

\bibitem[{\citenamefont{Mele and Nason}(1991)}]{Mele:1990cw}
\bibinfo{author}{\bibfnamefont{B.}~\bibnamefont{Mele}} \bibnamefont{and} \bibinfo{author}{\bibfnamefont{P.}~\bibnamefont{Nason}}, \bibinfo{journal}{Nucl. Phys. B} \textbf{\bibinfo{volume}{361}}, \bibinfo{pages}{626} (\bibinfo{year}{1991}), \bibinfo{note}{[Erratum: Nucl.Phys.B 921, 841--842 (2017)]}.

\bibitem[{\citenamefont{Cacciari and Greco}(1994)}]{Cacciari:1993mq}
\bibinfo{author}{\bibfnamefont{M.}~\bibnamefont{Cacciari}} \bibnamefont{and} \bibinfo{author}{\bibfnamefont{M.}~\bibnamefont{Greco}}, \bibinfo{journal}{Nucl. Phys. B} \textbf{\bibinfo{volume}{421}}, \bibinfo{pages}{530} (\bibinfo{year}{1994}), \eprint{hep-ph/9311260}.

\bibitem[{\citenamefont{Buza et~al.}(1998)\citenamefont{Buza, Matiounine, Smith, and van Neerven}}]{Buza:1996wv}
\bibinfo{author}{\bibfnamefont{M.}~\bibnamefont{Buza}}, \bibinfo{author}{\bibfnamefont{Y.}~\bibnamefont{Matiounine}}, \bibinfo{author}{\bibfnamefont{J.}~\bibnamefont{Smith}}, \bibnamefont{and} \bibinfo{author}{\bibfnamefont{W.~L.} \bibnamefont{van Neerven}}, \bibinfo{journal}{Eur. Phys. J. C} \textbf{\bibinfo{volume}{1}}, \bibinfo{pages}{301} (\bibinfo{year}{1998}), \eprint{hep-ph/9612398}.

\bibitem[{\citenamefont{Navas et~al.}(2024)}]{ParticleDataGroup:2024cfk}
\bibinfo{author}{\bibfnamefont{S.}~\bibnamefont{Navas}} \bibnamefont{et~al.} (\bibinfo{collaboration}{Particle Data Group}), \bibinfo{journal}{Phys. Rev. D} \textbf{\bibinfo{volume}{110}}, \bibinfo{pages}{030001} (\bibinfo{year}{2024}).

\bibitem[{\citenamefont{Bardeen et~al.}(1978)\citenamefont{Bardeen, Buras, Duke, and Muta}}]{PhysRevD.18.3998}
\bibinfo{author}{\bibfnamefont{W.~A.} \bibnamefont{Bardeen}}, \bibinfo{author}{\bibfnamefont{A.~J.} \bibnamefont{Buras}}, \bibinfo{author}{\bibfnamefont{D.~W.} \bibnamefont{Duke}}, \bibnamefont{and} \bibinfo{author}{\bibfnamefont{T.}~\bibnamefont{Muta}}, \bibinfo{journal}{Phys. Rev. D} \textbf{\bibinfo{volume}{18}}, \bibinfo{pages}{3998} (\bibinfo{year}{1978}).

\bibitem[{\citenamefont{Celiberto et~al.}(2015{\natexlab{b}})\citenamefont{Celiberto, Ivanov, Murdaca, and Papa}}]{Celiberto:2015mpa}
\bibinfo{author}{\bibfnamefont{F.~G.} \bibnamefont{Celiberto}}, \bibinfo{author}{\bibfnamefont{D.~{\relax Yu}.} \bibnamefont{Ivanov}}, \bibinfo{author}{\bibfnamefont{B.}~\bibnamefont{Murdaca}}, \bibnamefont{and} \bibinfo{author}{\bibfnamefont{A.}~\bibnamefont{Papa}}, \bibinfo{journal}{Acta Phys. Polon. Supp.} \textbf{\bibinfo{volume}{8}}, \bibinfo{pages}{935} (\bibinfo{year}{2015}{\natexlab{b}}), \eprint{1510.01626}.

\bibitem[{\citenamefont{Kotikov and Lipatov}(2000)}]{Kotikov:2000pm}
\bibinfo{author}{\bibfnamefont{A.~V.} \bibnamefont{Kotikov}} \bibnamefont{and} \bibinfo{author}{\bibfnamefont{L.~N.} \bibnamefont{Lipatov}}, \bibinfo{journal}{Nucl. Phys. B} \textbf{\bibinfo{volume}{582}}, \bibinfo{pages}{19} (\bibinfo{year}{2000}), \eprint{hep-ph/0004008}.

\bibitem[{\citenamefont{Del~Duca et~al.}(2003)\citenamefont{Del~Duca, Kilgore, Oleari, Schmidt, and Zeppenfeld}}]{DelDuca:2003ba}
\bibinfo{author}{\bibfnamefont{V.}~\bibnamefont{Del~Duca}}, \bibinfo{author}{\bibfnamefont{W.}~\bibnamefont{Kilgore}}, \bibinfo{author}{\bibfnamefont{C.}~\bibnamefont{Oleari}}, \bibinfo{author}{\bibfnamefont{C.~R.} \bibnamefont{Schmidt}}, \bibnamefont{and} \bibinfo{author}{\bibfnamefont{D.}~\bibnamefont{Zeppenfeld}}, \bibinfo{journal}{Phys. Rev. D} \textbf{\bibinfo{volume}{67}}, \bibinfo{pages}{073003} (\bibinfo{year}{2003}), \eprint{hep-ph/0301013}.

\bibitem[{\citenamefont{Ivanov and Papa}(2012)}]{Ivanov:2012iv}
\bibinfo{author}{\bibfnamefont{D.~{\relax Yu}.} \bibnamefont{Ivanov}} \bibnamefont{and} \bibinfo{author}{\bibfnamefont{A.}~\bibnamefont{Papa}}, \bibinfo{journal}{JHEP} \textbf{\bibinfo{volume}{07}}, \bibinfo{pages}{045} (\bibinfo{year}{2012}), \eprint{1205.6068}.

\bibitem[{\citenamefont{Buckley et~al.}(2015)\citenamefont{Buckley, Ferrando, Lloyd, Nordstr\"om, Page, R\"ufenacht, Sch\"onherr, and Watt}}]{Buckley:2014ana}
\bibinfo{author}{\bibfnamefont{A.}~\bibnamefont{Buckley}}, \bibinfo{author}{\bibfnamefont{J.}~\bibnamefont{Ferrando}}, \bibinfo{author}{\bibfnamefont{S.}~\bibnamefont{Lloyd}}, \bibinfo{author}{\bibfnamefont{K.}~\bibnamefont{Nordstr\"om}}, \bibinfo{author}{\bibfnamefont{B.}~\bibnamefont{Page}}, \bibinfo{author}{\bibfnamefont{M.}~\bibnamefont{R\"ufenacht}}, \bibinfo{author}{\bibfnamefont{M.}~\bibnamefont{Sch\"onherr}}, \bibnamefont{and} \bibinfo{author}{\bibfnamefont{G.}~\bibnamefont{Watt}}, \bibinfo{journal}{Eur. Phys. J. C} \textbf{\bibinfo{volume}{75}}, \bibinfo{pages}{132} (\bibinfo{year}{2015}), \eprint{1412.7420}.

\bibitem[{\citenamefont{Kniehl and Kramer}(2005)}]{Kniehl:2005de}
\bibinfo{author}{\bibfnamefont{B.~A.} \bibnamefont{Kniehl}} \bibnamefont{and} \bibinfo{author}{\bibfnamefont{G.}~\bibnamefont{Kramer}}, \bibinfo{journal}{Phys. Rev. D} \textbf{\bibinfo{volume}{71}}, \bibinfo{pages}{094013} (\bibinfo{year}{2005}), \eprint{hep-ph/0504058}.

\bibitem[{\citenamefont{Kneesch et~al.}(2008)\citenamefont{Kneesch, Kniehl, Kramer, and Schienbein}}]{Kneesch:2007ey}
\bibinfo{author}{\bibfnamefont{T.}~\bibnamefont{Kneesch}}, \bibinfo{author}{\bibfnamefont{B.~A.} \bibnamefont{Kniehl}}, \bibinfo{author}{\bibfnamefont{G.}~\bibnamefont{Kramer}}, \bibnamefont{and} \bibinfo{author}{\bibfnamefont{I.}~\bibnamefont{Schienbein}}, \bibinfo{journal}{Nucl. Phys. B} \textbf{\bibinfo{volume}{799}}, \bibinfo{pages}{34} (\bibinfo{year}{2008}), \eprint{0712.0481}.

\bibitem[{\citenamefont{Kniehl et~al.}(2020)\citenamefont{Kniehl, Kramer, Schienbein, and Spiesberger}}]{Kniehl:2020szu}
\bibinfo{author}{\bibfnamefont{B.~A.} \bibnamefont{Kniehl}}, \bibinfo{author}{\bibfnamefont{G.}~\bibnamefont{Kramer}}, \bibinfo{author}{\bibfnamefont{I.}~\bibnamefont{Schienbein}}, \bibnamefont{and} \bibinfo{author}{\bibfnamefont{H.}~\bibnamefont{Spiesberger}}, \bibinfo{journal}{Phys. Rev. D} \textbf{\bibinfo{volume}{101}}, \bibinfo{pages}{114021} (\bibinfo{year}{2020}), \eprint{2004.04213}.

\bibitem[{\citenamefont{Bowler}(1981)}]{Bowler:1981sb}
\bibinfo{author}{\bibfnamefont{M.~G.} \bibnamefont{Bowler}}, \bibinfo{journal}{Z. Phys. C} \textbf{\bibinfo{volume}{11}}, \bibinfo{pages}{169} (\bibinfo{year}{1981}).

\bibitem[{\citenamefont{Peterson et~al.}(1983)\citenamefont{Peterson, Schlatter, Schmitt, and Zerwas}}]{Peterson:1982ak}
\bibinfo{author}{\bibfnamefont{C.}~\bibnamefont{Peterson}}, \bibinfo{author}{\bibfnamefont{D.}~\bibnamefont{Schlatter}}, \bibinfo{author}{\bibfnamefont{I.}~\bibnamefont{Schmitt}}, \bibnamefont{and} \bibinfo{author}{\bibfnamefont{P.~M.} \bibnamefont{Zerwas}}, \bibinfo{journal}{Phys. Rev. D} \textbf{\bibinfo{volume}{27}}, \bibinfo{pages}{105} (\bibinfo{year}{1983}).

\bibitem[{\citenamefont{Kartvelishvili and Likhoded}(1985)}]{Kartvelishvili:1985ac}
\bibinfo{author}{\bibfnamefont{V.~G.} \bibnamefont{Kartvelishvili}} \bibnamefont{and} \bibinfo{author}{\bibfnamefont{A.~K.} \bibnamefont{Likhoded}}, \bibinfo{journal}{Yad. Fiz.} \textbf{\bibinfo{volume}{42}}, \bibinfo{pages}{1306} (\bibinfo{year}{1985}).

\bibitem[{\citenamefont{Bertone et~al.}(2017)\citenamefont{Bertone, Carrazza, Hartland, Nocera, and Rojo}}]{Bertone:2017tyb}
\bibinfo{author}{\bibfnamefont{V.}~\bibnamefont{Bertone}}, \bibinfo{author}{\bibfnamefont{S.}~\bibnamefont{Carrazza}}, \bibinfo{author}{\bibfnamefont{N.~P.} \bibnamefont{Hartland}}, \bibinfo{author}{\bibfnamefont{E.~R.} \bibnamefont{Nocera}}, \bibnamefont{and} \bibinfo{author}{\bibfnamefont{J.}~\bibnamefont{Rojo}} (\bibinfo{collaboration}{NNPDF}), \bibinfo{journal}{Eur. Phys. J. C} \textbf{\bibinfo{volume}{77}}, \bibinfo{pages}{516} (\bibinfo{year}{2017}), \eprint{1706.07049}.

\bibitem[{\citenamefont{Ball et~al.}(2021)}]{NNPDF:2021uiq}
\bibinfo{author}{\bibfnamefont{R.~D.} \bibnamefont{Ball}} \bibnamefont{et~al.} (\bibinfo{collaboration}{NNPDF}), \bibinfo{journal}{Eur. Phys. J. C} \textbf{\bibinfo{volume}{81}}, \bibinfo{pages}{958} (\bibinfo{year}{2021}), \eprint{2109.02671}.

\bibitem[{\citenamefont{Ball et~al.}(2022)}]{NNPDF:2021njg}
\bibinfo{author}{\bibfnamefont{R.~D.} \bibnamefont{Ball}} \bibnamefont{et~al.} (\bibinfo{collaboration}{NNPDF}), \bibinfo{journal}{Eur. Phys. J. C} \textbf{\bibinfo{volume}{82}}, \bibinfo{pages}{428} (\bibinfo{year}{2022}), \eprint{2109.02653}.

\bibitem[{\citenamefont{Celiberto and Papa}(2024)}]{Celiberto:2023rzw}
\bibinfo{author}{\bibfnamefont{F.~G.} \bibnamefont{Celiberto}} \bibnamefont{and} \bibinfo{author}{\bibfnamefont{A.}~\bibnamefont{Papa}}, \bibinfo{journal}{Phys. Lett. B} \textbf{\bibinfo{volume}{848}}, \bibinfo{pages}{138406} (\bibinfo{year}{2024}), \eprint{2308.00809}.

\bibitem[{\citenamefont{Celiberto et~al.}(2024{\natexlab{d}})\citenamefont{Celiberto, Gatto, and Papa}}]{Celiberto:2024mab}
\bibinfo{author}{\bibfnamefont{F.~G.} \bibnamefont{Celiberto}}, \bibinfo{author}{\bibfnamefont{G.}~\bibnamefont{Gatto}}, \bibnamefont{and} \bibinfo{author}{\bibfnamefont{A.}~\bibnamefont{Papa}}, \bibinfo{journal}{Eur. Phys. J. C} \textbf{\bibinfo{volume}{84}}, \bibinfo{pages}{1071} (\bibinfo{year}{2024}{\natexlab{d}}), \eprint{2405.14773}.

\bibitem[{\citenamefont{Celiberto and Gatto}(2025)}]{Celiberto:2024beg}
\bibinfo{author}{\bibfnamefont{F.~G.} \bibnamefont{Celiberto}} \bibnamefont{and} \bibinfo{author}{\bibfnamefont{G.}~\bibnamefont{Gatto}}, \bibinfo{journal}{Phys. Rev. D} \textbf{\bibinfo{volume}{111}}, \bibinfo{pages}{034037} (\bibinfo{year}{2025}), \eprint{2412.10549}.

\bibitem[{\citenamefont{Celiberto}(2025{\natexlab{a}})}]{Celiberto:2025dfe}
\bibinfo{author}{\bibfnamefont{F.~G.} \bibnamefont{Celiberto}}, \bibinfo{journal}{Phys. Rev. D} \textbf{\bibinfo{volume}{111}}, \bibinfo{pages}{L111501} (\bibinfo{year}{2025}{\natexlab{a}}), \eprint{2504.03949}.

\bibitem[{\citenamefont{Celiberto}(2025{\natexlab{b}})}]{Celiberto:2025ziy}
\bibinfo{author}{\bibfnamefont{F.~G.} \bibnamefont{Celiberto}}, \bibinfo{journal}{Phys. Rev. D} \textbf{\bibinfo{volume}{112}}, \bibinfo{pages}{074041} (\bibinfo{year}{2025}{\natexlab{b}}), \eprint{2507.09744}.

\bibitem[{\citenamefont{Celiberto et~al.}(2026{\natexlab{a}})\citenamefont{Celiberto, Giannini, Gon{\c{c}}alves, and Lima}}]{Celiberto:2025vra}
\bibinfo{author}{\bibfnamefont{F.~G.} \bibnamefont{Celiberto}}, \bibinfo{author}{\bibfnamefont{A.~V.} \bibnamefont{Giannini}}, \bibinfo{author}{\bibfnamefont{V.~P.} \bibnamefont{Gon{\c{c}}alves}}, \bibnamefont{and} \bibinfo{author}{\bibfnamefont{Y.~N.} \bibnamefont{Lima}}, \bibinfo{journal}{Phys. Rev. D} \textbf{\bibinfo{volume}{113}}, \bibinfo{pages}{054014} (\bibinfo{year}{2026}{\natexlab{a}}), \eprint{2511.18984}.

\bibitem[{\citenamefont{Celiberto}(2026{\natexlab{c}})}]{Celiberto:2026kks}
\bibinfo{author}{\bibfnamefont{F.~G.} \bibnamefont{Celiberto}}, \bibinfo{journal}{Phys. Rev. D} \textbf{\bibinfo{volume}{113}}, \bibinfo{pages}{114037} (\bibinfo{year}{2026}{\natexlab{c}}), \eprint{2604.11646}.

\bibitem[{\citenamefont{Celiberto}(2025{\natexlab{c}})}]{Celiberto:2025ipt}
\bibinfo{author}{\bibfnamefont{F.~G.} \bibnamefont{Celiberto}}, \bibinfo{journal}{Eur. Phys. J. C} \textbf{\bibinfo{volume}{85}}, \bibinfo{pages}{1395} (\bibinfo{year}{2025}{\natexlab{c}}), \eprint{2502.11136}.

\bibitem[{\citenamefont{Celiberto}(2026{\natexlab{d}})}]{Celiberto:2026rdk}
\bibinfo{author}{\bibfnamefont{F.~G.} \bibnamefont{Celiberto}}, \bibinfo{journal}{Phys. Rev. D} \textbf{\bibinfo{volume}{114}}, \bibinfo{pages}{014009} (\bibinfo{year}{2026}{\natexlab{d}}), \eprint{2605.01539}.

\bibitem[{\citenamefont{Celiberto}(2025{\natexlab{d}})}]{Celiberto:2025ogy}
\bibinfo{author}{\bibfnamefont{F.~G.} \bibnamefont{Celiberto}}, \bibinfo{journal}{Phys. Rev. D} \textbf{\bibinfo{volume}{112}}, \bibinfo{pages}{074023} (\bibinfo{year}{2025}{\natexlab{d}}), \eprint{2506.00776}.

\bibitem[{\citenamefont{Celiberto}(2026{\natexlab{e}})}]{Celiberto:2026qiz}
\bibinfo{author}{\bibfnamefont{F.~G.} \bibnamefont{Celiberto}}, \bibinfo{journal}{Phys. Rev. D} \textbf{\bibinfo{volume}{114}}, \bibinfo{pages}{054038} (\bibinfo{year}{2026}{\natexlab{e}}), \eprint{2606.20063}.

\bibitem[{\citenamefont{Bolognino et~al.}(2021)\citenamefont{Bolognino, Celiberto, Fucilla, Ivanov, and Papa}}]{Bolognino:2021mrc}
\bibinfo{author}{\bibfnamefont{A.~D.} \bibnamefont{Bolognino}}, \bibinfo{author}{\bibfnamefont{F.~G.} \bibnamefont{Celiberto}}, \bibinfo{author}{\bibfnamefont{M.}~\bibnamefont{Fucilla}}, \bibinfo{author}{\bibfnamefont{D.~{\relax Yu}.} \bibnamefont{Ivanov}}, \bibnamefont{and} \bibinfo{author}{\bibfnamefont{A.}~\bibnamefont{Papa}}, \bibinfo{journal}{Phys. Rev. D} \textbf{\bibinfo{volume}{103}}, \bibinfo{pages}{094004} (\bibinfo{year}{2021}), \eprint{2103.07396}.

\bibitem[{\citenamefont{Celiberto}(2026{\natexlab{f}})}]{Celiberto:2026_nst}
\bibinfo{author}{\bibfnamefont{F.~G.} \bibnamefont{Celiberto}} (\bibinfo{year}{2026}{\natexlab{f}}), \bibinfo{note}{work in progress}.

\bibitem[{\citenamefont{Epele et~al.}(2026)\citenamefont{Epele, Hekhorn, Helenius, Paukkunen, and Zurita}}]{Epele:2026dia}
\bibinfo{author}{\bibfnamefont{M.}~\bibnamefont{Epele}}, \bibinfo{author}{\bibfnamefont{F.}~\bibnamefont{Hekhorn}}, \bibinfo{author}{\bibfnamefont{I.}~\bibnamefont{Helenius}}, \bibinfo{author}{\bibfnamefont{H.}~\bibnamefont{Paukkunen}}, \bibnamefont{and} \bibinfo{author}{\bibfnamefont{P.}~\bibnamefont{Zurita}}, in \emph{\bibinfo{booktitle}{{33rd International Workshop on Deep Inelastic Scattering and Related Subjects}}} (\bibinfo{year}{2026}), \eprint{2609.10327}.

\bibitem[{\citenamefont{Salam}(1998)}]{Salam:1998tj}
\bibinfo{author}{\bibfnamefont{G.~P.} \bibnamefont{Salam}}, \bibinfo{journal}{JHEP} \textbf{\bibinfo{volume}{07}}, \bibinfo{pages}{019} (\bibinfo{year}{1998}), \eprint{hep-ph/9806482}.

\bibitem[{\citenamefont{Ciafaloni et~al.}(2004)\citenamefont{Ciafaloni, Colferai, Salam, and Stasto}}]{Ciafaloni:2003kd}
\bibinfo{author}{\bibfnamefont{M.}~\bibnamefont{Ciafaloni}}, \bibinfo{author}{\bibfnamefont{D.}~\bibnamefont{Colferai}}, \bibinfo{author}{\bibfnamefont{G.~P.} \bibnamefont{Salam}}, \bibnamefont{and} \bibinfo{author}{\bibfnamefont{A.~M.} \bibnamefont{Stasto}}, \bibinfo{journal}{Phys. Lett. B} \textbf{\bibinfo{volume}{587}}, \bibinfo{pages}{87} (\bibinfo{year}{2004}), \eprint{hep-ph/0311325}.

\bibitem[{\citenamefont{Ciafaloni et~al.}(2003)\citenamefont{Ciafaloni, Colferai, Salam, and Stasto}}]{Ciafaloni:2003rd}
\bibinfo{author}{\bibfnamefont{M.}~\bibnamefont{Ciafaloni}}, \bibinfo{author}{\bibfnamefont{D.}~\bibnamefont{Colferai}}, \bibinfo{author}{\bibfnamefont{G.~P.} \bibnamefont{Salam}}, \bibnamefont{and} \bibinfo{author}{\bibfnamefont{A.~M.} \bibnamefont{Stasto}}, \bibinfo{journal}{Phys. Rev. D} \textbf{\bibinfo{volume}{68}}, \bibinfo{pages}{114003} (\bibinfo{year}{2003}), \eprint{hep-ph/0307188}.

\bibitem[{\citenamefont{Sabio~Vera}(2005)}]{SabioVera:2005tiv}
\bibinfo{author}{\bibfnamefont{A.}~\bibnamefont{Sabio~Vera}}, \bibinfo{journal}{Nucl. Phys. B} \textbf{\bibinfo{volume}{722}}, \bibinfo{pages}{65} (\bibinfo{year}{2005}), \eprint{hep-ph/0505128}.

\bibitem[{\citenamefont{Khachatryan et~al.}(2016)}]{Khachatryan:2016udy}
\bibinfo{author}{\bibfnamefont{V.}~\bibnamefont{Khachatryan}} \bibnamefont{et~al.} (\bibinfo{collaboration}{CMS}), \bibinfo{journal}{JHEP} \textbf{\bibinfo{volume}{08}}, \bibinfo{pages}{139} (\bibinfo{year}{2016}), \eprint{1601.06713}.

\bibitem[{\citenamefont{Sabio~Vera}(2006)}]{Vera:2006un}
\bibinfo{author}{\bibfnamefont{A.}~\bibnamefont{Sabio~Vera}}, \bibinfo{journal}{Nucl. Phys. B} \textbf{\bibinfo{volume}{746}}, \bibinfo{pages}{1} (\bibinfo{year}{2006}), \eprint{hep-ph/0602250}.

\bibitem[{\citenamefont{Sabio~Vera and Schwennsen}(2007)}]{Vera:2007kn}
\bibinfo{author}{\bibfnamefont{A.}~\bibnamefont{Sabio~Vera}} \bibnamefont{and} \bibinfo{author}{\bibfnamefont{F.}~\bibnamefont{Schwennsen}}, \bibinfo{journal}{Nucl. Phys. B} \textbf{\bibinfo{volume}{776}}, \bibinfo{pages}{170} (\bibinfo{year}{2007}), \eprint{hep-ph/0702158}.

\bibitem[{\citenamefont{Marquet and Royon}(2009)}]{Marquet:2007xx}
\bibinfo{author}{\bibfnamefont{C.}~\bibnamefont{Marquet}} \bibnamefont{and} \bibinfo{author}{\bibfnamefont{C.}~\bibnamefont{Royon}}, \bibinfo{journal}{Phys. Rev. D} \textbf{\bibinfo{volume}{79}}, \bibinfo{pages}{034028} (\bibinfo{year}{2009}), \eprint{0704.3409}.

\bibitem[{\citenamefont{Cieri et~al.}(2015)\citenamefont{Cieri, Coradeschi, and de~Florian}}]{Cieri:2015rqa}
\bibinfo{author}{\bibfnamefont{L.}~\bibnamefont{Cieri}}, \bibinfo{author}{\bibfnamefont{F.}~\bibnamefont{Coradeschi}}, \bibnamefont{and} \bibinfo{author}{\bibfnamefont{D.}~\bibnamefont{de~Florian}}, \bibinfo{journal}{JHEP} \textbf{\bibinfo{volume}{06}}, \bibinfo{pages}{185} (\bibinfo{year}{2015}), \eprint{1505.03162}.

\bibitem[{\citenamefont{Alioli et~al.}(2021)\citenamefont{Alioli, Broggio, Gavardi, Kallweit, Lim, Nagar, Napoletano, and Rottoli}}]{Alioli:2020qrd}
\bibinfo{author}{\bibfnamefont{S.}~\bibnamefont{Alioli}}, \bibinfo{author}{\bibfnamefont{A.}~\bibnamefont{Broggio}}, \bibinfo{author}{\bibfnamefont{A.}~\bibnamefont{Gavardi}}, \bibinfo{author}{\bibfnamefont{S.}~\bibnamefont{Kallweit}}, \bibinfo{author}{\bibfnamefont{M.~A.} \bibnamefont{Lim}}, \bibinfo{author}{\bibfnamefont{R.}~\bibnamefont{Nagar}}, \bibinfo{author}{\bibfnamefont{D.}~\bibnamefont{Napoletano}}, \bibnamefont{and} \bibinfo{author}{\bibfnamefont{L.}~\bibnamefont{Rottoli}}, \bibinfo{journal}{JHEP} \textbf{\bibinfo{volume}{04}}, \bibinfo{pages}{041} (\bibinfo{year}{2021}), \eprint{2010.10498}.

\bibitem[{\citenamefont{Becher and Neumann}(2021)}]{Becher:2020ugp}
\bibinfo{author}{\bibfnamefont{T.}~\bibnamefont{Becher}} \bibnamefont{and} \bibinfo{author}{\bibfnamefont{T.}~\bibnamefont{Neumann}}, \bibinfo{journal}{JHEP} \textbf{\bibinfo{volume}{03}}, \bibinfo{pages}{199} (\bibinfo{year}{2021}), \eprint{2009.11437}.

\bibitem[{\citenamefont{Neumann}(2021)}]{Neumann:2021zkb}
\bibinfo{author}{\bibfnamefont{T.}~\bibnamefont{Neumann}}, \bibinfo{journal}{Eur. Phys. J. C} \textbf{\bibinfo{volume}{81}}, \bibinfo{pages}{905} (\bibinfo{year}{2021}), \eprint{2107.12478}.

\bibitem[{\citenamefont{Ferrera and Pires}(2017)}]{Ferrera:2016prr}
\bibinfo{author}{\bibfnamefont{G.}~\bibnamefont{Ferrera}} \bibnamefont{and} \bibinfo{author}{\bibfnamefont{J.}~\bibnamefont{Pires}}, \bibinfo{journal}{JHEP} \textbf{\bibinfo{volume}{02}}, \bibinfo{pages}{139} (\bibinfo{year}{2017}), \eprint{1609.01691}.

\bibitem[{\citenamefont{Ju and Sch\"onherr}(2021)}]{Ju:2021lah}
\bibinfo{author}{\bibfnamefont{W.-L.} \bibnamefont{Ju}} \bibnamefont{and} \bibinfo{author}{\bibfnamefont{M.}~\bibnamefont{Sch\"onherr}}, \bibinfo{journal}{JHEP} \textbf{\bibinfo{volume}{10}}, \bibinfo{pages}{088} (\bibinfo{year}{2021}), \eprint{2106.11260}.

\bibitem[{\citenamefont{Monni et~al.}(2020)\citenamefont{Monni, Rottoli, and Torrielli}}]{Monni:2019yyr}
\bibinfo{author}{\bibfnamefont{P.~F.} \bibnamefont{Monni}}, \bibinfo{author}{\bibfnamefont{L.}~\bibnamefont{Rottoli}}, \bibnamefont{and} \bibinfo{author}{\bibfnamefont{P.}~\bibnamefont{Torrielli}}, \bibinfo{journal}{Phys. Rev. Lett.} \textbf{\bibinfo{volume}{124}}, \bibinfo{pages}{252001} (\bibinfo{year}{2020}), \eprint{1909.04704}.

\bibitem[{\citenamefont{Buonocore et~al.}(2022)\citenamefont{Buonocore, Grazzini, Haag, and Rottoli}}]{Buonocore:2021akg}
\bibinfo{author}{\bibfnamefont{L.}~\bibnamefont{Buonocore}}, \bibinfo{author}{\bibfnamefont{M.}~\bibnamefont{Grazzini}}, \bibinfo{author}{\bibfnamefont{J.}~\bibnamefont{Haag}}, \bibnamefont{and} \bibinfo{author}{\bibfnamefont{L.}~\bibnamefont{Rottoli}}, \bibinfo{journal}{Eur. Phys. J. C} \textbf{\bibinfo{volume}{82}}, \bibinfo{pages}{27} (\bibinfo{year}{2022}), \eprint{2110.06913}.

\bibitem[{\citenamefont{Wiesemann et~al.}(2020)\citenamefont{Wiesemann, Rottoli, and Torrielli}}]{Wiesemann:2020gbm}
\bibinfo{author}{\bibfnamefont{M.}~\bibnamefont{Wiesemann}}, \bibinfo{author}{\bibfnamefont{L.}~\bibnamefont{Rottoli}}, \bibnamefont{and} \bibinfo{author}{\bibfnamefont{P.}~\bibnamefont{Torrielli}}, \bibinfo{journal}{Phys. Lett. B} \textbf{\bibinfo{volume}{809}}, \bibinfo{pages}{135718} (\bibinfo{year}{2020}), \eprint{2006.09338}.

\bibitem[{\citenamefont{Ebert et~al.}(2021)\citenamefont{Ebert, Michel, Stewart, and Tackmann}}]{Ebert:2020dfc}
\bibinfo{author}{\bibfnamefont{M.~A.} \bibnamefont{Ebert}}, \bibinfo{author}{\bibfnamefont{J.~K.~L.} \bibnamefont{Michel}}, \bibinfo{author}{\bibfnamefont{I.~W.} \bibnamefont{Stewart}}, \bibnamefont{and} \bibinfo{author}{\bibfnamefont{F.~J.} \bibnamefont{Tackmann}}, \bibinfo{journal}{JHEP} \textbf{\bibinfo{volume}{04}}, \bibinfo{pages}{102} (\bibinfo{year}{2021}), \eprint{2006.11382}.

\bibitem[{\citenamefont{Re et~al.}(2021)\citenamefont{Re, Rottoli, and Torrielli}}]{Re:2021con}
\bibinfo{author}{\bibfnamefont{E.}~\bibnamefont{Re}}, \bibinfo{author}{\bibfnamefont{L.}~\bibnamefont{Rottoli}}, \bibnamefont{and} \bibinfo{author}{\bibfnamefont{P.}~\bibnamefont{Torrielli}}, \bibinfo{journal}{JHEP} \textbf{\bibinfo{volume}{09}}, \bibinfo{pages}{108} (\bibinfo{year}{2021}), \eprint{2104.07509}.

\bibitem[{\citenamefont{Chen et~al.}(2022)\citenamefont{Chen, Gehrmann, Glover, Huss, Monni, Re, Rottoli, and Torrielli}}]{Chen:2022cgv}
\bibinfo{author}{\bibfnamefont{X.}~\bibnamefont{Chen}}, \bibinfo{author}{\bibfnamefont{T.}~\bibnamefont{Gehrmann}}, \bibinfo{author}{\bibfnamefont{E.~W.~N.} \bibnamefont{Glover}}, \bibinfo{author}{\bibfnamefont{A.}~\bibnamefont{Huss}}, \bibinfo{author}{\bibfnamefont{P.~F.} \bibnamefont{Monni}}, \bibinfo{author}{\bibfnamefont{E.}~\bibnamefont{Re}}, \bibinfo{author}{\bibfnamefont{L.}~\bibnamefont{Rottoli}}, \bibnamefont{and} \bibinfo{author}{\bibfnamefont{P.}~\bibnamefont{Torrielli}}, \bibinfo{journal}{Phys. Rev. Lett.} \textbf{\bibinfo{volume}{128}}, \bibinfo{pages}{252001} (\bibinfo{year}{2022}), \eprint{2203.01565}.

\bibitem[{\citenamefont{Neumann and Campbell}(2023)}]{Neumann:2022lft}
\bibinfo{author}{\bibfnamefont{T.}~\bibnamefont{Neumann}} \bibnamefont{and} \bibinfo{author}{\bibfnamefont{J.}~\bibnamefont{Campbell}}, \bibinfo{journal}{Phys. Rev. D} \textbf{\bibinfo{volume}{107}}, \bibinfo{pages}{L011506} (\bibinfo{year}{2023}), \eprint{2207.07056}.

\bibitem[{\citenamefont{Bizon et~al.}(2018)\citenamefont{Bizon, Monni, Re, Rottoli, and Torrielli}}]{Bizon:2017rah}
\bibinfo{author}{\bibfnamefont{W.}~\bibnamefont{Bizon}}, \bibinfo{author}{\bibfnamefont{P.~F.} \bibnamefont{Monni}}, \bibinfo{author}{\bibfnamefont{E.}~\bibnamefont{Re}}, \bibinfo{author}{\bibfnamefont{L.}~\bibnamefont{Rottoli}}, \bibnamefont{and} \bibinfo{author}{\bibfnamefont{P.}~\bibnamefont{Torrielli}}, \bibinfo{journal}{JHEP} \textbf{\bibinfo{volume}{02}}, \bibinfo{pages}{108} (\bibinfo{year}{2018}), \eprint{1705.09127}.

\bibitem[{\citenamefont{Billis et~al.}(2021)\citenamefont{Billis, Dehnadi, Ebert, Michel, and Tackmann}}]{Billis:2021ecs}
\bibinfo{author}{\bibfnamefont{G.}~\bibnamefont{Billis}}, \bibinfo{author}{\bibfnamefont{B.}~\bibnamefont{Dehnadi}}, \bibinfo{author}{\bibfnamefont{M.~A.} \bibnamefont{Ebert}}, \bibinfo{author}{\bibfnamefont{J.~K.~L.} \bibnamefont{Michel}}, \bibnamefont{and} \bibinfo{author}{\bibfnamefont{F.~J.} \bibnamefont{Tackmann}}, \bibinfo{journal}{Phys. Rev. Lett.} \textbf{\bibinfo{volume}{127}}, \bibinfo{pages}{072001} (\bibinfo{year}{2021}), \eprint{2102.08039}.

\bibitem[{\citenamefont{Caola et~al.}(2022)\citenamefont{Caola, Chen, Duhr, Liu, Mistlberger, Petriello, Vita, and Weinzierl}}]{Caola:2022ayt}
\bibinfo{author}{\bibfnamefont{F.}~\bibnamefont{Caola}}, \bibinfo{author}{\bibfnamefont{W.}~\bibnamefont{Chen}}, \bibinfo{author}{\bibfnamefont{C.}~\bibnamefont{Duhr}}, \bibinfo{author}{\bibfnamefont{X.}~\bibnamefont{Liu}}, \bibinfo{author}{\bibfnamefont{B.}~\bibnamefont{Mistlberger}}, \bibinfo{author}{\bibfnamefont{F.}~\bibnamefont{Petriello}}, \bibinfo{author}{\bibfnamefont{G.}~\bibnamefont{Vita}}, \bibnamefont{and} \bibinfo{author}{\bibfnamefont{S.}~\bibnamefont{Weinzierl}}, in \emph{\bibinfo{booktitle}{{2022 Snowmass Summer Study}}} (\bibinfo{year}{2022}), \eprint{2203.06730}.

\bibitem[{\citenamefont{Kallweit et~al.}(2020)\citenamefont{Kallweit, Re, Rottoli, and Wiesemann}}]{Kallweit:2020gva}
\bibinfo{author}{\bibfnamefont{S.}~\bibnamefont{Kallweit}}, \bibinfo{author}{\bibfnamefont{E.}~\bibnamefont{Re}}, \bibinfo{author}{\bibfnamefont{L.}~\bibnamefont{Rottoli}}, \bibnamefont{and} \bibinfo{author}{\bibfnamefont{M.}~\bibnamefont{Wiesemann}}, \bibinfo{journal}{JHEP} \textbf{\bibinfo{volume}{12}}, \bibinfo{pages}{147} (\bibinfo{year}{2020}), \eprint{2004.07720}.

\bibitem[{\citenamefont{Boer et~al.}(2012)\citenamefont{Boer, den Dunnen, Pisano, Schlegel, and Vogelsang}}]{Boer:2011kf}
\bibinfo{author}{\bibfnamefont{D.}~\bibnamefont{Boer}}, \bibinfo{author}{\bibfnamefont{W.~J.} \bibnamefont{den Dunnen}}, \bibinfo{author}{\bibfnamefont{C.}~\bibnamefont{Pisano}}, \bibinfo{author}{\bibfnamefont{M.}~\bibnamefont{Schlegel}}, \bibnamefont{and} \bibinfo{author}{\bibfnamefont{W.}~\bibnamefont{Vogelsang}}, \bibinfo{journal}{Phys. Rev. Lett.} \textbf{\bibinfo{volume}{108}}, \bibinfo{pages}{032002} (\bibinfo{year}{2012}), \eprint{1109.1444}.

\bibitem[{\citenamefont{Boer and Pisano}(2015)}]{Boer:2014lka}
\bibinfo{author}{\bibfnamefont{D.}~\bibnamefont{Boer}} \bibnamefont{and} \bibinfo{author}{\bibfnamefont{C.}~\bibnamefont{Pisano}}, \bibinfo{journal}{Phys. Rev. D} \textbf{\bibinfo{volume}{91}}, \bibinfo{pages}{074024} (\bibinfo{year}{2015}), \eprint{1412.5556}.

\bibitem[{\citenamefont{Echevarria et~al.}(2015)\citenamefont{Echevarria, Kasemets, Mulders, and Pisano}}]{Echevarria:2015uaa}
\bibinfo{author}{\bibfnamefont{M.~G.} \bibnamefont{Echevarria}}, \bibinfo{author}{\bibfnamefont{T.}~\bibnamefont{Kasemets}}, \bibinfo{author}{\bibfnamefont{P.~J.} \bibnamefont{Mulders}}, \bibnamefont{and} \bibinfo{author}{\bibfnamefont{C.}~\bibnamefont{Pisano}}, \bibinfo{journal}{JHEP} \textbf{\bibinfo{volume}{07}}, \bibinfo{pages}{158} (\bibinfo{year}{2015}), \bibinfo{note}{[Erratum: JHEP05,073(2017)]}, \eprint{1502.05354}.

\bibitem[{\citenamefont{Gutierrez-Reyes et~al.}(2019)\citenamefont{Gutierrez-Reyes, Leal-Gomez, Scimemi, and Vladimirov}}]{Gutierrez-Reyes:2019rug}
\bibinfo{author}{\bibfnamefont{D.}~\bibnamefont{Gutierrez-Reyes}}, \bibinfo{author}{\bibfnamefont{S.}~\bibnamefont{Leal-Gomez}}, \bibinfo{author}{\bibfnamefont{I.}~\bibnamefont{Scimemi}}, \bibnamefont{and} \bibinfo{author}{\bibfnamefont{A.}~\bibnamefont{Vladimirov}}, \bibinfo{journal}{JHEP} \textbf{\bibinfo{volume}{11}}, \bibinfo{pages}{121} (\bibinfo{year}{2019}), \eprint{1907.03780}.

\bibitem[{\citenamefont{Bacchetta et~al.}(2020)\citenamefont{Bacchetta, Celiberto, Radici, and Taels}}]{Bacchetta:2020vty}
\bibinfo{author}{\bibfnamefont{A.}~\bibnamefont{Bacchetta}}, \bibinfo{author}{\bibfnamefont{F.~G.} \bibnamefont{Celiberto}}, \bibinfo{author}{\bibfnamefont{M.}~\bibnamefont{Radici}}, \bibnamefont{and} \bibinfo{author}{\bibfnamefont{P.}~\bibnamefont{Taels}}, \bibinfo{journal}{Eur. Phys. J. C} \textbf{\bibinfo{volume}{80}}, \bibinfo{pages}{733} (\bibinfo{year}{2020}), \eprint{2005.02288}.

\bibitem[{\citenamefont{Celiberto}(2021{\natexlab{b}})}]{Celiberto:2021zww}
\bibinfo{author}{\bibfnamefont{F.~G.} \bibnamefont{Celiberto}}, \bibinfo{journal}{Nuovo Cim.} \textbf{\bibinfo{volume}{C44}}, \bibinfo{pages}{36} (\bibinfo{year}{2021}{\natexlab{b}}), \eprint{2101.04630}.

\bibitem[{\citenamefont{Bacchetta et~al.}(2024)\citenamefont{Bacchetta, Celiberto, and Radici}}]{Bacchetta:2024fci}
\bibinfo{author}{\bibfnamefont{A.}~\bibnamefont{Bacchetta}}, \bibinfo{author}{\bibfnamefont{F.~G.} \bibnamefont{Celiberto}}, \bibnamefont{and} \bibinfo{author}{\bibfnamefont{M.}~\bibnamefont{Radici}}, \bibinfo{journal}{Eur. Phys. J. C} \textbf{\bibinfo{volume}{84}}, \bibinfo{pages}{576} (\bibinfo{year}{2024}), \eprint{2402.17556}.

\bibitem[{\citenamefont{Anedda et~al.}(2026)\citenamefont{Anedda, Bertone, Bozzi, and Cerutti}}]{Anedda:2026cox}
\bibinfo{author}{\bibfnamefont{S.}~\bibnamefont{Anedda}}, \bibinfo{author}{\bibfnamefont{V.}~\bibnamefont{Bertone}}, \bibinfo{author}{\bibfnamefont{G.}~\bibnamefont{Bozzi}}, \bibnamefont{and} \bibinfo{author}{\bibfnamefont{M.}~\bibnamefont{Cerutti}}, \bibinfo{journal}{Eur. Phys. J. C} \textbf{\bibinfo{volume}{86}}, \bibinfo{pages}{1024} (\bibinfo{year}{2026}), \eprint{2605.28216}.

\bibitem[{\citenamefont{Celiberto et~al.}(2025)\citenamefont{Celiberto, Delle~Rose, Fucilla, Gatto, and Papa}}]{Celiberto:2024mdt}
\bibinfo{author}{\bibfnamefont{F.~G.} \bibnamefont{Celiberto}}, \bibinfo{author}{\bibfnamefont{L.}~\bibnamefont{Delle~Rose}}, \bibinfo{author}{\bibfnamefont{M.}~\bibnamefont{Fucilla}}, \bibinfo{author}{\bibfnamefont{G.}~\bibnamefont{Gatto}}, \bibnamefont{and} \bibinfo{author}{\bibfnamefont{A.}~\bibnamefont{Papa}}, \bibinfo{journal}{PoS} \textbf{\bibinfo{volume}{DIS2024}}, \bibinfo{pages}{126} (\bibinfo{year}{2025}), \eprint{2408.08757}.

\bibitem[{\citenamefont{Celiberto et~al.}(2026{\natexlab{b}})\citenamefont{Celiberto, Delle~Rose, and Papa}}]{Celiberto:2025edg}
\bibinfo{author}{\bibfnamefont{F.~G.} \bibnamefont{Celiberto}}, \bibinfo{author}{\bibfnamefont{L.}~\bibnamefont{Delle~Rose}}, \bibnamefont{and} \bibinfo{author}{\bibfnamefont{A.}~\bibnamefont{Papa}}, \bibinfo{journal}{PoS} \textbf{\bibinfo{volume}{EPS-HEP2025}}, \bibinfo{pages}{352} (\bibinfo{year}{2026}{\natexlab{b}}), \eprint{2510.22445}.

\bibitem[{\citenamefont{Celiberto and Lonigro}(2026{\natexlab{a}})}]{Celiberto:2026mjs}
\bibinfo{author}{\bibfnamefont{F.~G.} \bibnamefont{Celiberto}} \bibnamefont{and} \bibinfo{author}{\bibfnamefont{F.}~\bibnamefont{Lonigro}}, in \emph{\bibinfo{booktitle}{{14th Large Hadron Collider Physics Conference}}} (\bibinfo{year}{2026}{\natexlab{a}}), \eprint{2609.01203}.

\bibitem[{\citenamefont{Celiberto and Lonigro}(2026{\natexlab{b}})}]{Celiberto:2026dwf}
\bibinfo{author}{\bibfnamefont{F.~G.} \bibnamefont{Celiberto}} \bibnamefont{and} \bibinfo{author}{\bibfnamefont{F.}~\bibnamefont{Lonigro}}, in \emph{\bibinfo{booktitle}{{33rd International Workshop on Deep Inelastic Scattering and Related Subjects}}} (\bibinfo{year}{2026}{\natexlab{b}}), \eprint{2609.04395}.

\end{thebibliography}

\end{document}